\documentclass[journal]{IEEEtran} 
\usepackage[dvips]{graphicx}
\usepackage{enumerate}
\usepackage{amsmath}
\usepackage{amssymb}
\usepackage{cite}
\usepackage{booktabs}

\usepackage{bbding}
\usepackage{color}
\usepackage{amsthm}
\usepackage{multirow}
\usepackage{subfigure} 
\usepackage[ruled,linesnumbered]{algorithm2e}
\def\QEDclosed{\mbox{\rule[0pt]{1.3ex}{1.3ex}}}
\def\QED{\QEDclosed}
\def\proof{\emph{Proof:} }
\def\endproof{\hspace*{\fill}~\QED\par\endtrivlist\unskip}

\newcommand{\bea}{\begin{eqnarray}}
\newcommand{\eea}{\end{eqnarray}}
\newcommand{\tabincell}[2]{\begin{tabular}{@{}#1@{}}#2\end{tabular}}
\ifCLASSINFOpdf
\else
\fi
\begin{document}

\title{Secure Control of Networked Control Systems Using Dynamic Watermarking}

\author{Dajun Du,
        Changda Zhang,
        Xue Li,
        Minrui Fei,
        Taicheng Yang,
        Huiyu Zhou
\thanks{The work of D. Du, C. Zhang, X. Li, M. Fei, T. Yang, and H. Zhou was supported by the National Science Foundation of China under Grant Nos. 92067106, 61773253, 61633016 and 61533010, 111 Project under Grant No.D18003.}
\thanks{D. Du, C. Zhang, X. Li, and M. Fei are with Shanghai Key Laboratory of Power Station Automation Technology, School of Mechatronic Engineering and Automation, Shanghai University, Shanghai 200444, China.}
\thanks{T. Yang is with Department of Engineering and Design, University of Sussex, Brighton BN1 9QT, U.K.}
\thanks{H. Zhou is with School of Informatics, University of Leicester, Leicester LE1 7RH, U.K.}}

%
%

\markboth{IEEE TRANSACTIONS ON CYBERNETICS}%
{Shell \MakeLowercase{\textit{et al.}}: Bare Demo of IEEEtran.cls for IEEE Journals}
%



\maketitle

\begin{abstract}
We here investigate secure control of networked control systems developing a new dynamic watermarking (DW) scheme. Firstly, the weaknesses of the conventional DW scheme are revealed, and the tradeoff between the effectiveness of false data injection attack (FDIA) detection and system performance loss is analysed. Secondly, we propose a new DW scheme, and its attack detection capability is interrogated using the additive distortion power of a closed-loop system. Furthermore, the FDIA detection effectiveness of the closed-loop system is analysed using auto/cross covariance of the signals, where the positive correlation between the FDIA detection effectiveness and the watermarking intensity is measured. Thirdly, the tolerance capacity of FDIA against the closed-loop system is investigated, and theoretical analysis shows that the system performance can be recovered from FDIA using our new DW scheme. Finally, experimental results from a networked inverted pendulum system demonstrate the validity of our proposed scheme.
\end{abstract}


\IEEEpeerreviewmaketitle

\section{Introduction}
With the rapid popularization and application of network technologies, the coming decades may witness the extensive deployment of networked control systems (NCSs). Such systems are often embedded by physical plants and digital devices (e.g., digital filter and controller), which are linked by communication networks. Smart grids \cite{SEG}, Internet of Things \cite{IOT} and networked robots \cite{ROB} are examples of NCSs. Yet, nowadays it is easier to access the network by malicious users, and NCSs are vulnerable to cyber attacks such as denial-of-service attack \cite{COF}, replay attack \cite{RAT,SAET}, and false data injection attack (FDIA) \cite{UEG,FDI,SDF}. Insecure and attacked NCSs may suffer fateful consequences including huge economic losses, just as the attacks on the Iran nuclear plant \cite{RAT} and Ukraine electric grid \cite{UEG} demonstrated. In this context, it is not surprising that secure control of NCSs has attracted widespread attention with emerging attack detection \cite{SAD} and resilient control \cite{RCD}.

Based on the use of the probing signals, attack detection can be roughly classified into passive and active detection. Passive detection does not inject the probing signal into the system, which is  sometimes invalid for some sophisticated attacks such as replay attack \cite{PWM1} and FDIA \cite{OLA}. To solve the problem, active detection is developed by injecting the probing signal into the system, and watermarking-based detection is a kind of typical active detection. It is motivated by the traditional digital watermarking (i.e., the digital code) that is embedded in electronic documents, which is employed to preserve the valuable information \cite{DTW}. According to signal sources of generation, watermarking-based detection can be roughly divided into two categories: additive and multiplicative watermarking. A typical example of additive watermarking is termed as physical watermarking or dynamic watermarking (DW). Their basic concept is that the control law is actively encrypted by injecting certain probing signal (e.g., an independent identically distributed \cite{PWM1,PWM2,PWM3} or stationary \cite{PWM4} Gaussian signal), and specific tests (e.g., $\chi^2$ test \cite{PWM1} or DW tests \cite{DWM1,DWM2,DWM3}) are performed to infer a malicious activity. Physical watermarking \cite{PWM1} is designed particularly for preventing replay attack, which is also driven to investigate the tradeoff between the watermarking intensity and replay attack detection effectiveness. As an evolution of physical watermarking, DW \cite{DWM1} is designed to yield the security property, which has been further improved for general linear time-invariant systems \cite{DWM4} and time-varying systems \cite{DWM5}.

However, physical watermarking and DW accompany non-zero system performance loss for attack detection. Indeed, the lower watermarking intensity is cautiously chosen in the system to ensure less system performance loss, where the worse attack detection effectiveness is determined. Against these limitations, multiplicative watermarking has been developed, and a typical example is termed as sensor multiplicative watermarking \cite{SMW1,SMW2,SMW3}. In such watermarking scheme, each sensor output is separately encrypted by an infinite/finite impulse response filter \cite{SMW1}, and the encrypted sensor output is decrypted by an equalizing filter \cite{SMW2,SMW3}. With the advantage of zero system performance loss from the watermarking scheme, sensor multiplicative watermarking is also originally designed for replay attack \cite{SMW1} and later developed for FDIA (e.g., routing attack \cite{SMW2} as well as man-in-the-middle attack \cite{SMW3}). But then, to the best of our knowledge, there is no theoretical guarantee of the security property using DW for sensor multiplicative watermarking. Therefore, to this end, the design of a new comprehensive watermarking scheme combining the security property of DW and the advantage of sensor multiplicative watermarking is developed.

At a high level, resilient control has been investigated in several research works, which can be roughly classified into detection-independent and detection-dependent control. Inspired from fault-tolerant control \cite{FTCC,RFTC,ANO}, detection-independent control examines the tolerance capacity of attacks against systems; e.g., the characterization of maximum perturbation from FDIA is posed as reachable set computation \cite{TO1} and an FDIA tolerance principle is established based on adaptively truncating the injection channels of attacks \cite{ASM}. Note that even though the tolerance capacity of attacks against systems is examined, the system experiences performance loss, which calls for detection-dependent control. The idea of detection-dependent control is that once there is an attack to be alarmed, mechanism-depending detection can be used to recover system performance. For example, modification of the attacked signals \cite{DRM} and the compensation mechanisms \cite{ADMM,WNCS,DAD} are conducted, where the malicious system outputs are discarded. However, to the best of our knowledge, few research studies concerns with resilient control based on the watermarking schemes. Therefore, an additional desire is to design a mechanism for the proposed watermarking scheme to recover the system performance under attacks.

Motivated by the above observations, this paper looks at a new DW scheme. Specifically, the following challenges will be addressed:
\begin{enumerate}
  \item [(1)]
  What are the security weaknesses of the conventional DW scheme?
  \item [(2)]
  How to design a new DW scheme against the security weaknesses? What are the security property and the attack detection performance of the new DW scheme?
  \item [(3)]
  What is the recovery capability of NCSs based on the new DW scheme?
\end{enumerate}

To deal with these challenges, this paper explores a new DW scheme under cyber attacks. The main contributions of this paper are summarized as follows:
\begin{enumerate}
  \item [(1)]
  The security weaknesses of the conventional DW scheme are revealed, and a tradeoff between the FDIA detection effectiveness and system performance loss is explored. In addition, when a low watermarking intensity has to be chosen, there exists an FDIA that cannot be effectively detected and causes instability by the conventional DW scheme.
  \item [(2)]
  A new DW scheme is proposed by integrating the watermarking as symmetric-key encryption and new DW testing and compensation, where attacks can be detected using the additive distortion power of the closed-loop system. Furthermore, FDIA detection effectiveness of the closed-loop system is analysed using auto/cross covariance of signals, and the positive correlation between FDIA detection effectiveness and watermarking intensity is analysed.
  \item [(3)]
  The tolerance capacity of FDIA against the closed-loop system is investigated, and it is shown that system performance can be recovered from FDIA, where the quantitative relationship between the new DW scheme and the system performance is revealed.
\end{enumerate}

To clearly present the contributions of this paper, compared with existing results in the literatures, a comparative analysis is listed in Tab.~\ref{Tabcomp}. It is shown from Tab.~\ref{Tabcomp} that the existing results have only solved part of the problems of attack detection and resilient control, but the proposed new DW scheme can generate a comprehensive solution of the problems by enhancing the attack detection effectiveness, developing the security property, guaranteeing zero system performance loss from watermarking, and providing the attack tolerance capacity as well as system performance recovery.
\begin{table}[t]
\centering
\caption{Comparative Analysis between the Contributions of This Paper and the Existing Results in the Literatures}
\label{Tabcomp}
\begin{tabular}{lccccc}
  \toprule
  ~& ADEW$^1$ & SPAW$^2$ & ZPLW$^3$ & ATCA$^4$ & SPR$^5$\\
  \midrule
  \tabincell{l}{CDW$^6$\\ \cite{DWM1,PWM1,PWM2,PWM3,PWM4,DWM2,DWM3,DWM4,DWM5}} & \CheckmarkBold   & \CheckmarkBold & \XSolidBrush     &\XSolidBrush      & \XSolidBrush \\
  \tabincell{l}{SMW$^7$\\ \cite{SMW1,SMW2,SMW3}} & \CheckmarkBold   & \XSolidBrush   & \CheckmarkBold   &\XSolidBrush      & \XSolidBrush \\
  \tabincell{l}{DIC$^8$\\ \cite{TO1,ASM}} & \XSolidBrush     & \XSolidBrush & \CheckmarkBold &\CheckmarkBold &\XSolidBrush\\
  \tabincell{l}{DDC$^9$\\ \cite{DRM,ADMM,WNCS,DAD}} & \XSolidBrush     & \XSolidBrush & \CheckmarkBold               &\CheckmarkBold    & \CheckmarkBold\\
  \tabincell{l}{New DW\\ (this paper)} & \CheckmarkBold   & \CheckmarkBold & \CheckmarkBold   & \CheckmarkBold   & \CheckmarkBold \\
  \bottomrule
  \multicolumn{6}{l}{$^1$ Attack detection effectiveness enhanced by watermarking.}\\
  \multicolumn{6}{l}{$^2$ Security property analysis from CDW.}\\
  \multicolumn{6}{l}{$^3$ Zero system performance loss from watermarking.}\\
  \multicolumn{6}{l}{$^4$ Attack tolerance capacity analysis. $^5$ System performance recovery.}\\
  \multicolumn{6}{l}{$^6$ Conventional DW. $^7$ Sensor multiplicative watermarking.}\\
  \multicolumn{6}{l}{$^8$ Detection-independent control. $^9$ Detection-dependent control.}\\
\end{tabular}
\end{table}

The remainder of this paper is organized as follows. Section \uppercase\expandafter{\romannumeral+2} contains problem formulation, focusing on the security property and weaknesses of the conventional DW scheme. Section \uppercase\expandafter{\romannumeral+3} presents secure control of NCSs based on a new DW scheme, where the design of the new DW scheme, security property and processes are presented. Experimental results for an inverted pendulum system are given in Section \uppercase\expandafter{\romannumeral+4}, followed by conclusion shown in Section \uppercase\expandafter{\romannumeral+5}.

\emph{Notation:} $\mathbb{E}[\cdot]$ represents the expectation of a vector or matrix, $A_{i\cdot}$ ($A_{\cdot i}$) is used for the sub-matrix of a given matrix $A$ formed from row (column) $i$, $tr(\cdot)$ denotes the trace of a matrix, $\rho(\cdot)$ denotes the spectral radius of a matrix, $\left\|\cdot\right\|$ denotes the Euclidian norm of a vector or the spectral norm of a matrix, $\left\|\cdot\right\|_{F}$ denotes the Frobenius norm of a matrix, $\left|\cdot\right|$ denotes the absolute value of a real number, and $\sup \mathbb{R}$ ($\inf \mathbb{R}$) denotes the supremum (infimum) of a real set $\mathbb{R}\subseteq \mathcal{R}$. The zero vector is denoted as $0_{n}\in \mathcal{R}^{n}$, whilst $0_{n\times m} \in \mathcal{R}^{n \times m}$ and $ I_{n\times n} \in \mathcal{R}^{n \times n}$ indicate respectively the zero and identity matrices. For simplicity, the subscripts are often omitted if their dimensions are clear. Table~\ref{TabNota} summarizes the notations most frequently used throughout the remainder of the paper.

\begin{table}[t]
\centering
\caption{Table of Notations}
\label{TabNota}
\begin{tabular}{rcl}
  \hline
  $T$ & $\triangleq$ & Time window size for detection \\
  $\mathcal{W}_d$, $\mathcal{V}_d$ & $\triangleq$  & Conventional DW (asymptotic) tests 1, 2\\
  $\varphi_{d,1}$, $\varphi_{d,2}$ & $\triangleq$  & Conventional DW statistical tests 1, 2\\
  $\vartheta_{d,1}$, $\vartheta_{d,2}$ & $\triangleq$  & Thresholds for DW statistical tests 1, 2\\
  $\mathcal{D}_{r_d}$                  & $\triangleq$  & Additional distortion on residual $r_d$\\
  $x_a$                                & $\triangleq$  & State of the false data injection attack\\
  $J^{o}$ & $\triangleq$  & System performance under no attack \\
  ${\left. {{J^o}} \right|_{{w_d}(k) = 0}}$ & $\triangleq$  & $J^{o}$ under no DW scheme \\
  $\mathcal{W}_i$, $\mathcal{V}$ & $\triangleq$  & New DW (asymptotic) tests 1, 2\\
  $\varphi_{1,i}$, $\varphi_{2}$ & $\triangleq$  & New DW statistical tests 1, 2\\
  $\vartheta_{1,i}$, $\vartheta_{2}$ & $\triangleq$  & Thresholds for new DW statistical tests 1, 2\\
  $\mathcal{D}_{r}$                  & $\triangleq$  & Additional distortion on residual $r$\\
  ${\left. {{J^o}} \right|_{{w_y}(k) = 0}}$ & $\triangleq$  & $J^{o}$ under no new DW scheme\\
  $\varepsilon$ & $\triangleq$  & Detection results being $0$ or $1$\\
  $h$           & $\triangleq$  & Delay of healthy system output\\
  $e$                  & $\triangleq$  & Estimation error\\
  \hline
\end{tabular}
\end{table}

\section{Problem Formulation}

\subsection{Security Property Analysis of the Conventional DW Scheme}
\begin{figure}[!t]
  \centering
  \includegraphics[width=0.485\textwidth]{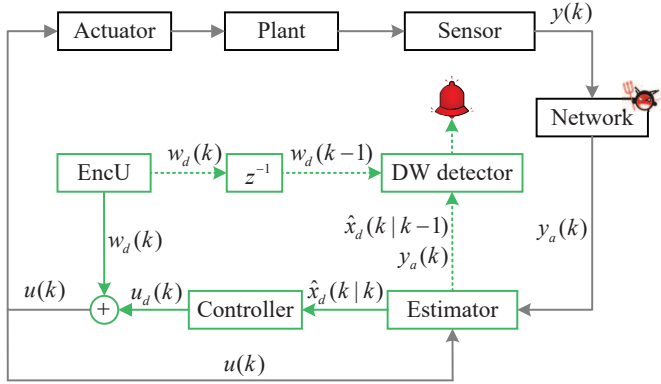}
  \caption{Framework of secure control of NCSs based on the conventional Dynamic Watermarking (DW) scheme \cite{DWM1,PWM1}.}
  \label{fig1}
\end{figure}
The framework of secure control of NCSs based on the conventional DW scheme is illustrated in Fig.~\ref{fig1}. The system output $y(k)$  is first sampled and transmitted to the estimator via a network, which may be attacked, notated as $y_a(k)$. Note that when there exist cyber attacks, $y_a(k) \ne y(k)$; otherwise $y_a(k)=y(k)$. Then, the estimator has the state estimate $\hat x_d(k|k)$ by using $y_a(k)$, which is used to compute the controller signal $u_d(k)$ in the controller node. Furthermore, a watermarking signal $w_d(k)$ from the watermarking generator EncU is injected into $u_d(k)$, resulting in $u(k)$, which is applied to the plant by the actuator. Meanwhile, using $w_d(k-1)$ from the EncU and $\hat x_d(k|k-1)$ and $y_a(k)$ from the estimator, the DW detector detects whether or not $y_a(k)$ is attacked; if an attack takes place, the alarm will be triggered.

To analyse the security property and weaknesses, the discrete-time linear time-invariant plant is considered as follows
\bea
\left\{ \begin{gathered}
  x(k+1) = Ax(k) + B u(k) + \Gamma n(k) \hfill \\
  y(k) = C x(k) + v(k) \hfill \\
\end{gathered}  \right.
\label{eq2A1}
\eea
where $x(k) \in \mathcal{R}^{m_x}$ is the system state; $u(k) \in \mathcal{R}$ is the control input; $y(k) \in \mathcal{R}^{m_y}$ is the system output; the process noise $n(k) \in \mathcal{R}^{m_n}$ and measurement noise $v(k) \in \mathcal{R}^{m_y}$ are independent identically distributed (i.i.d.) zero-mean white Gaussians with covariance matrices $\Sigma_n $ and $\Sigma_v$ respectively, and mutually independent. $A$, $B$, $\Gamma$, $C$ are constant matrices with appropriate dimensions.

Next, to estimate the system state for measuring control input and attack detection, the estimator adopts the steady-state Kalman filter by
\begin{align}
  &\hat x_d(k|k) = \hat x_d(k|k - 1) + L({y_a}(k) - C\hat x_d(k|k - 1)) \hfill \label{eq2A2},\\
  &\hat x_d(k + 1|k) = A\hat x_d(k|k) + Bu(k) \hfill \label{eq2A3}
\end{align}
where $L= {P}{C^T} \Sigma_o^{-1}$ is the steady-state Kalman gain; $\Sigma_o = {C{P}{C^T} + {\Sigma _v}} $, and $P$ is the unique positive definite solution of the Riccati equation \cite{PWM1}.

According to $\hat x_d(k|k)$, the linear quadratic Gaussian controller is implemented to minimize the objective function $J$ with matrices $Q,R \geqslant 0$ \cite{PWM1}. It is well known that if there is no attacks (i.e., $y_a(k) \equiv y(k)$), the solution of the minimization problem on $J$ will lead to a fixed gain controller:
\bea
u_d(k) = K \hat x_d(k|k)
\label{eq2A4}
\eea
where ${K} =  - {\left( {B^T{S}{B} + R} \right)^{ - 1}}B^T{S}{A}$ is the controller gain, and $S$ is the unique positive definite solution of the Riccati equation \cite{PWM1}. Furthermore, $u_d(k)$ is injected with a watermarking signal $w_d(k)$ by EncU, i.e.,
\bea
u(k) = u_d(k) + w_d(k)
\label{eq2A5}
\eea
where $w_d(k)$ is drawn from an i.i.d. Gaussian distribution with zero mean and variance $\sigma^2_{w_d}$, and $w_d(k)$ is chosen to be also independent of $u_d(k)$ \cite{PWM1}.

Furthermore, back to the side of the estimator, its signals $\hat x_d(k|k-1)$ and $y_a(k)$ are transferred to the DW detector to evaluate whether or not the attack takes place. Therefore, using $w_d(k-1)$, $\hat x_d(k|k-1)$ and $y_a(k)$, two DW tests \cite{DWM1} are operated in the DW detector, i.e.,
\begin{enumerate}
  \item
  DW Test 1: checking whether or not
  \bea
  \mathop {\lim }\limits_{T \to \infty } \mathcal{W}_d(T) = 0
  \label{eq2A6}
  \eea
  where $\mathcal{W}_d(T) := \frac{1}{T}\sum\nolimits_{k = 1}^T {{w_d}(k - 1)L{r_d}(k)}$; $T$ is the time window size; $r_d(k)$ is the residual, i.e., $r_d(k) := y_a(k) - C\hat x_d(k|k-1)$.
  \item
  DW Test 2: checking whether or not
  \bea
  \mathop {\lim }\limits_{T \to \infty } \mathcal{V}_d(T) = 0
  \label{eq2A7}
  \eea
  where $\mathcal{V}_d(T) := \frac{1}{T}\sum\nolimits_{k = 1}^T {L{r_d}(k){{\left( {L{r_d}(k)} \right)}^T}}  - L{\Sigma _o}{L^T}$; $\Sigma_o$ is the same as the one in (\ref{eq2A2}) and also the covariance matrix of the normal residual.
\end{enumerate}

\emph{Remark 1:} From (\ref{eq2A6}) and (\ref{eq2A7}), it is clear that applying directly DW Tests 1 and 2 in real world is unrealistic owing to the limit $T \to \infty$. To solve the problem, two DW statistical tests need to be revisited in a finite $T$. For instance, two indicators have been defined in \cite{DWM2} as ${\varphi _{d,1}}(k) := {\left\| \mathcal{W}^{k}_d(T) \right\|_F}$ and ${\varphi _{d,2}}(k) := \left| tr \left(\mathcal{V}^{k}_d(T) \right) \right|$, where $\mathcal{W}^{k}_d(T)$ and $\mathcal{V}^{k}_d(T)$ are $\mathcal{W}_d(T)$ and $\mathcal{V}_d(T)$ calculated at the current time window $\{k-T+1,k-T+2,\ldots,k\}$ respectively. Furthermore, to evaluate whether or not the attack takes place, let ${\vartheta _{d,1}}$, ${\vartheta _{d,2}}$ be the preset thresholds; when an attack takes place, we expect ${\varphi _{d,1}}(k) \geqslant {\vartheta _{d,1}}$ or ${\varphi _{d,2}}(k) \geqslant {\vartheta _{d,2}}$ so that the attack alarm can be made.
\endproof

The above discussion refers to a framework of secure control of NCSs based on the conventional DW scheme described by (\ref{eq2A1})--(\ref{eq2A7}), then the security property will be analysed.

Note that to clearly distinguish between the normal (i.e., attack-free) system where $y_a(k) \equiv y(k)$ and the system under attacks, we denote ${y^o}(k)$, ${{\hat x}^o_{d}}(k|k - 1)$, $r^o_{d}(k)= y^o(k)-C\hat x^o_{d}(k|k-1)$, $J^{o}$ as the attack-free  counterpart of $y_a(k)$, ${{\hat x}}_d(k|k - 1)$, $r_d(k)$, $J$. Furthermore, to quantify the additional distortion caused by the attacker on systems, we define
\bea
\mathcal{D}_{r_d}(k) := L r_d(k) - L r^o_{d}(k).
\label{eq2A8}
\eea
According to $\{ \mathcal{D}_{r_d} \} $, the \emph{additive distortion power} \cite{DWM1} of systems is defined by
\bea
\mathop {\lim \sup }\limits_{T \to \infty } \frac{1}{T}\sum\limits_{k = 1}^T {{{\left\| \mathcal{D}_{r_d}(k) \right\|}^2}}.
\label{eq2A9}
\eea

Using the above definition of additive distortion power (\ref{eq2A9}), the security property of the conventional DW scheme is provided.

\textbf{Security property.} (cf. \cite[Th. 5]{DWM1}) If $y_a(k)$ passes the tests (\ref{eq2A6}) and (\ref{eq2A7}), then
\bea
\mathop {\lim \sup }\limits_{T \to \infty } \frac{1}{T}\sum\limits_{k = 1}^T {{{\left\| \mathcal{D}_{r_d}(k) \right\|}^2}}=0
\label{eq2A10}
\eea
which means that the additive distortion power of the systems that bypass the DW Tests 1 and 2 is restricted to be zero.
\endproof

\emph{Remark 2:} Eq. (\ref{eq2A10}) interprets the theoretical foundation of attack detection. In other words, if an attacker (by, e.g., injecting false data into the network) forces $y_a(k) \ne y(k)$ and dissatisfies (\ref{eq2A10}), then the attacker will be detected by the conventional DW scheme.
\endproof

\subsection{Security Weaknesses Analysis of the Conventional DW Scheme}
The above has presented a framework of secure control of NCSs based on the conventional DW scheme and analysed its security property. However, with the FDIA, there remain limitations on attack detection effectiveness, and system performance loss from attacks and watermarking. The details of the FDIA and limitations are analysed below.

Firstly, to analyse limitations of the conventional DW scheme, the following FDIA (cf. \cite{FDI,DWM5}) is given by
\bea
{x_a}(k + 1) = A_a{x_a}(k), y_a(k)=C{x_a}(k), k \in [k_0^a,\infty)
\label{eq2B1}
\eea
where $x_a(k) \in \mathcal{R}^{m_{x}}$ is the state of the FDIA; $k_0^a$ is the initial instant of the FDIA; $A_a$ is a constant matrix with $\rho (A_a)<1$; $C$ is the same as the one in (\ref{eq2A1}). Then, under the FDIA (\ref{eq2B1}), the closed-loop system based on the conventional DW scheme described by (\ref{eq2A1})--(\ref{eq2A7}) can be modelled as
\bea
{\zeta _d}(k + 1) = {\mathcal{A}_0}{\zeta _d}(k) + {\Lambda _d}{\psi _d}(k)
\label{eq2B2}
\eea
where variables $\zeta_d (k) := \left[ {x(k);x_a(k)-\hat x_d(k|k-1);{x_{a}}(k)} \right]$, $\psi_d(k) :=\left[ {n(k);v(k);{w_d}(k)} \right]$; the matrices $\mathcal{A}_0$ and $\Lambda_d$ are given by
\[{\mathcal{A}_0} := \left[ {\begin{array}{*{20}{c}}
  A&{\rm H} \\
  0&\Xi
\end{array}} \right],{\Lambda _d} := \left[ {\begin{array}{*{20}{c}}
  \Gamma &0&B \\
  0&0&{ - B} \\
  0&0&0
\end{array}} \right]\]
and ${\rm H} := \left[ { - BK(I + LC),BK} \right]$, $\Xi  := \left[ {{\Phi _1},{\Phi _2};0,{A_a}} \right]$, ${\Phi _1} := (A + BK)(I - LC)$, and ${\Phi _2} := {A_a} - (A + BK)$.

\emph{Remark 3:} When $A_a$ satisfies $\rho(A_a)<1$ for the FDIA (\ref{eq2B1}), $y_a(k)$ looks like normal $y(k)$; otherwise $\rho(A_a)>1$ means that $y_a(k)$ will diverge, and it can easily judge whether the FDIA (\ref{eq2B1}) takes place. Furthermore, it is reasonable to choose $\rho(A_a)<1$ for the FDIA (\ref{eq2B1}) because it can be seen from the following limitations 1 and 2 that $\rho(A_a)<1$ yields stealthiness and destructiveness of the FDIA (\ref{eq2B1}). An example of the FDIA (\ref{eq2B1}) can be seen in \cite[Th. II.3]{DWM5}, where $A_a=A+BK$ and $\rho(A+BK)<1$ in (\ref{eq2B1}).
\endproof

Secondly, to quantify attack detection effectiveness whilst considering the ergodic theorem \cite{NET} of stationary processes, the cross covariance of watermarking and residual and the auto covariance of residual are used to approximate the temporal averaging in (\ref{eq2A6}) and (\ref{eq2A7}) respectively, i.e.,
\bea
\mathbb{E}\left[ {{w_d}(k - 1)L{r_d}(k)} \right] \approx \mathop {\lim }\limits_{T \to \infty } {\mathcal{W}_d}(T),
\label{eqQADE1}
\eea
\bea
\mathbb{E}\left[ {L r_d(k)\left( L r_d(k) \right)^{T}} \right] \approx \mathop {\lim }\limits_{T \to \infty } {\mathcal{V}_d}(T)+L\Sigma_o L^{T}.
\label{eqQADE2}
\eea

Finally, according to the defined FDIA (\ref{eq2B1}) and closed-loop system (\ref{eq2B2}) and quantities (\ref{eqQADE1}), (\ref{eqQADE2}), the following three limitations are revealed.

\textbf{Limitation 1--FDIA Detection Effectiveness Limited by Watermarking.} For the system (\ref{eq2B2}), the FDIA (\ref{eq2B1}) will result in
\begin{subequations}
\begin{align}
&\mathbb{E}\left[ {w_d(k - 1)L r_d(k)} \right] = -\sigma _{w_d}^2LCB,
\label{eq2B3a}\\
&\mathbb{E}\left[ {L r_d(\infty )\left( L r_d(\infty ) \right)^{T}} \right] = LC{M_d}{C^T}{L^T}
\label{eq2B3b}
\end{align}
\end{subequations}
where $M_d = {\Phi _1}{M_d}\Phi _1^T + \sigma _{w_d}^2B{B^T}$, and ${\Phi _1}$ is the same as the one  shown in (\ref{eq2B2}). The proof is given in Section \uppercase\expandafter{\romannumeral+1}.A of the supplementary materials.
\endproof

\textbf{Limitation 2--System Performance Loss from the FDIA.} For the system (\ref{eq2B2}), if $\rho (A)>1$, then the FDIA (\ref{eq2B1}) will result in
\bea
\mathop {\lim }\limits_{k \to \infty } x(k) = \infty.
\label{eq2B4}
\eea
The proof is given in Section \uppercase\expandafter{\romannumeral+1}.A of the supplementary materials.
\endproof

\emph{Remark 4:} From limitation 1, it seems that the FDIA (\ref{eq2B1}) cannot bypass DW Tests (\ref{eq2A6}) and (\ref{eq2A7}), provided $\sigma^2_{w_d} \ne 0$ in (\ref{eq2B3a}) and (\ref{eq2B3b}). However, the DW statistical tests ${\varphi _{d,1}}(k)$ and ${\varphi _{d,2}}(k)$ are practically used, where loose (i.e., big) detection thresholds ${\vartheta _{d,1}}$, ${\vartheta _{d,2}}$ can be set. Note that when a small value of $\sigma_{w_d}^2$ has to be chosen, ${\varphi _{d,1}}(k)<{\vartheta _{d,1}}$ or ${\varphi _{d,2}}(k)<{\vartheta _{d,2}}$ so that the DW statistical tests are bypassed by the FDIA (\ref{eq2B1}) as presented in Section \uppercase\expandafter{\romannumeral+4}. Limitation 2 points out that if the plant (\ref{eq2A1}) is open-loop unstable (i.e., $\rho (A)>1$), the system state $x(k)$ in (\ref{eq2B2}) will diverge (or cross the limit in the real world) under the FDIA (\ref{eq2B1}) even though a well-designed controller (\ref{eq2A4}) is used. Therefore, two of the following tasks are to enhance FDIA detection effectiveness and to recovery the system performance.
\endproof

\textbf{Limitation 3--System Performance Loss From Watermarking.} (cf. \cite[Th. 3]{PWM1}) For the system (\ref{eq2B2}), if there is no attacks (i.e., $y_a(k) \equiv y(k)$), then the attack-free system performance is
\bea
{J^o} = {\left. {{J^o}} \right|_{{w_d}(k) = 0}} + \Delta {J^o}
\label{eq2B5}
\eea
where $\Delta {J^o} = \sigma _{{w_d}}^2tr({B^T}SB + R)$; ${\left. {{J^o}} \right|_{w_d(k) = 0}}$ is the attack-free system performance without the conventional DW scheme; ${{{\Delta {J^o}} \mathord{\left/
 {\vphantom {{\Delta {J^o}} {\left. {{J^o}} \right|}}} \right.
 \kern-\nulldelimiterspace} {\left. {{J^o}} \right|}}_{{w_d}(k) = 0}}$ is attack-free system performance loss from watermarking.
\endproof

\emph{Remark 5:} The tradeoff between FDIA detection effectiveness and system performance loss from watermarking is analysed as follows. Limitation 3 highlights that the inevitable cost paid for the above security property and FDIA detection effectiveness in limitation 1 of the conventional DW scheme is the attack-free system performance loss from watermarking. An intuitive example \cite{PWM1} is detection of replay attacks, while the example for detection of the FDIA (\ref{eq2B1}) is given in Section \uppercase\expandafter{\romannumeral+4}. In the example of \cite{PWM1}, one needs to pay the cost of $91\%$ system performance loss from watermarking to collect about $35\%$ detection rate at each step. Therefore, one of the following tasks is to decrease or eliminate the attack-free system performance loss from watermarking.
\endproof

To deal with the security property and limitations 1-3, a new DW scheme will be designed by considering the following three aspects:
\begin{enumerate}
  \item[a)]
  The new DW scheme should develop the security property of the conventional DW scheme;
  \item[b)]
  The new DW scheme should be able to detect effectively the FDIA (\ref{eq2B1}), while the FDIA detection effectiveness should be explored. The system performance loss from watermarking should be zero;
  \item[c)]
  The new DW scheme should be able to recovery system performance, while the relationship between the new DW scheme and its system performance should be explored.
\end{enumerate}

\section{Secure Control of NCSs Based on A New Dynamic Watermarking Scheme}
The security property and weaknesses of the conventional DW scheme have been thoroughly analysed. To solve these problems, a new DW scheme integrating watermarking as symmetric-key encryption and new testing and compensation mechanism is firstly designed. Then, we investigate the security property and attempt to overcome the security weaknesses using the new DW scheme.
\begin{figure}[!t]
  \centering
  \includegraphics[width=0.48\textwidth]{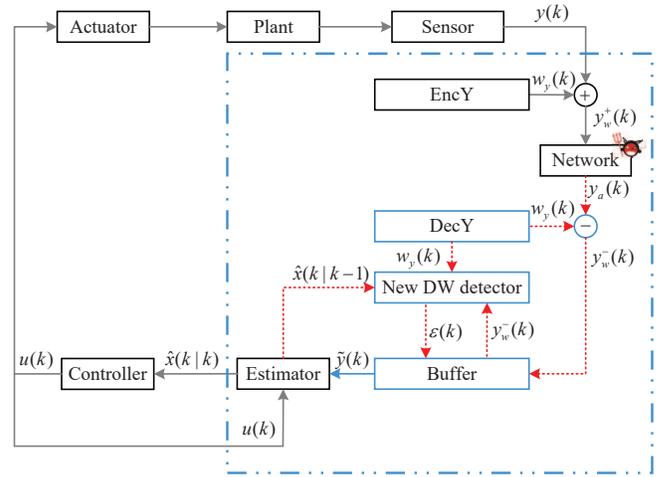}
  \caption{Framework of secure control of NCSs based on the new DW scheme integrating watermarking as symmetric-key encryption and new testing and compensation mechanism.}
  \label{fig2}
\end{figure}

\subsection{A New DW Scheme}
The framework of secure control of NCSs based on the new DW scheme integrating the watermarking as symmetric-key encryption and new testing and compensation mechanism is shown in Fig.~\ref{fig2}. The new DW scheme is analogous to digital watermarking \cite{DTW} (i.e., the digital code) that is embedded in electronic documents to preserve the valuable information. Therefore, motivated by digital watermarking and watermarking-as-key encryption \cite{DWM5} and symmetric key \cite{IMC}, the system output $y(k)$ is first sampled and encrypted with a watermarking signal $w_y(k)$ (as a key) by EncY (i.e., $y^{+}_w(k)$). After $y^{+}_w(k)$ is transmitted over the network, which may be attacked and becomes $y_a(k)$, it is decrypted with $w_y(k)$ by DecY and saved in the buffer (i.e., $y^{-}_w(k)$). Furthermore, using $y^{-}_w(k)$ from the buffer, $w_y(k)$ from the DecY and $\hat x(k|k-1)$ from the estimator, the new DW detector checks whether or not $y^{-}_w(k)$ is attacked and gives the corresponding detection results $\varepsilon (k)$. According to $\varepsilon (k)$, a compensation mechanism is used to update the input of the estimator (i.e., $\tilde y(k)$). Finally, the controller calculates $u(k)$ using the state estimate $\hat x(k|k)$ from the estimator. The following detail analysis is presented.

For the plant (\ref{eq2A1}), to solve the problem of information leakage and improve the confidentiality of signals, $y(k)$ is \emph{encrypted} by EncY in Fig.~\ref{fig2}, it follows that
\bea
y^{+}_{w}(k) = y(k) + w_y(k)
\label{eq3A1}
\eea
where $w_y(k)$, as a key, is a watermarking signal drawn from an i.i.d. Gaussian distribution with zero mean and covariance matrix ${\Sigma _{{w_y}}} = diag\{ \sigma _{{w_{y,1}}}^2, \ldots ,\sigma _{{w_{{y,m_y}}}}^2\} $, and independent of $y(k)$. Then, $y^{+}_{w}(k)$ is transmitted via the network,  if attacked and then becomes $y_a(k)$. Note that when there exist cyber attacks, $y_a(k) \ne y^{+}_{w}(k)$; otherwise $y_a(k)=y^{+}_{w}(k)$.

Next, to eliminate the side effect of the encrypted signal after transmission over the network, $y_a(k)$ is \emph{decrypted} by the DecY, i.e.,
\bea
y^{-}_{w}(k) = y_{a}(k)  - w_y(k).
\label{eq3A2}
\eea
Note that when there exist cyber attacks, $y^{-}_{w}(k) \ne y(k)$; otherwise $y^{-}_{w}(k) =y(k)$. As a consequence, the watermarking signal $w_y(k)$ and encryption (\ref{eq3A1}) and decryption (\ref{eq3A2}) comprise the watermarking as symmetric-key encryption.

\begin{algorithm}[!t]
\caption{Online Algorithm for Producing Same $w_y(k)$ in EncY and DecY}\label{alg1}
Initialization: The same seeds and sequence numbers for EncY and DecY are set, e.g., both seeds are set as 1 and sequence numbers are set as $k^{+}_{w}=k^{-}_{w}=1$\;
\While{$k=1,2,\cdots$}{
EncY receives $y(k)$ and then generates $w_y(k^{+}_{w})$ by using the seed and $k^{+}_{w}$\;
EncY encrypts $y(k)$ with $w_y(k^{+}_{w})$ by using (\ref{eq3A1})\;
$k^{+}_{w} \leftarrow k^{+}_{w}+1$\;
DecY receives $y_{a}(k)$ and then generates $w_y(k^{-}_{w})$ by using the seed and $k^{-}_{w}$\;
DecY decrypts $y_{a}(k)$ with $w_y(k^{-}_{w})$ by using (\ref{eq3A2})\;
$k^{-}_{w} \leftarrow k^{-}_{w}+1$\;
}
\end{algorithm}
\emph{Remark 6:} For the decryption in (\ref{eq3A2}), there is a practical key consistency problem: how to produce the same $w_y(k)$ in (\ref{eq3A1}) and (\ref{eq3A2}) at the same instant $k$? The watermarking signal $w_y(k)$ cannot simply be sent via the network because the attacker may observe and distort it \emph{en route}.  Alternatively, an online algorithm for producing $w_y(k)$ in EncY and DecY is designed based on secure mechanism \cite{IMC} and Marsaglia-Bray method \cite{MBM}, which can guarantee the same $w_y(k)$ on both sides of the network.
\endproof

\emph{Remark 7:} Compared with the conventional DW scheme shown in Fig.~\ref{fig1} where the control input is encrypted using the method of (\ref{eq2A5}), the new DW scheme encrypts the system output using the method of (\ref{eq3A1}). When decryption is added to the conventional DW scheme on the controller side, limitation 3 on system performance loss from watermarking may be overcome, but limitation 1 on FDIA detection effectiveness will still remain. Therefore, the watermarking as symmetric-key encryption (\ref{eq3A1}) and (\ref{eq3A2}), together with the following the new DW Tests, are integrated in the new DW scheme and expected to cope with limitations 1 and 3.
\endproof

Furthermore, $y^{-}_{w}(k)$ is saved in the buffer, which will be detected to determine whether or not this signal is attacked. Therefore, $y^{-}_{w}(k)$ is transferred to the new DW detector, where the following two designed tests are operated using ${{\hat x}}(k|k - 1)$ from the estimator and $w_y(k)$ from the DecY, i.e.,
\begin{enumerate}
  \item
  \textbf{New DW Test 1:} checking whether or not
  \bea
  \mathop {\lim }\limits_{T \to \infty } \mathcal{W}_i(T) = 0
  \label{eq3A3}
  \eea
  where $\mathcal{W}_i(T) := \frac{1}{T}\sum\nolimits_{k = 1}^T {{w_{y,i}}(k)L r(k)}$, $i=1,\cdots,m_y$; $w_{y,i}$ is the $i$th component of $w_{y}$; $T$ is the time window size; $L$ has been given in (\ref{eq2A2}); due to the decryption (\ref{eq3A2}), the new residual $r(k)$ is defined by $r(k) := y^{-}_{w}(k) - C{{\hat x}}(k|k - 1)$.
  \endproof
  \item
  \textbf{New DW Test 2:} checking whether or not
  \bea
  \mathop {\lim }\limits_{T \to \infty } \mathcal{V}(T) = 0
  \label{eq3A4}
  \eea
  where $\mathcal{V}(T) := \frac{1}{T}\sum\nolimits_{k = 1}^T {Lr(k){{\left( {Lr(k)} \right)}^T}}  - L{\Sigma _o}{L^T}$; $\Sigma_o$ is the same as the one in (\ref{eq2A2}).
  \endproof
\end{enumerate}

\emph{Remark 8:} Compared with the conventional DW Tests (\ref{eq2A6}) and (\ref{eq2A7}), there are two differences in signals for the new DW Tests (\ref{eq3A3}) and (\ref{eq3A4}): the watermarking signal is changed from $w_d(k-1)$ to $w_y(k)$, and the residual signal is changed from $r_d(k)$ to $r(k)$. Note that the identity is that the tests are formed using the time averaging of products of (i) watermarking signal and residual (i.e., (\ref{eq2A6}) and (\ref{eq3A3})) and (ii) two identical residual (i.e., (\ref{eq2A7}) and (\ref{eq3A4})). Therefore, it is expected that the new DW Tests (\ref{eq3A3}) and (\ref{eq3A4}), together with the above watermarking as symmetric-key encryption, can develop the security property of the conventional DW scheme, and overcome limitations 1 and 3.
\endproof

\emph{Remark 9:} Note that in the new DW Tests 1 and 2, we require $T \to \infty$, which is similar to the conventional DW Tests 1 and 2. To solve the problem, following the definitions made in Remark 1 and \cite{DWM2}, the new DW Tests 1, 2 need to be converted to the statistical tests ${\varphi _{1,i}}(k), {\varphi _2}(k)$ for practical applications, i.e., ${\varphi _{1,i}}(k) := {\left\| \mathcal{W}^{k}_{i}(T) \right\|_F}$, ${\varphi _2}(k) := \left| {tr\left( \mathcal{V}^{k}(T) \right)} \right|$, where $\mathcal{W}^{k}_{i}(T)$ and $\mathcal{V}^{k}(T)$ are $\mathcal{W}_{i}(T)$ and $\mathcal{V}(T)$ calculated within the current time window $\{k-T+1,k-T+2,\ldots,k\}$, respectively. Furthermore, let ${\vartheta _{1,i}},{\vartheta _2}$ be the preset thresholds and if $y^{-}_{w}(k)$ is attacked, we expect ${\varphi _{1,i}}(k) \geqslant {\vartheta _{1,i}}$ or ${\varphi _2}(k) \geqslant {\vartheta _2}$ thereby the detection result is set as $\varepsilon (k)=0$; otherwise, $\varepsilon (k) = 1$. The detection results will be used for compensation below.
\endproof

After $y^{-}_{w}(k)$ has been detected, there may be two cases for the attacked signals:
\begin{enumerate}
  \item[1)]
  If there is no compensation, then $y^{-}_{w}(k)$ will be directly sent to the estimator whether the signal is attacked or not, i.e.,
  \bea
  \tilde y(k) = y_{w}^{-}(k),\ {\rm no\ compensation}.
  \label{eq3A6}
  \eea
  \item[2)]
  The above (\ref{eq3A6}) could cause system performance degradation even instability. To improve the system resilience under attacks, a compensation mechanism is employed, where $y^{-}_{w}(k)$ will be sent to the estimator based on the detection results $\varepsilon (k)$, i.e.,
  \bea
  \tilde y(k) = \left\{ \begin{gathered}
  y_{w}^{-}(k),if\ \varepsilon (k) = 1 \hfill \\
  \tilde y(k - h'(k)),if\ \varepsilon (k) = 0 \hfill \\
  \end{gathered}  \right.
  \label{eq3A5}
  \eea
  where $h'(k) := k-\max \{ k'|\varepsilon (k') = 1, k' \leqslant k\} >0$  is the duration of last successful buffer update. For simplicity, we only consider the situation where the missed detection\footnote{Missed detection is that even though an attack takes place, the attack alarm is not triggered (i.e., $\varepsilon (k)=1$).} can be ignored (i.e., $y_{w}^{-}(k)=y(k),if\ \varepsilon (k) = 1$), then (\ref{eq3A5}) becomes
  \bea
  \tilde y(k) = \left\{ \begin{gathered}
  y(k),if\ \varepsilon (k) = 1 \hfill \\
  \tilde y(k - h'(k)),if\ \varepsilon (k) = 0 \hfill \\
  \end{gathered}  \right.
  \label{eq3A5r2}
  \eea
  Define
  \bea
  h(k) := \left\{ \begin{gathered}
  0,if\ \varepsilon (k) = 1 \hfill \\
  h'(k),if\ \varepsilon (k) = 0 \hfill \\
  \end{gathered}  \right.
  \label{eq3A5r1}
  \eea
  According to (\ref{eq3A5r1}), (\ref{eq3A5r2}) can be re-written as
  \bea
  \tilde y(k) = y(k-h(k)),\ {\rm with\ compensation}
  \label{eq3A7}
  \eea
  where $h(k)$ is the delay of healthy system output.
\end{enumerate}

\emph{Remark 10:} Compared with the conventional DW scheme and the scheme (\ref{eq3A6}) without compensation, the compensation (\ref{eq3A7}) based on the detection results $\varepsilon (k)$ is used to discard the attacked $y^{-}_{w}(k)$ (e.g., injected with false data) and to enable the latest $y(k-h(k))$. It is expected to overcome the limitation 2, i.e., the recovery of system performance.
\endproof

Then, $\hat x(k|k)$  can be computed by
\bea
\hat x(k|k) = \hat x(k|k - 1) + L(\tilde y(k) - C\hat x(k|k - 1)).
\label{eq3A8}
\eea
Using $\hat x(k|k)$, since there is no watermarking signal on the controller side, the control input (\ref{eq2A5}) can be re-written as
\bea
u(k)=K\hat x(k|k)
\label{eq3A9}
\eea
where $K$ is the same as (\ref{eq2A4}), which can update $\hat x(k + 1|k)$ by
\bea
\hat x(k + 1|k) = A\hat x(k|k) + Bu(k).
\label{eq3A10}
\eea

Finally, under the FDIA (\ref{eq2B1}), to identify the system based on the new DW scheme without or with compensation, we produce the corresponding close-loop system models as follows:

\begin{enumerate}
  \item[1)]
  On the one hand, under the FDIA (\ref{eq2B1}), the system based on the new DW scheme without compensation, described in (\ref{eq2A1}), (\ref{eq3A1})-(\ref{eq3A4}), (\ref{eq3A6}) and (\ref{eq3A8})-(\ref{eq3A10}), can be modelled as
  \bea
  {\zeta }(k + 1) = {\mathcal{A}_{0}}{\zeta}(k) + {\Lambda_{0}}\psi (k),\ {\rm no\ compensation}
  \label{eq3A11}
  \eea
  where marks ${\zeta}(k) := \left[ {x(k);x_a(k)-\hat x(k|k-1)};x_a(k) \right]$, $\psi (k) := \left[ {n(k);v(k);w_y(k)} \right]$; the matrices $\mathcal{A}_0$ and $\Lambda_0$ are given by
  \[\begin{aligned}
  &{\mathcal{A}_0}\ {\rm in}\ (12),\ {\rm and} \\
  &{\Lambda _0} := \left[ {\begin{array}{*{20}{c}}
  \Gamma &0&{BKL} \\
  0&0&{(A + BK)L} \\
  0&0&0
  \end{array}} \right].
  \end{aligned}\]
  \item[2)]
  On the other hand, under the FDIA (\ref{eq2B1}), the system based on the new DW scheme with compensation, described by (\ref{eq2A1}), (\ref{eq3A1})-(\ref{eq3A4}), and (\ref{eq3A7})-(\ref{eq3A10}), can be constructed as
  \begin{align}
  &\bar \zeta(k + 1) = {A_0}\bar \zeta(k) + {A_1}E\bar \zeta(k - h (k)) + {\Gamma _0}\bar \psi (k),
  \nonumber\\
  &{\rm with\ compensation} \label{eq3A12}
  \end{align}
  where the variables ${\bar \zeta}(k) := \left[ {x(k);\hat x(k-1|k-1)} \right]$, $\bar \psi (k) := \left[ {n(k);v(k - h (k))} \right]$; the matrices $E$, $A_0$, $A_1$ and $\Gamma _0$ are given by
  \[\begin{aligned}
  &E := \left[ {I,0} \right],\ {\rm and}\\
  &{A_0} := \left[ {\begin{array}{*{20}{c}}
  A&{BK(I - LC)(A + BK)} \\
  0&{(I - LC)(A + BK)}
  \end{array}} \right],  \\
  &{A_1} := \left[ {\begin{array}{*{20}{c}}
  {BKLC} \\
  {LC}
  \end{array}} \right],{\Gamma _0} := \left[ {\begin{array}{*{20}{c}}
  \Gamma &{BKL} \\
  0&L
  \end{array}} \right].
  \end{aligned}\]
\end{enumerate}

\emph{Remark 11:} In the designed secure controller, the attack detection method, instead of the control strategy, usually brings the difference of computational time complexity. Therefore, compared with the tests used in \cite{DRM,ADMM,WNCS,DAD}, computational time complexity of the new DW Tests (\ref{eq3A3}) and (\ref{eq3A4}) is analysed as follows. For instance, considering that $r(k)\in \mathcal{R}^{m_y}$, $x(k)\in \mathcal{R}^{m_x}$ and the finite $T$, computational time complexity of the residual-based test $\sum\nolimits_{s = k - T - 1}^k {{r^T}(s)r(s)}$ used in \cite{DAD} is $O(m^2_y)$. However, computational time complexity of the new DW Tests (\ref{eq3A3}) and (\ref{eq3A4}) is $O(1)+O(m^2_x+m_xm_y) \approx O(m^2_x+m_xm_y)$, where $O(1)$ and $O(m^2_x+m_xm_y)$ are spent by watermarking generation, and calculation of $\mathcal{W}_{i}(T)$ and $\mathcal{V}(T)$, respectively. To clearly compare computational time complexity of the new DW Tests with the residual-based test, a ratio $\wp  := {{m_y^2} \mathord{\left/
{\vphantom {{m_y^2} {\left( {m_x^2 + {m_x}{m_y}} \right)}}} \right.
\kern-\nulldelimiterspace} {\left( {m_x^2 + {m_x}{m_y}} \right)}}$ is defined and shown in Fig.~A.2 of Section II.B in the supplementary materials. It can be seen from Fig.~A.2 that if $m_y < m_x$, then computational time complexity of the new DW Tests is usually more than that of the residual-based test used in \cite{DAD}, as shown in the experimental results of Section \uppercase\expandafter{\romannumeral+4}.B; if $m_y$ is much larger than $m_x$, computational time complexity of the new DW Tests will be less than that of the residual-based test used in \cite{DAD}, which could take place in the power system.
\endproof

\subsection{Security Property Analysis of the New DW Scheme}
We have developed a framework for secure control of NCSs based on the new DW scheme, and then the security property of the new DW scheme is analysed using the following Theorem 1.

\emph{Theorem 1:} For the system (\ref{eq3A11}) or (\ref{eq3A12}), if $y_{w}^{-}(k)$ passes the new DW Tests 1 and 2, then
\bea
\mathop {\lim \sup }\limits_{T \to \infty } \frac{1}{T}\sum\limits_{k = 1}^T {{{\left\| {{\mathcal{D}_r}(k)} \right\|}^2}}=0
\label{eq3B1}
\eea
where $\mathcal{D}_r(k) := L r(k) - L{r^o}(k)$; and $r^o(k):={y_w^{o-}}(k)-C{{\hat x}^o}(k|k - 1)$, ${y_w^{o-}}(k)$, ${{\hat x}^o}(k|k - 1)$ are the attack-free (i.e., $y_w^{-}(k) \equiv y(k)$) counterparts of $r(k)$, ${y}^{-}_{w}(k)$, ${{\hat x}}(k|k - 1)$.
\endproof

\proof The proof is given in Section \uppercase\expandafter{\romannumeral+1}.B of the supplementary materials.
\endproof

\emph{Remark 12:} The security property of the conventional DW scheme indicates that the additive distortion power of the closed-loop systems is restricted to be zero when DW Tests 1 and 2 are bypassed. Theorem 1 reveals that the additive distortion power of the closed-loop systems is restricted to be zero when the new DW Tests 1 and 2 are bypassed. In addition, (\ref{eq3B1}) explains the theoretical basis of attack detection as (\ref{eq2A10}). Therefore, by integrating the watermarking as symmetric-key encryption and the new DW Tests 1 and 2, the new DW scheme develops the security property of the conventional DW scheme.
\endproof

\subsection{FDIA Detection Effectiveness Analysis of the New DW Scheme}
The above discussion is about the security property of the new DW scheme, and FDIA detection effectiveness of the new DW scheme can be evaluated in the following Theorem 2 and Corollary 1.

\emph{Theorem 2: } For the system (\ref{eq3A11}), if $\rho(A)>1$, then the FDIA (\ref{eq2B1}) will result in
\begin{subequations}
\begin{align}
&\mathbb{E}\left[ {w_{y,i}(k)Lr(k)} \right] = -\sigma _{w_{y,i}}^2{L_{\cdot i}},
\label{eq3C1a} \\
&\mathbb{E}\left[ {Lr(\infty )\left( Lr(\infty ) \right)^{T}} \right] = L\left( CM{C^T} + \Sigma_{w_y} \right){L^T},
\label{eq3C1b}\\
&\mathop {\lim }\limits_{k \to \infty } x(k) = \infty
\label{eq3C1c}
\end{align}
\end{subequations}
where $M = {\Phi _1}{M}\Phi _1^T + (A+BK)L\Sigma_{w_y}{\left( (A+BK)L \right)^T}$, ${\Phi _1}$ is the same as the one of (\ref{eq2B2}).
\endproof

\proof The proof is given in Section \uppercase\expandafter{\romannumeral+1}.C of the supplementary materials.
\endproof

\emph{Corollary 1:} For the system (\ref{eq3A12}), the FDIA (\ref{eq2B1}) will result in (\ref{eq3C1a}).
\endproof

\proof The proof is omitted.
\endproof

\emph{Remark 13:} Limitation 1 earlier has revealed the positive correlation between the FDIA detection effectiveness of the conventional DW scheme and the value of $\sigma^2_{w_{d}}$. However, due to limitation 3, a small $\sigma_{w_d}^2$ needs to be selected for yielding small system performance loss due to watermarking. This makes it possible that the FDIA (\ref{eq2B1}) may bypass the conventional DW scheme. Similarly, Theorem 2 and Corollary 1 highlight that FDIA detection effectiveness of the new DW scheme with or without compensation is positively correlated with the value of $\sigma^2_{w_{y,i}}$. Note that the key difference is that a big $\sigma^2_{w_{y,i}}$ can be selected in the new DW scheme because ${J^o} = {\left. {{J^o}} \right|_{{w_y}(k) = 0}}$ thanks to the watermarking as symmetric-key encryption (\ref{eq3A1}) and (\ref{eq3A2}), where ${\left. {{J^o}} \right|_{{w_y}(k) = 0}}$ is  the attack-free system's performance without the new DW scheme. That is, limitation 3 on system performance loss due to watermarking is overcome by the new DW scheme. This makes it possible that the FDIA (\ref{eq2B1}) will be detected by the new DW scheme with an enabled big enough $\sigma^2_{w_{y,i}}$. Therefore, by integrating the watermarking as symmetric-key encryption (\ref{eq3A1}) and (\ref{eq3A2}) and the new DW Tests (\ref{eq3A3}) and (\ref{eq3A4}), limitations 1 and 3 are overcome by the new DW scheme.
\endproof

\emph{Remark 14:} As stated above, the basis of the new DW scheme can be attributed to the overcoming of limitation 3 on system performance loss from watermarking. The conventional DW scheme brings system performance loss from watermarking, because the control signal is only encrypted by the watermarking signal and without decryption. Unlike the conventional DW scheme, the watermarking as symmetric-key encryption (\ref{eq3A1}) and (\ref{eq3A2}) used in the new DW scheme can prevent watermarking signals from exciting the normal system operation, so the limitation 3 can be avoided.
\endproof

\subsection{Performance Analysis of NCSs based on the New DW Scheme}
Now, we will further analyse why the new DW scheme needs to be employed for the recovery of system performance, and quantify the recovery capability of system performance based on the new DW scheme.

\emph{1) Why the New DW Scheme Needs to Be Used for Recovery of System Performance?} Theorem 2 has presented that when the compensation is ignored, if $\rho(A)>0$ in (\ref{eq2A1}), then the FDIA (\ref{eq2B1}) emerging in $[k_0^{a},\infty)$ will influence the system stability. Therefore, it is necessary to use the new DW scheme against the FDIA (\ref{eq2B1}) emerging in $[k_0^{a},\infty)$ for recovery of system performance. However, if we use the compensation (\ref{eq3A7}) to cope with the FDIA (\ref{eq2B1}) emerging in $[k_0^{a},\infty)$, $h(k)$ will go infinity and according to time-delay system theory \cite{ORA}, the system will be unstable. The system will lose its stability even though the compensation is used. But then, on the one hand, in practice the operation time of the FDIA (\ref{eq2B1}) must be the union of subsets of $[k_0^{a},\infty)$; on the other hand, sometimes the attackers achieve their goal in a short time without worrying about the attack detection. In this situation, is it necessary to use the new DW scheme for the recovery of system performance? To answer this question, based on the switched system theory, under the FDIA (\ref{eq2B1}) with the subsets of $[k_0^{a},\infty)$, the following stability analysis is presented.

The first task is to construct the corresponding switched system model. The early work (\ref{eq3A11}) has modelled the subsystem based on the new DW scheme without compensation under the FDIA (\ref{eq2B1}), then the attack-free subsystem based on the new DW scheme without compensation can be given by setting $h(k) \equiv 0$ in (\ref{eq3A12}) and adopting the same variables $\zeta(k)$ and $\psi(k)$ in (\ref{eq3A11}), i.e.,
\bea
\zeta(k+1) = \mathcal{A}_1 \zeta(k) + \Lambda_1 \psi(k)
\label{eq3D1}
\eea
where the matrices $\Lambda_1$ and $\mathcal{A}_1$ are given by
\[\begin{gathered}
  {\Lambda _1} := \left[ {\begin{array}{*{20}{c}}
  \Gamma &{BKL}&0 \\
  0&{-(A + BK)L}&0 \\
  0&0&0
\end{array}} \right], \hfill \\
  {\mathcal{A}_1} := \left[ {\begin{array}{*{20}{c}}
  {A + BKLC}&{ - BK(I - LC)}&0 \\
  {-(A + BK)LC}&{{\Phi _1}}&0 \\
  0&0&0
\end{array}} \right]. \hfill \\
\end{gathered} \]
Therefore, combining (\ref{eq3A11}) and (\ref{eq3D1}), under the FDIA (\ref{eq2B1}) with the subsets of $[k_0^{a},\infty)$, the system based on the new DW scheme can be modelled as
\bea
{\zeta }(k + 1) = {\mathcal{A}_{s(k)}}{\zeta}(k) + {\Lambda_{s(k)}}\psi (k)
\label{eq3D2}
\eea
where $s(k) = 0\ or\ 1$ is the switching signal: when an attack takes place, there is $s(k) = 0$; otherwise $s(k) = 1$. We know ${\mathcal{A} _0}$ is unstable and ${\mathcal{A} _1}$ is stable. Then, there exist real numbers $\lambda _{+}>1$, $0<\lambda _{-}<1$, $g_0$ and $g_1$ satisfy
\begin{gather}
\left\| {{{\left( {\mathcal{A}_0 } \right)}^k}} \right\| \leqslant {\left( {\lambda _ -  } \right)^{{g _0}}}{\left( {\lambda _ +  } \right)^k},\left\| {{{\left( {\mathcal{A}_1 } \right)}^k}} \right\| \leqslant {\left( {\lambda _ -  } \right)^{{g _1}}}{\left( {\lambda _ -  } \right)^k}.
\label{eq3D3}
\end{gather}
Meanwhile, we denote the total activation time of the unstable subsystem (or the stable subsystem) by $\mathcal{T} _{0}$ (or $\mathcal{T} _{1}$). We denote the switching of $s$ by ${N _s }(0,k) \leqslant {N _0} + \frac{k}{\tau  }$ where $\tau $ is the average dwell time. According to the definitions of attack frequency and duration made in \cite{ISS}, the larger $\mathcal{T} _{0}$ and $\tau $ stand for a longer attack duration and a smaller attack frequency, respectively.

Finally, stability analysis of the switched system (\ref{eq3D2}) is demonstrated in the following Theorem 3.

\emph{Theorem 3:} For the given constant $\lambda^{\dag}$, if the following conditions are satisfied:
\begin{align}
&\mathbb{E}\left[ {\left\| {\psi (k)} \right\|} \right] < \infty,
\label{eq3D4}\\
&\left\{ \begin{gathered}
  \tau  \in {\mathbb{R}_ + },if\;g \geqslant 0 \hfill \\
  \tau  \geqslant \tau _{ave} = \frac{g}{{{\lambda ^\dag } - {\lambda ^*}}},if\;g < 0 \hfill \\
\end{gathered}  \right.
\label{eq3D5}\\
&\mathop {\inf }\limits_{k \geqslant 0} \left[ {{{{\mathcal{T}_1 }} \mathord{\left/
 {\vphantom {{{\mathcal{T}_1 }} {{\mathcal{T}_0 }}}} \right.
 \kern-\nulldelimiterspace} {{\mathcal{T}_0 }}}} \right] \geqslant {{\left( {\ln {\lambda _{+}} - {\lambda ^*}\ln {\lambda _{-}}} \right)} \mathord{\left/
 {\vphantom {{\left( {\ln {\lambda _{+}} - {\lambda ^*}\ln {\lambda _{-}}} \right)} {\left( {({\lambda ^*} - 1)\ln{\lambda _{-}}} \right)}}} \right.
 \kern-\nulldelimiterspace} {\left( {({\lambda ^*} - 1)\ln{\lambda _{-}}} \right)}}
\label{eq3D6}
\end{align}
where $0 < {\lambda ^\dag } < {\lambda ^ * } < 1$ and $g = \min \{ {g_0},{g_1}\}$ from (\ref{eq3D3}), then there always exists a finite constant $\tau$ such that the switched system (\ref{eq3D2}) is uniformly bounded.
\endproof

\proof The proof is given in Section \uppercase\expandafter{\romannumeral+1}.D of the supplementary materials.
\endproof

\emph{Remark 15:} The tolerance capacity of FDIAs for NCSs based on the new DW scheme without compensation is analysed as follows. Theorem 3 points out that under FDIA (\ref{eq2B1}), the state of the system based on the new DW scheme without compensation is uniformly bounded when the noise is bounded (\ref{eq3D4}), the attack frequency is small (\ref{eq3D5}) and the attack duration is short (\ref{eq3D6}).  As shown in Section \uppercase\expandafter{\romannumeral+4}, under the FDIA (\ref{eq2B1}) with a short successive duration, when the initial state of the FDIA (\ref{eq2B1}) is properly chosen, the condition (\ref{eq3D6}) may be violated and the system will lose its stability. Therefore, it is necessary to use the new DW scheme against the FDIA (\ref{eq2B1}) with the subsets of $[k_0^{a},\infty)$.
\endproof

\emph{2) How Well Does System Performance Recover from the New DW Scheme?} We have proved that it is necessary to use the new DW scheme against the FDIA (\ref{eq2B1}) for the recovery of system performance. Then, the recovery capability of the system based on the new DW scheme is presented in the following, which is described by the quantitative relationship between the system performance and the maximally allowable delay of healthy system output.

Firstly, the system performance is described by the power of the estimation errors, i.e.,
\bea
\mathop {\lim \sup }\limits_{T \to \infty } \frac{1}{T}\sum\limits_{k = 0}^{T - 1} {{{\left\| {e(k)} \right\|}^2}}
\label{eq3D7}
\eea
where the estimation errors $e(k):=x(k)-\hat x(k|k)$.

According to the defined system performance, the relationship between the system performance and the new DW scheme is presented in the following Theorem 4 and Corollary 2.

\emph{Theorem 4: } Under the zero-initial condition, if given an integer $\bar h \geqslant h(k)$, there exists a real number $\beta$, real symmetric matrices $Z_i>0\ (i=1,2,3)$, and matrices $W_i\ (i=1,2,\ldots,5)$ satisfying
\bea
\left[ {\begin{array}{*{20}{c}}
  \Theta &\Omega \\
   * &\Upsilon
\end{array}} \right] < 0
\label{eq3D8}
\eea
where
\[\begin{gathered}
  \Theta := \left[ {\begin{array}{*{20}{c}}
  {{\Theta _{11}}}&{{E^T}{W_2} - W_1^T}&{{E^T}{W_3}} \\
   * &{ - W_2^T - {W_2} - {Z_3}}&{ - {W_3}} \\
   * & * &{ - \beta I}
\end{array}} \right], \hfill \\
  {\Theta _{11}} := W_1^TE + {E^T}{W_1} - {Z_1} + {E^T}{Z_3}E, \hfill \\
  \Omega  :=  \hfill \\
  \left[ {\begin{array}{*{20}{c}}
  {\bar hW_1^T}&{\bar h{{({A_0} - I)}^T}{E^T}W_4^T}&{{{(E - {E^c}{A_0})}^T}}&{A_0^TW_5^T} \\
  {\bar hW_2^T}&{\bar hA_1^T{E^T}W_4^T}&{{{( - {E^c}{A_1})}^T}}&{A_1^TW_5^T} \\
  {\bar hW_3^T}&{\bar h\Gamma _0^T{E^T}W_4^T}&{{{( - {E^c}{\Gamma _0})}^T}}&{\Gamma _0^TW_5^T}
\end{array}} \right], \hfill \\
  \Upsilon  :=  \hfill \\
  diag\left\{ { - \bar h{Z_2},\bar h\left( {{Z_2} - W_4^T - {W_4}} \right), - I,{Z_1} - W_5^T - {W_5}} \right\} \hfill \\
\end{gathered}\]
and $E^{c} := \left[ {0,I} \right]$, then the system (\ref{eq3A12}) is asymptotically stable and the power of the estimation errors satisfies
\begin{gather}
\mathop {\lim \sup }\limits_{T \to \infty } \frac{1}{T}\sum\limits_{k = 0}^{T - 1} {{{\left\| {e(k)} \right\|}^2}}  = \beta \left( {tr\left( {{\Sigma _n}} \right) + tr\left( {{\Sigma _v}} \right)} \right).
\label{eq3D9}
\end{gather}
\endproof

\proof The proof is given in Section \uppercase\expandafter{\romannumeral+1}.E of the supplementary materials.
\endproof

\emph{Corollary 2: } Under the zero-initial condition, if given an integer $\bar h \geqslant h(k)$, there exists a real number $\beta$, real symmetric matrices $Z_i>0\ (i=1,2,3)$, and matrices $W_i\ (i=1,2,\ldots,5)$ satisfying
\bea
\left[ {\begin{array}{*{20}{c}}
  \Theta &\Omega \mathcal{C} \\
   * &\Upsilon
\end{array}} \right] < 0
\label{eq3D10}
\eea
where $\Theta$, $\Omega$, $\Upsilon$, and $E^{c}$ has been given in Theorem 4, $\mathcal{C} := diag\{ I,I,{C^T},I\} $, then the system (\ref{eq3A12}) is asymptotically stable and the power of the estimation errors satisfies
\begin{gather}
\mathop {\lim \sup }\limits_{T \to \infty } \frac{1}{T}\sum\limits_{k = 0}^{T - 1} {{{\left\| {Ce(k)} \right\|}^2}}  = \beta \left( {tr\left( {{\Sigma _n}} \right) + tr\left( {{\Sigma _v}} \right)} \right).
\label{eq3D11}
\end{gather}
\endproof

\proof The proof is omitted.
\endproof

\emph{Remark 16:} Limitation 2 and Theorem 3 have revealed that the FDIA (\ref{eq2B1}) emerging in $[k_0^a,\infty)$ or the union of subsets of $[k_0^a,\infty)$ can destroy system stability without compensation. Theorem 4 reveals that by using the new DW scheme, the system performance can be recovered from unstable mode under the FDIA (\ref{eq2B1}) to stable mode (specifically, lying on noise level (\ref{eq3D9}) or (\ref{eq3D11}) with respect to process and measurement noises), and the relationship between the system performance and the maximally allowable delay $\bar h$ of healthy system output is quantified. Therefore, by collating all the compensations, limitation 2 can be overcome when $\bar h$ is not violated.
\endproof

\emph{Remark 17:} Note that over detection\footnote{Over detection is that after the real attacks stop, attack alarms (i.e., $\varepsilon (k)=0$) will continue for some time depending on the value of the window size $T$.} could lead to $h(k)>\bar h$, causing the system to be unstable (i.e., infinite estimation errors). This shows the cost to pay for the \emph{compensation mechanism} (\ref{eq3A7}) against limitation 2 on system performance loss from attacks: i.e., it is highly recommended that a small time window size $T$ should be selected when the proposed compensation mechanism is used in the real world, as done in Section \uppercase\expandafter{\romannumeral+4}.
\endproof

\section{Experiments}
The platform of the networked inverted pendulum visual servo system (NIPVSS) \cite{NIP} is employed to validate the new DW scheme. We first give the structure and corresponding parameters of the platform. Then, the real-time experiments are carried out on the platform.

\subsection{Experiment Platform Setup}
\begin{figure}[!t]
  \centering
  \includegraphics[width=0.488\textwidth]{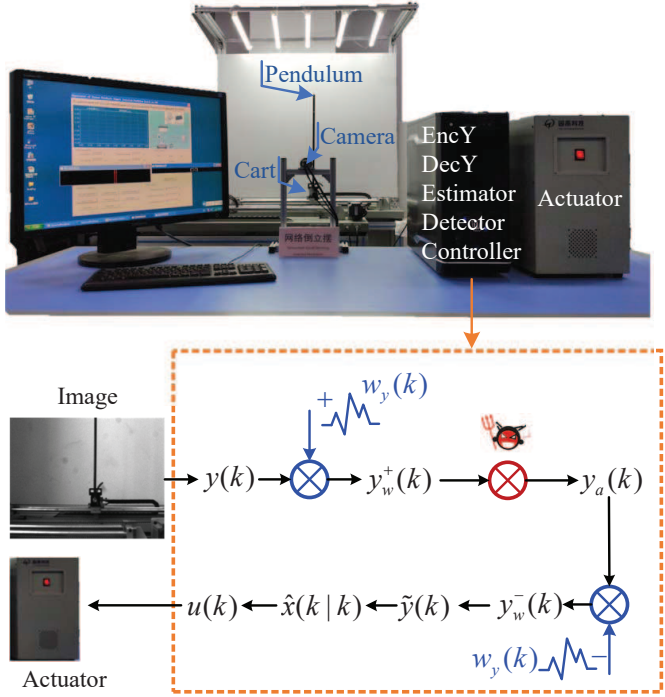}
  \caption{Experimental platform of the NIPVSS with the new DW scheme.}
  \label{fig3}
\end{figure}

The experimental platform of the NIPVSS based on the new DW scheme is shown in Fig.~\ref{fig3}. The single acA640-120gm monochrome camera at 100 Hz @ $640 \times 480$ pixels with light sources acting as the sensor captures the images of the inverted pendulum. Then the images are sent to the computer through the 1 Gbps wired network, which runs Microsoft Visual Studio 2010 in Windows XP with an Intel Core i5 processor (3.2 GHz) and 4 GB RAM. After having received the images, the computer processes the images to obtain the state information of the cart position and the pendulum angle. Furthermore, in the computer, the state information is encrypted and decrypted based on the new DW scheme and Algorithm 1, then the current results are used to formulate the estimator, the new DW detector and the controller. Finally, the control signal is applied to the inverted pendulum by use of the motion control box serving as the actuator.

By use of linearization and setting the sampling period as 10 ms, $x(k) := \left[ {\alpha(k);\theta(k);\dot \alpha(k);\dot \theta(k)} \right] \in \mathcal{R}^{4}$ is set as the state of the cart position, pendulum angle, cart velocity and pendulum angular velocity, and $u(k) = \ddot \alpha (k) \in \mathcal{R}$ is set as the control input. Considering the process noise $n(k)$ with the covariance\footnote{We cannot determine the real covariance matrix of the process noise. Our solution is to preset the process noise covariance matrix and form the estimator using prior knowledge. If the system operates stably under the process noise and corresponding estimator, then we will use them.} $\Sigma_{n}=diag\{10^{-5},10^{-5}\}$, the discrete-time model of the inverted pendulum as described in (\ref{eq2A1}) is
\[\begin{gathered}
  A = \left[ {\begin{array}{*{20}{c}}
  1&0&0.0100&0 \\
  0&1.0015&0&0.0100 \\
  0&0&1&0 \\
  0&0.2945&0&1.0015
\end{array}} \right],B = \left[ {\begin{array}{*{20}{c}}
  {0} \\
  {0.0002} \\
  {0.0100} \\
  {0.0300}
\end{array}} \right], \hfill \\
  \Gamma=[0,0;0,0;1,0;0,1]. \hfill \\
\end{gathered}\]

By use of the camera as the sensor, the states of the cart position and the pendulum angel can be retrieved, i.e., $C=[I,0]$ in (\ref{eq2A1}). The measurement noise $v(k) \in \mathcal{R}^{2}$ in (\ref{eq2A1}) with the covariance matrix $\Sigma_{v}=diag\{2.7\times 10^{-7},5.5\times 10^{-6}\}$ from the computational error \cite{NIP} is also introduced, which is analysed in Section \uppercase\expandafter{\romannumeral+2}.A of the supplementary materials.

The watermarking $w_{y,i}(k)$ shown in (\ref{eq3A1}) and (\ref{eq3A2}) is generated by the C++ function \emph{gaussrand(mean, std)}\footnote{The detailed process of \emph{gaussrand(mean, std)} can be found in https://encyclopedia.thefreedictionary.com/Marsaglia+polar+method.} based on Algorithm 1, where ``mean'' and ``std'' are the mean value (i.e. zero) and standard deviation of $w_{y,i}(k)$, respectively. The seed of function \emph{rand()} in \emph{gaussrand(mean, std)} takes the value $1$.

The estimator parameters in (\ref{eq3A8}) and ($\ref{eq3A10}$) are $A$, $B$, $C$ and
\[
L = [0.2951,0;0,0.1673;5.1094,0;0,1.5290].
\]

We set $Q=diag\{10,10,10,10\}$, $R=1$, and the controller gain shown in ($\ref{eq3A9}$) is
\[K=[2.8889,-36.6415,4.9141,-7.3267].\]

It should be noted that in the platform, the entry angle is due to the finite field of vision of the camera, and the limit position due to the fixed range of cart movement. Specifically, we have $|\theta | < 0.8$ rad, $|\alpha | < 0.3$ m. Once the angle crosses the entry level or the cart position is over the limit, the system will trigger self termination (i.e., the servo is put OFF). Moreover, there is a situation where if the cart velocity or the pendulum angular velocity is too large, the cart position will be put BACK to the original point (i.e., zero) and then the control action is ended. These are shown in real-time experiments below.

\subsection{FDIA Detection Effectiveness of NIPVSSs Based on the Conventional and New DW Schemes}
We select the FDIA (\ref{eq2B1}) emerging at $k \geqslant 2$, where ${A_a} = diag\{ 0.1,0.1,0.1,0.1\}$ and ${x_a}(2) = \left[ {10^{-7};0;0;10^{-7}} \right]$. We set $\sigma^2_{w_d}=\sigma^2_{w_{y,i}}=0.0001$ for the conventional and new DW schemes. The time window size is set as $T=5$, the detection thresholds for the conventional and new DW schemes are set as ${\vartheta _{d,1}} = 0.0002$, ${\vartheta _{d,2}} = 0.0015$, and ${\vartheta _{1,i}} = {\vartheta _2} = 0.0007$, experimentally.

\begin{figure}[!t]
  \centering
  \subfigure[Cart position $\alpha(k)$]{\includegraphics[width=0.48\textwidth]{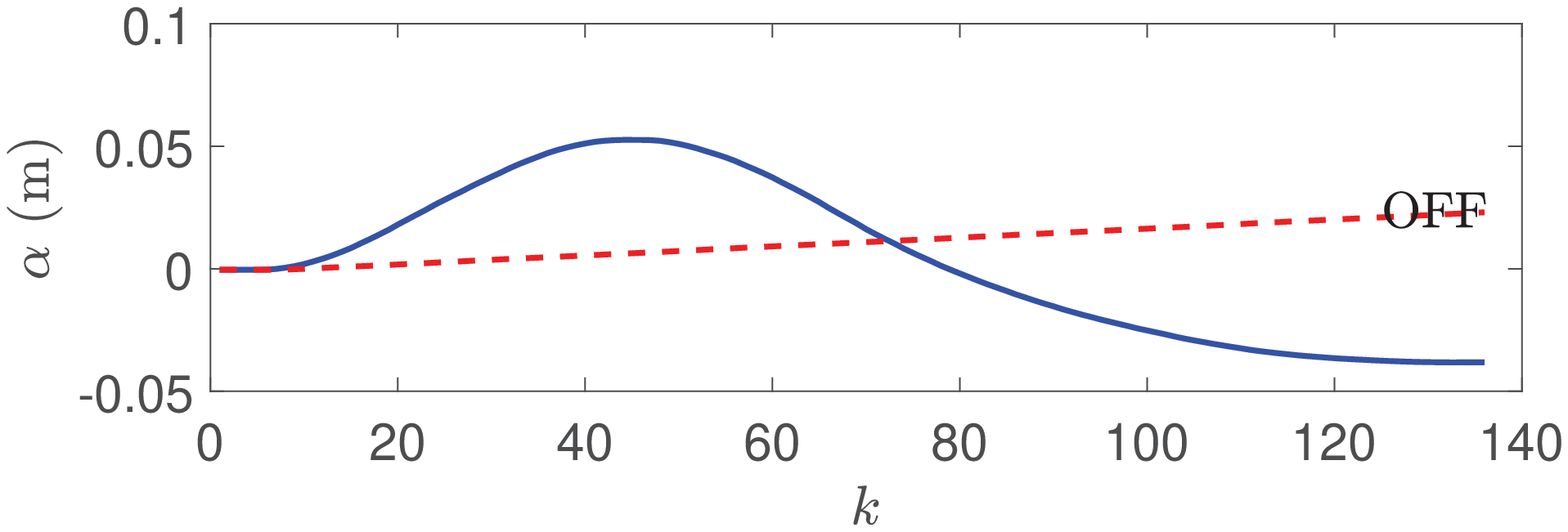}}  \\ \vspace{-0.15in}
  \subfigure[Pendulum angle $\theta(k)$]{\includegraphics[width=0.48\textwidth]{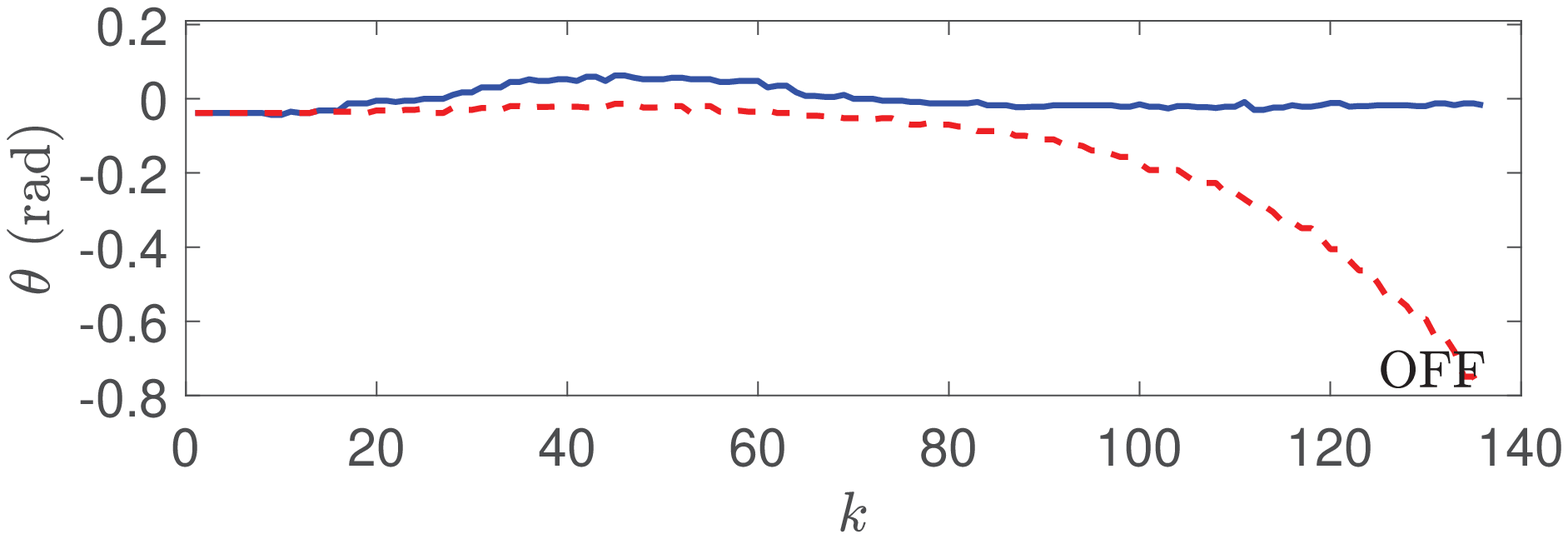}} \\ \vspace{-0.15in}
  \subfigure[DW statistical test 1 $\varphi _{d,1}(k)$]{\includegraphics[width=0.48\textwidth]{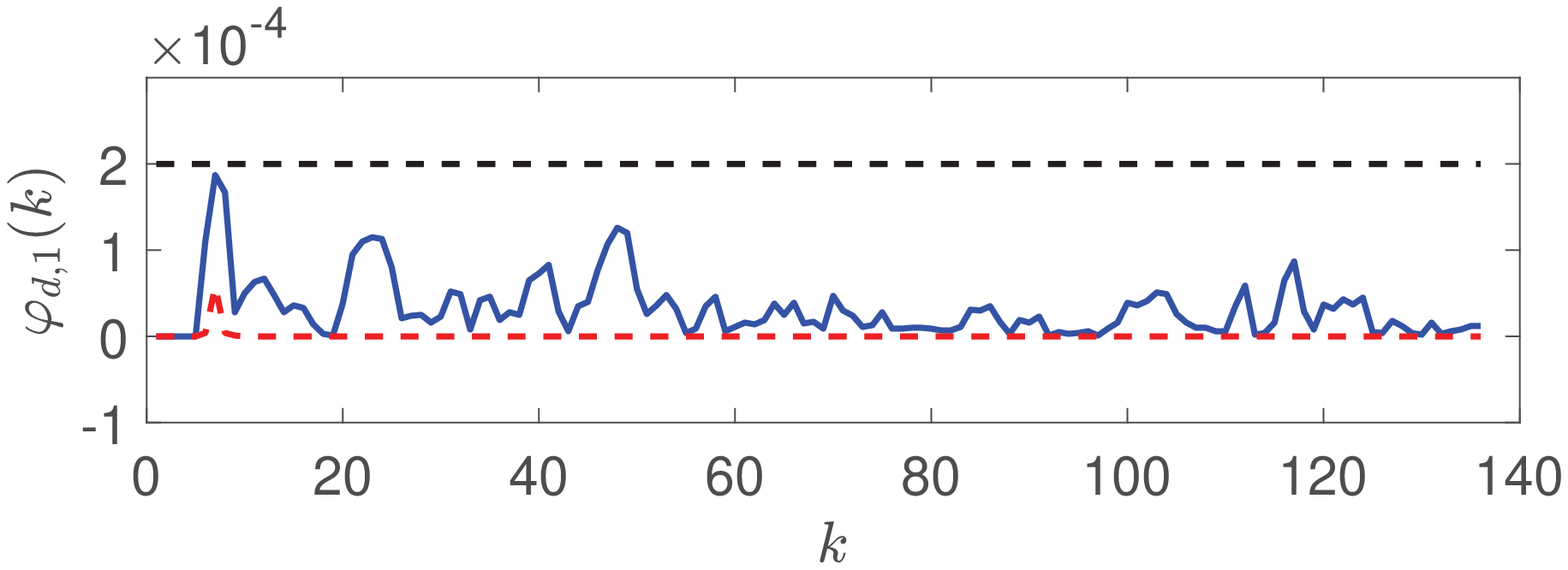}} \\ \vspace{-0.15in}
  \subfigure[DW statistical test 2 $\varphi _{d,2}(k)$]{\includegraphics[width=0.48\textwidth]{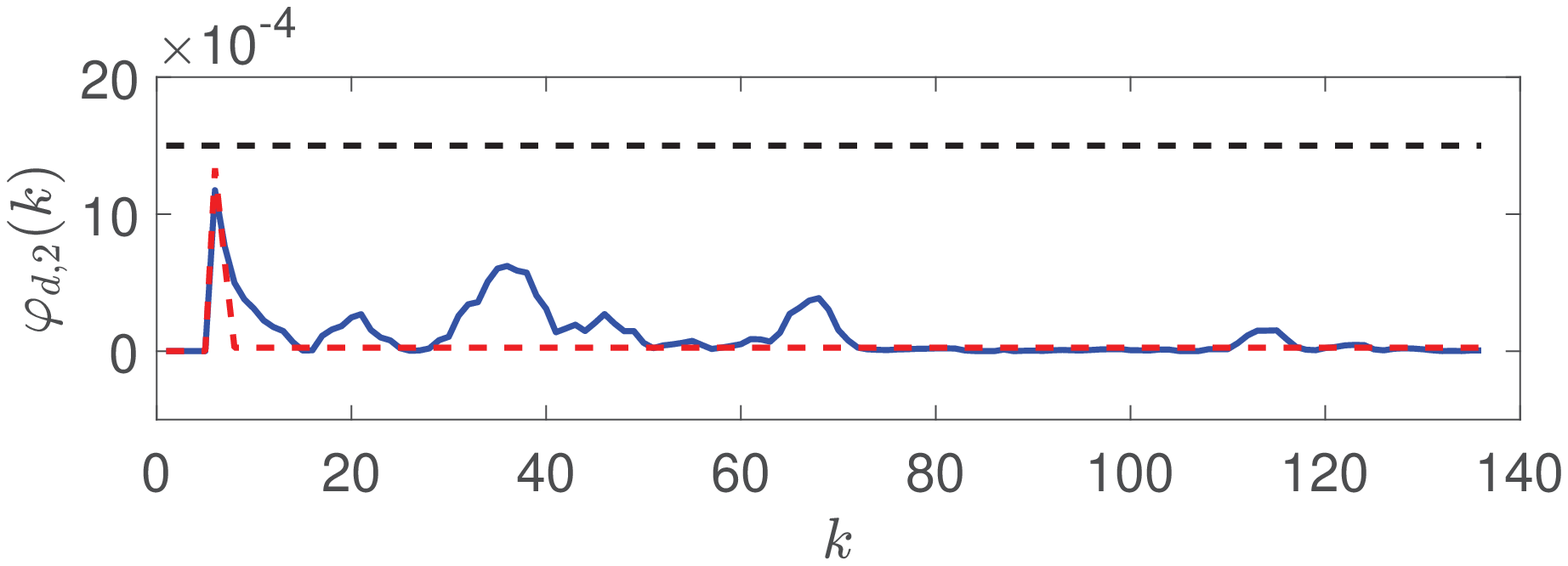}} \\
  \caption{States and DW statistical tests of NIPVSSs under the FDIA (\ref{eq2B1}) emerging at $k \geqslant 2$ based on the conventional DW scheme with $\sigma^2_{w_d} = 0.0001$. Blue line: Normal system. Red line: System under the FDIA. Black line: Detection thresholds. ``OFF'' denotes that the servo is put off.}
  \label{fig4}
\end{figure}

\begin{figure}[!t]
  \centering
  \subfigure[Cart position $\alpha(k)$]{\includegraphics[width=0.48\textwidth]{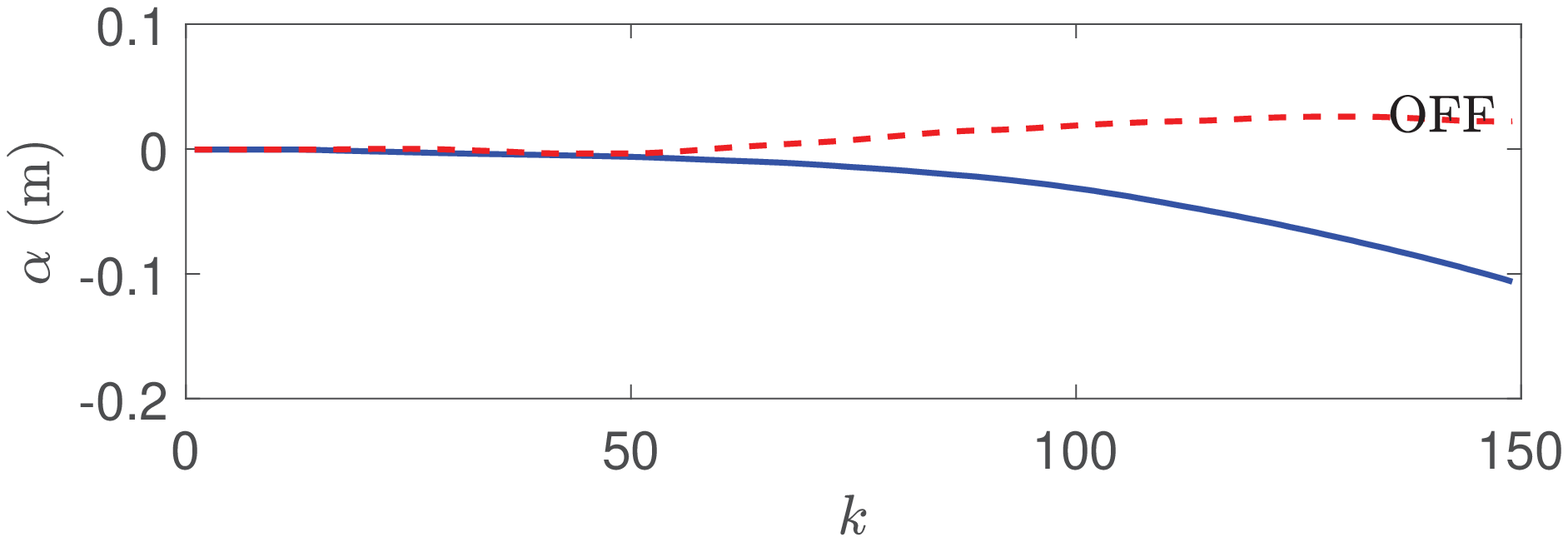}} \\ \vspace{-0.15in}
  \subfigure[Pendulum angle $\theta(k)$]{\includegraphics[width=0.48\textwidth]{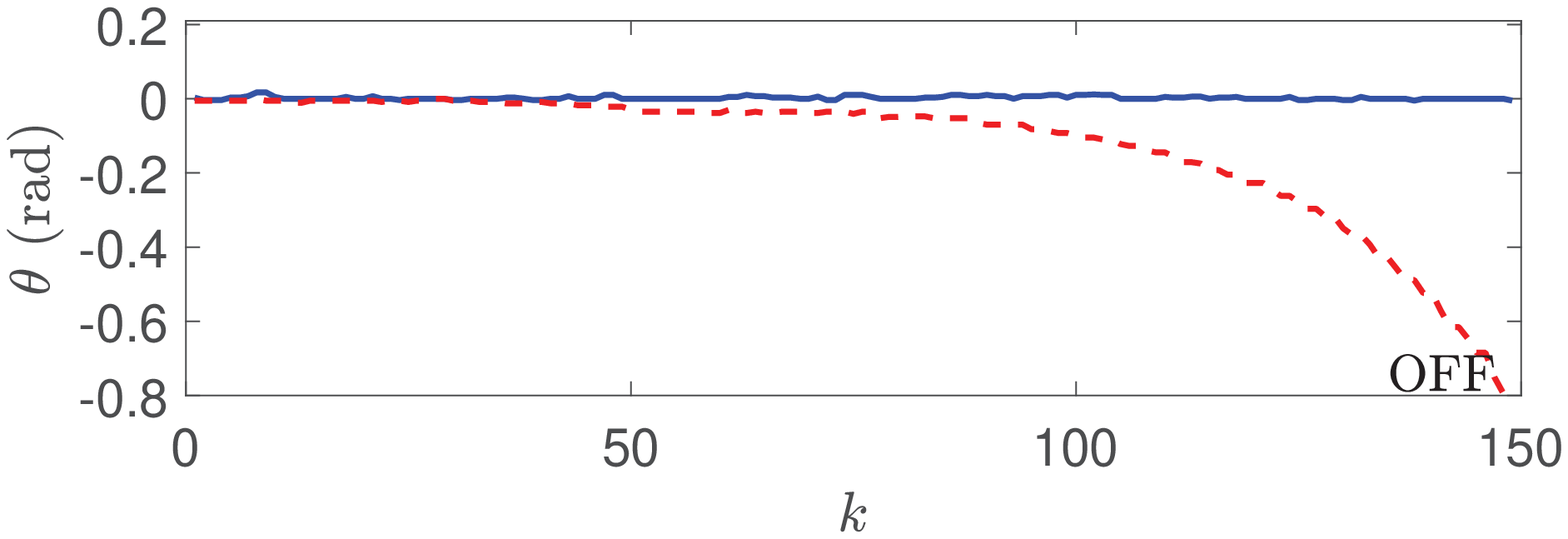}} \\ \vspace{-0.15in}
  \subfigure[New DW statistical test 1 $\varphi _{1,1}(k)$ and $\varphi _{1,2}(k)$]{\includegraphics[width=0.48\textwidth]{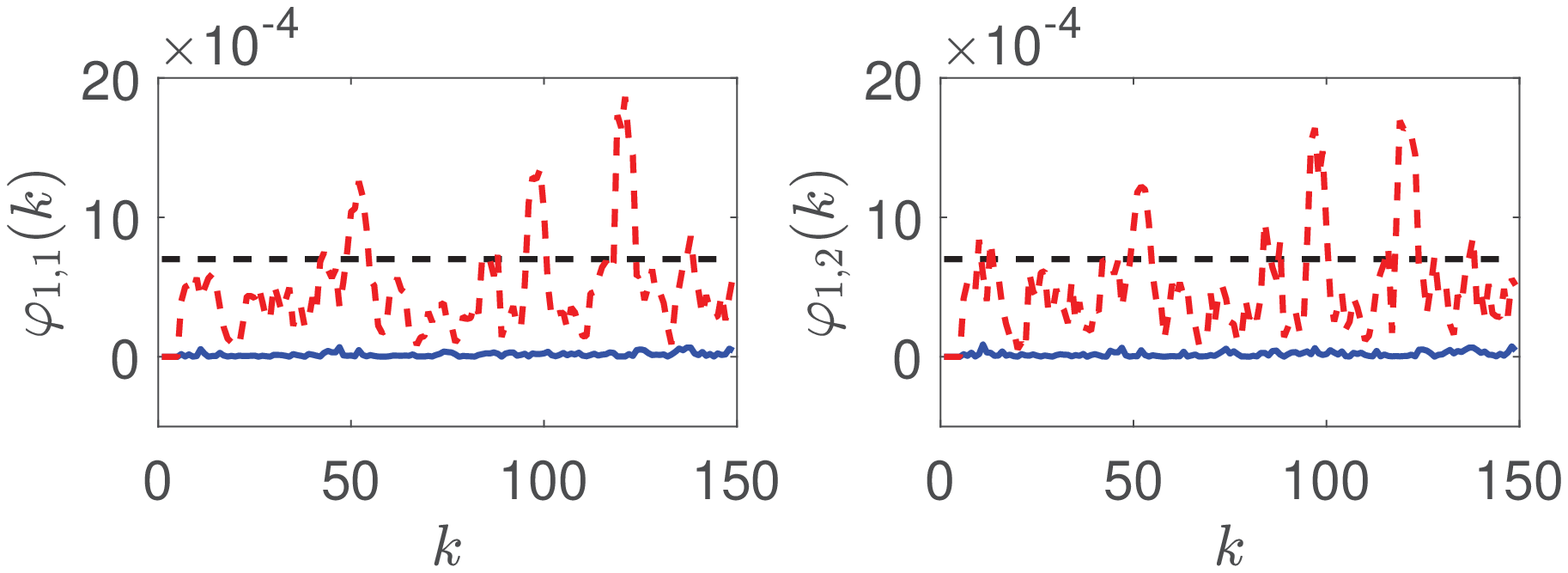}} \\ \vspace{-0.15in}
  \subfigure[New DW statistical test 2 $\varphi _{2}(k)$]{\includegraphics[width=0.48\textwidth]{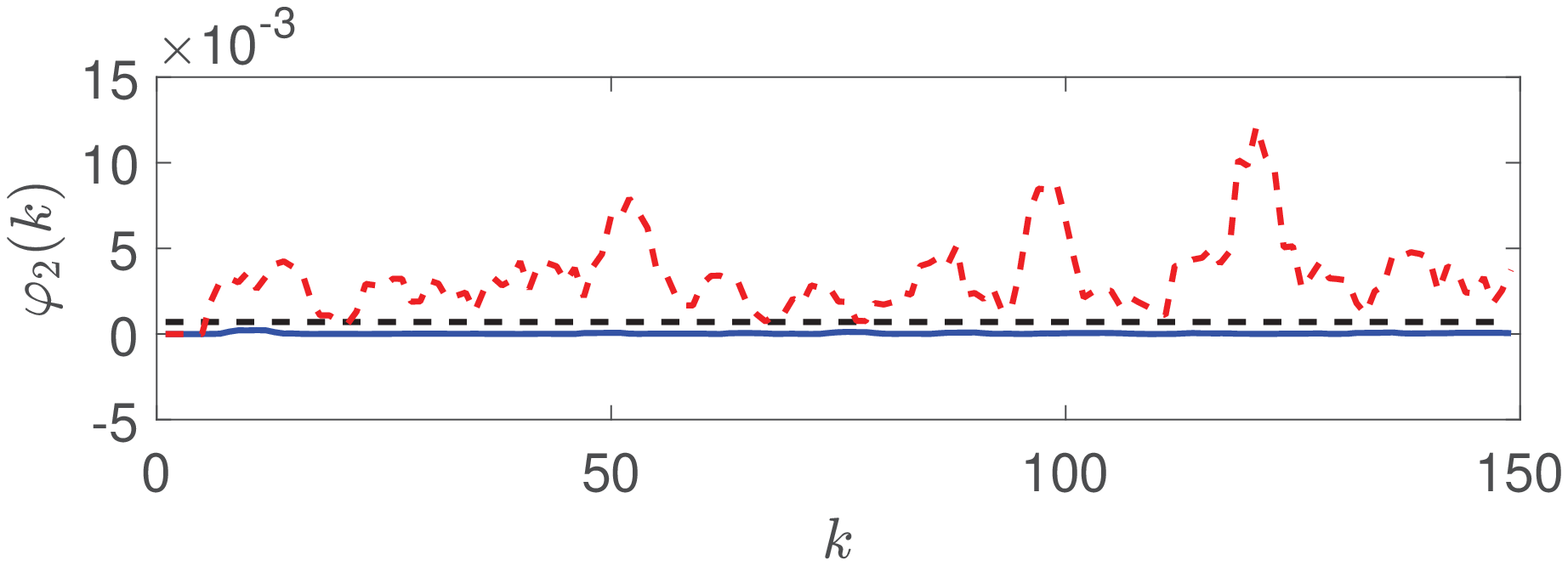}} \\
  \caption{States and the new DW statistical tests of NIPVSSs under the FDIA (\ref{eq2B1}) emerging at $k \geqslant 2$ based on the new DW scheme with $\sigma^2_{w_{y,i}} = 0.0001$. Blue line: Normal system. Red line: System under the FDIA. Black line: Detection thresholds. ``OFF'' denotes that the servo is put off.}
  \label{fig5}
\end{figure}

Figs.~\ref{fig4} and \ref{fig5} present the experiment results of applying the FDIA to NIPVSSs based on the conventional and new DW schemes. When there is FDIA, the states can be separated into two situations. However, the conventional DW scheme fails to detect the FDIA, but the new DW scheme succeeds in detecting the FDAI. Table~\ref{Tab1} shows the FDIA detection effectiveness and system performance loss from watermarking of NIPVSSs based on the conventional and new DW schemes, where it can be seen that compared with the conventional DW scheme, the new DW scheme provides much better FDIA detection effectiveness and zero system performance loss from watermarking.
\begin{table}[t]
\centering
\caption{FDIA Detection Effectiveness and System Performance Loss from Watermarking of NIPVSSs with $\sigma^2_{w_d} = \sigma^2_{w_{y,i}} = 0.0001$}
\label{Tab1}
\begin{tabular}{lcc}
\toprule
  Indices ($\sigma^2_d = \sigma^2_{w_{y,i}} = 0.0001$) &Normal &under FDIA \\
\midrule
  CDW$^1$: ${\left\| {\mathbb{E}\left[ {w_d(k - 1)Lr_d(k)} \right]} \right\|_F}$                                              & 0         & 3.4459E-8\\
  CDW: $\left| {tr\left( {\mathbb{E}\left[ {Lr_d(\infty ){{\left( {Lr_d(\infty )} \right)}^T}} \right]} \right)} \right|$ & 2.5660E-5 & 4.1303E-8\\
  CDW: System performance loss                                                                                           & 33.67\%   & -        \\
  New DW: ${\left\| {\mathbb{E}\left[ {{w_{y,1}}(k)Lr(k)} \right]} \right\|_F}$                                              & 0         & 5.1179E-4\\
  New DW: ${\left\| {\mathbb{E}\left[ {{w_{y,2}}(k)Lr(k)} \right]} \right\|_F}$                                              & 0         & 1.5381E-4\\
  New DW: $\left| {tr\left( {\mathbb{E}\left[ {Lr(\infty ){{\left( {Lr(\infty )} \right)}^T}} \right]} \right)} \right|$     & 2.5660E-5 & 3.4152E-1\\
  New DW: System performance loss                                                                                           & 0         & -        \\
\bottomrule
\multicolumn{3}{l}{$^1$ Conventional DW.}\\
\end{tabular}
\end{table}

Computational time complexity of the new DW tests (\ref{eq3A3}) and (\ref{eq3A4}) and the residual-based test \cite{DAD} is also compared in the experiments of Fig.~\ref{fig5}, which is listed in Tab.~A.I of Section II.B in the supplementary materials. Tab.~A.I shows that the proposed new DW Tests require more time than the residual-based test when $m_y=2 < m_x=4$. However, the microsecond-level running time of the new DW Tests is negligible in comparison the control cycle (10 ms). Therefore, computational time in the use of the new DW Tests is not significant.

To investigate the FDIA under different initial instants, we perform the same experiments of Figs.~\ref{fig4} and \ref{fig5} again, with the FDIA emerging at $k \geqslant 400$. Experiment results are shown in Figs. A.3 and A.4 of Section~\uppercase\expandafter{\romannumeral2}.C in the supplementary materials, which indicates that when the initial instant of the FDIA is close to the initial instant of system operation, the conventional DW scheme with $\sigma^2_{w_d}=0.0001$ cannot detect the FDIA while the new DW scheme with $\sigma^2_{w_{y,i}}=0.0001$ can detect the FDIA; when the initial instant of the FDIA is far from the initial instant of system operation, both the conventional DW scheme with $\sigma^2_{w_d}=0.0001$ and the new DW scheme with $\sigma^2_{w_{y,i}}=0.0001$ can detect the FDIA.

Further analysis of FDIA detection effectiveness on the conventional and the new DW schemes with different watermarking density is carried out. Due to space constraints and their similarity to Figs.~\ref{fig4} and \ref{fig5}, the outcomes from further analysis are omitted here. The results show that (i) the conventional DW scheme cannot detect the FDIA even though $\sigma ^2_{w_d} = 10$ where the system states have severe shaking; (ii) a much larger watermarking density brings better FDIA detection effectiveness by the new DW scheme.

\subsection{Performance of NIPVSSs Based on the New DW Scheme}
To present the performance of NIPVSSs based on the new DW scheme under the FDIA, we select the FDIA (\ref{eq2B1}) taking place at $k=100, 101, 102, 103$, whose parameters are set as ${A_a} = diag\{ 0.1,0.1,0.1,0.1\}, {x_a}(100) = \left[ {2;2;2;2} \right]$. We set $\sigma^2_{w_{y,i}} = 0.0001$ for the new DW scheme. The time window size is set as $T=5$, and the detection thresholds for the new DW scheme are set as ${\vartheta _{1,i}} = {\vartheta _2} = 0.0007$.
\subsubsection{Performance of NIPVSSs Under No Attacks}
We firstly examine the impact of the conventional and new DW schemes with different watermarking densities on the performance of NIPVSSs under no attacks, shown in Fig.~A.5 of Section \uppercase\expandafter{\romannumeral2}.D in the supplementary materials. It is shown that when the watermarking density increases to $10$, the conventional DW scheme brings $33671.78\%$ system performance loss where the system states have severe shaking, but zero system performance loss is yielded by the new DW scheme.

\subsubsection{Performance of NIPVSSs Based on the New DW Scheme without Compensation under the FDIA}
When the compensation is ignored, the stability analysis is presented in Theorem 3. There always exist non-singular matrices $V_0$ and $V_1$ that satisfy
\[{\mathcal{A}_0} = {V_0}{D_0}V_0^{ - 1},{\mathcal{A}_1} = {V_1}{D_1}V_1^{ - 1}\]
where matrices ${D_0} = diag\{ {\lambda _{1}}({\mathcal{A}_0}), \ldots ,{\lambda _{8}}({\mathcal{A}_0})\}$ and ${D_1} = diag\{ {\lambda _{1}}({\mathcal{A}_1}), \ldots ,{\lambda _{8}}({\mathcal{A}_1})\}$. We select
\[\begin{gathered}
  {\lambda _ + } = \max \left( {\mathcal{M} \left( {{\lambda _i}({\mathcal{A}_0})} \right)} \right) = 5.4250, \hfill \\
  {\lambda _ - } = \max \left( {\mathcal{M} \left( {{\lambda _i}({\mathcal{A}_1})} \right)} \right) = 0.9895, \hfill \\
  {g_0} = \frac{{\ln \left( {{{{\mathcal{S} _{\max }}({V_0})} \mathord{\left/
 {\vphantom {{{\mathcal{S} _{\max }}({V_0})} {{\mathcal{S} _{\min }}({V_0})}}} \right.
 \kern-\nulldelimiterspace} {{\mathcal{S} _{\min }}({V_0})}}} \right)}}{{\ln {\lambda _ - }}} =  - 3879.8947, \hfill \\
  {g_1} = \frac{{\ln \left( {{{{\mathcal{S} _{\max }}({V_1})} \mathord{\left/
 {\vphantom {{{\mathcal{S} _{\max }}({V_1})} {{\mathcal{S} _{\min }}({V_1})}}} \right.
 \kern-\nulldelimiterspace} {{\mathcal{S} _{\min }}({V_1})}}} \right)}}{{\ln {\lambda _ - }}} =  - 614.4731 \hfill \\
\end{gathered} \]
where $\mathcal{M}(\cdot)$ is the modulus of a complex number, $\mathcal{S}_{\min}(\cdot)$ ($\mathcal{S}_{\max}(\cdot)$) is the minimal (maximal) singular value of a matrix. Hence $g = \min \{ {g_0},{g_1}\}  = {g_0}$. Now, by taking $\lambda^{*}=0$, the infimum of the value of $\mathop {\inf }\limits_{k \geqslant 0} \left[ {{{{\mathcal{T}_1}} \mathord{\left/
 {\vphantom {{{\mathcal{T}_1}} {{\mathcal{T}_0}}}} \right.
 \kern-\nulldelimiterspace} {{\mathcal{T}_0}}}} \right]$ can be obtained as follows
\[\mathop {\inf }\limits_{1 > {\lambda ^ * } \geqslant 0} \mathop {\inf }\limits_{k \geqslant 0} \left[ {{{{\mathcal{T}_1}} \mathord{\left/
 {\vphantom {{{\mathcal{T}_1}} {{\mathcal{T}_0}}}} \right.
 \kern-\nulldelimiterspace} {{\mathcal{T}_0}}}} \right]\mathop  = \limits^{{\lambda ^ * } = 0} 159.4495.\]
The experiment is carried out at $k \in [0,140]$, and here $\mathcal{T}_0=4$, $\mathcal{T}_1=141-\mathcal{T}_0=137$. Since ${{{\mathcal{T}_1}} \mathord{\left/
 {\vphantom {{{\mathcal{T}_1}} {{\mathcal{T}_0}}}} \right.
 \kern-\nulldelimiterspace} {{\mathcal{T}_0}}} = 34.25 < 159.4495$, the condition (\ref{eq3D6}) is not satisfied. This means that it is possible for the FDIA (\ref{eq2B1}) taking place at $k=100,101,102,103$ to influence the stability of the NIPVSS.

\begin{figure}[!t]
  \centering
  \subfigure[Cart position $\alpha(k)$]{\includegraphics[width=0.48\textwidth]{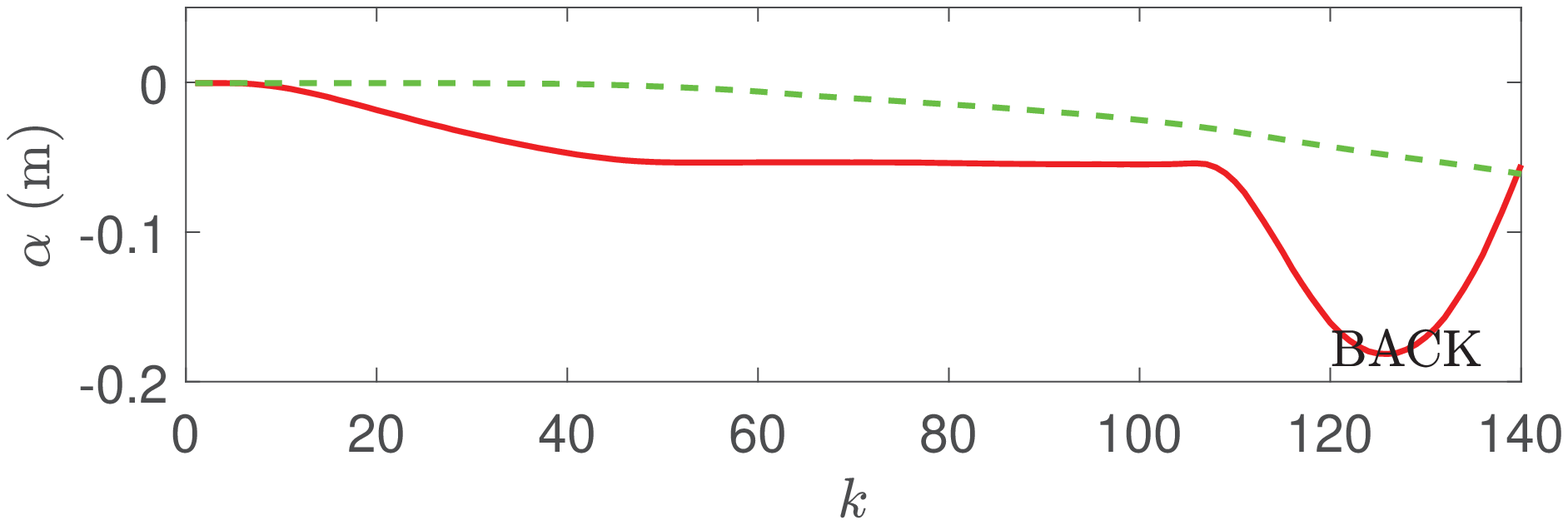}}  \\ \vspace{-0.15in}
  \subfigure[Pendulum angle $\theta(k)$]{\includegraphics[width=0.48\textwidth]{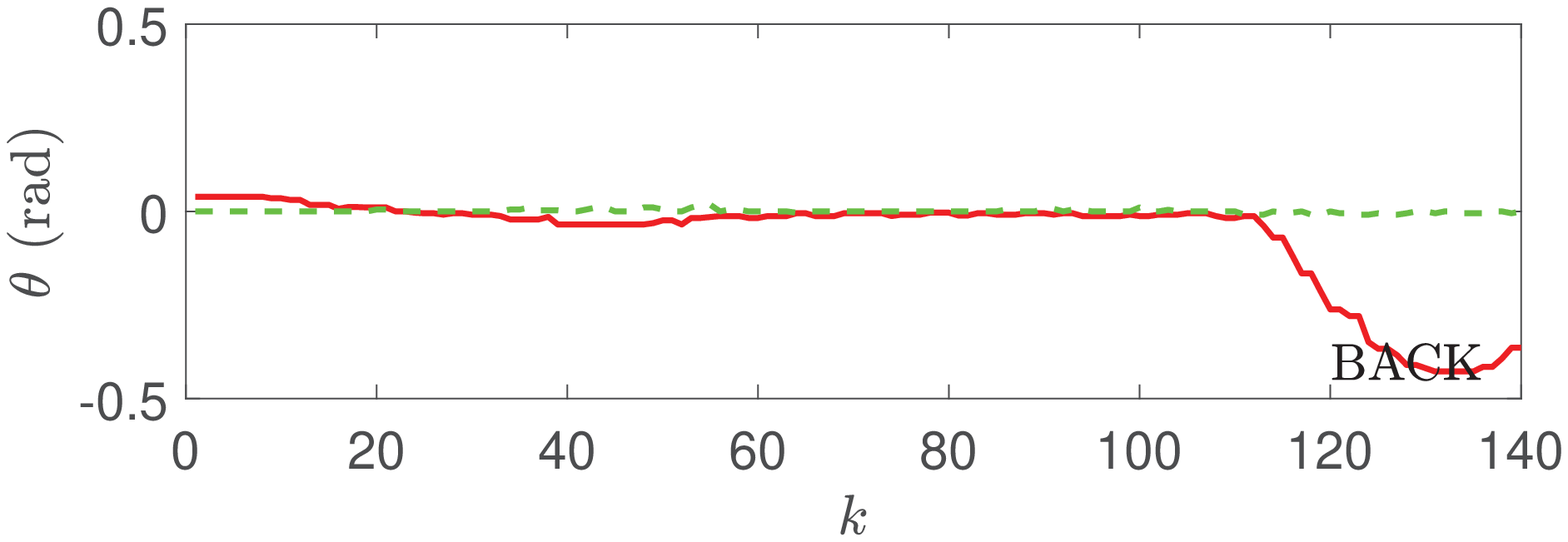}} \\ \vspace{-0.15in}
  \subfigure[New DW statistical test 1 $\varphi _{1,1}(k)$ and $\varphi _{1,2}(k)$]{\includegraphics[width=0.48\textwidth]{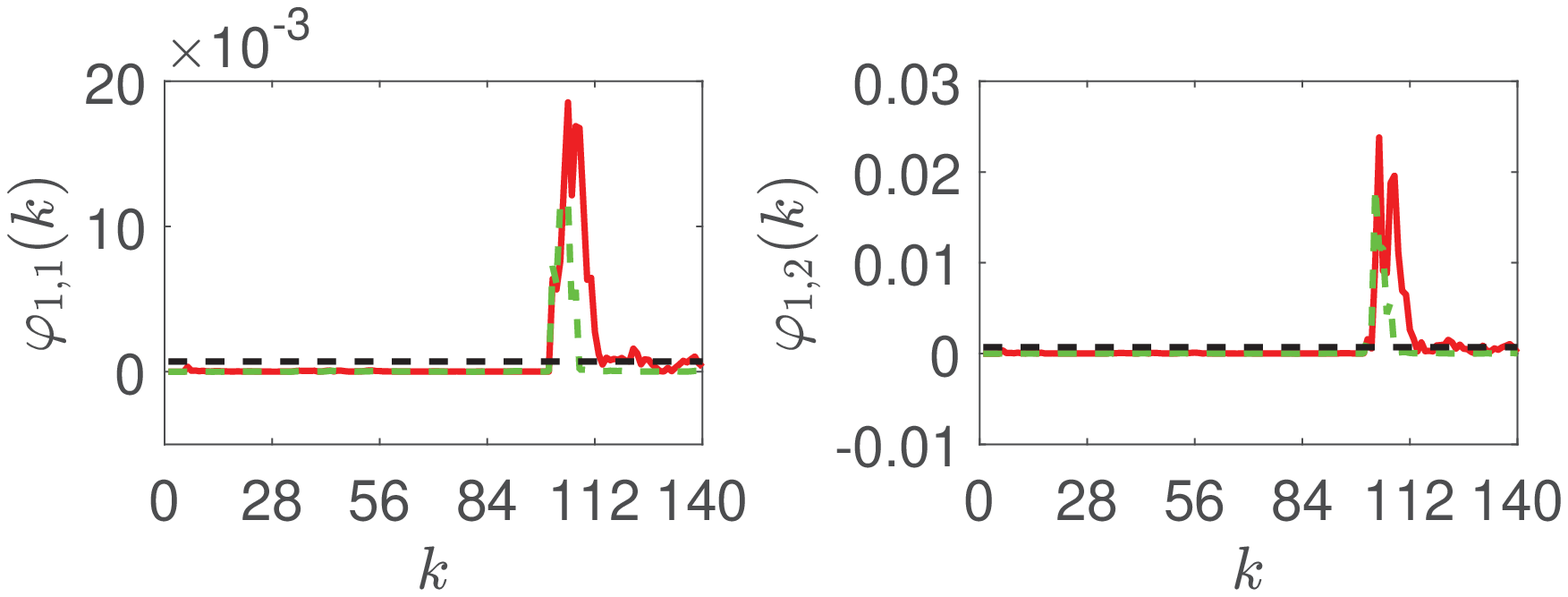}} \\ \vspace{-0.15in}
  \subfigure[New DW statistical test 2 $\varphi _{2}(k)$]{\includegraphics[width=0.48\textwidth]{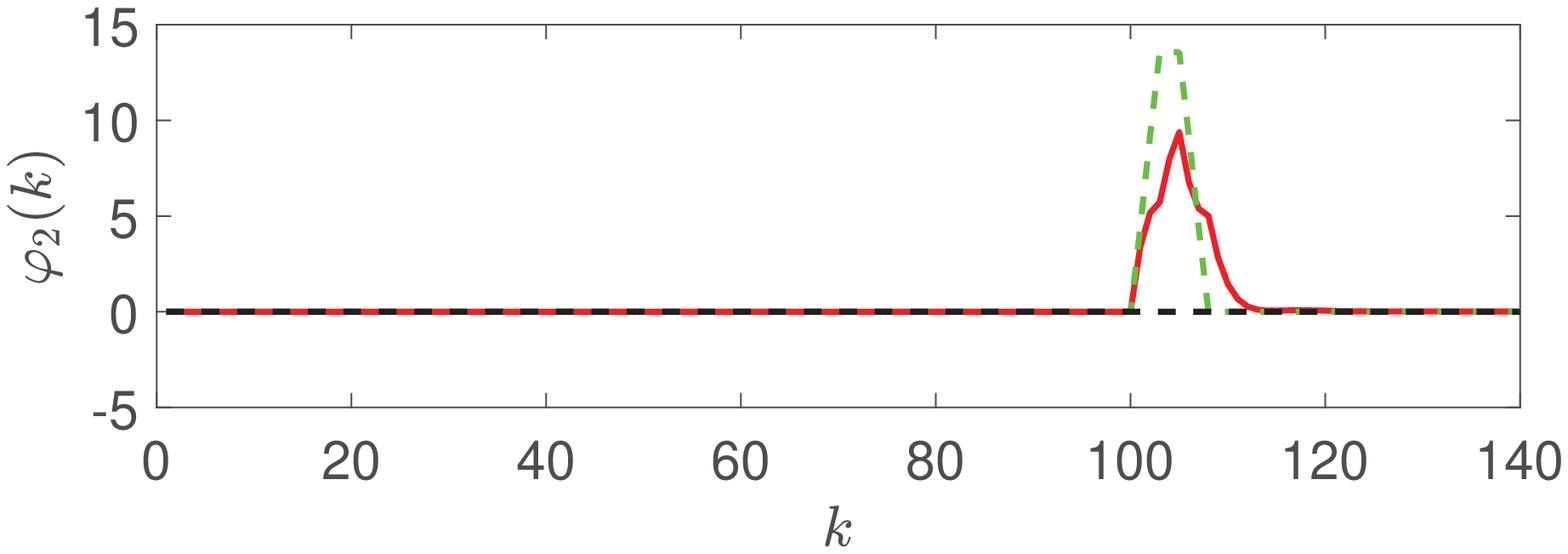}} \\
  \caption{States and the new DW statistical tests of NIPVSSs under the FDIA (\ref{eq2B1}) taking place at $k=100,101,102,103$ based on the new DW scheme with $\sigma^2_{w_{y,i}} = 0.0001$. Red line: System without compensation. Green line: System with compensation. Black line: Detection thresholds. ``BACK'' indicates that the cart position is put back to zero.}
  \label{fig6}
\end{figure}

Fig.~\ref{fig6} presents the experimental results of applying the FDIA (\ref{eq2B1}) taking place at $k=100,101,102,103$ to the NIPVSS without compensation. As shown as the red lines in Fig.~\ref{fig6}, the states of the NIPVSS become divergent without compensation.

\subsubsection{Performance of NIPVSSs Based on the New DW Scheme with Compensation under the FDIA}
Fig.~\ref{fig6} also presents the experimental results of applying the FDIA (\ref{eq2B1}) taking place at $k=100,101,102,103$ to the NIPVSS with compensation. As shown as the green lines in Fig.~\ref{fig6} (a), (b), the states of the NIPVSS keep stable. Note that the detection indicators $\varphi _{1,1}(k)$ and $\varphi _{1,2}(k)$ shown in Fig.~\ref{fig6} (c) validate Corollary 1, i.e., with the compensation used, the new DW scheme has a sufficient ability to detect the FDIA (\ref{eq2B1}).

\begin{figure}[!t]
  \centering
  \includegraphics[width=0.488\textwidth]{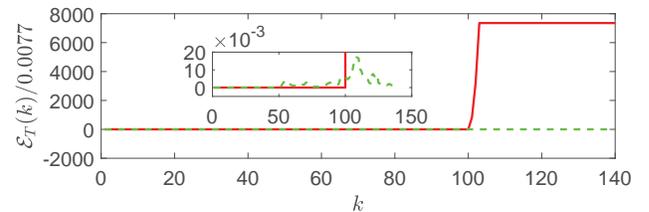}
  \caption{Values of $\mathcal{E}_{T}(k)/0.0077$ without or with compensation. Red line: without compensation. Green line: with compensation.}
  \label{fig7}
\end{figure}

By solving Corollary 2, when the compensation is used, we can obtain $\bar h=4$ with $\beta=2.9800\times 10^{2}$. The corresponding system performance is
\[\mathop {\lim \sup }\limits_{T \to \infty } \frac{1}{T}\sum\limits_{k = 0}^{T - 1} {{{\left\| {Ce(k)} \right\|}^2}}  = 0.0077.\]
To validate Corollary 2, we obtain the system performance in a finite time window $T=5$, i.e.,
\[
\mathcal{E}_{T}(k)=\frac{1}{T}\sum\nolimits_{s = k - T + 1}^k {{{\left\| {Ce(s)} \right\|}^2}}.
\]
Fig.~\ref{fig7} presents the values of $\mathcal{E}_{T}(k)/0.0077$ without and with compensation. It can be seen from Fig.~\ref{fig7} that: 1) the system performance $\mathcal{E}_{T}(k)/0.0077$ is recovered with a very big transient value about 8000 (unstable mode) to a very small transient value about 0.02 (stable mode), and 2) the value of $\mathcal{E}_{T}(k)/0.0077$ with compensation is less than $0.02$, which means that the real system performance is far better than the theoretical system performance.

To further investigate the compensated variables based on the new DW scheme, we define the detection indictor on $\tilde r(k) := \tilde y(k) - C\hat x(k|k-1)$ as follows
\[{{\tilde \varphi }_2}(k) := \left| {tr\left( {\frac{1}{T}\sum\limits_{p = k - T + 1}^k {L\tilde r(p){{\left( {L\tilde r(p)} \right)}^T}}  - L{\Sigma _o}{L^T}} \right)} \right|.\]
Fig.~A.6 of the supplementary materials presents the values of ${{\tilde \varphi }_2}(k)$, where it can be seen that the detection indictor ${{\tilde \varphi }_2}(k)$ is less than the detection threshold by use of compensation.

\section{Conclusion}
A new DW scheme was proposed for NCSs secure control. Firstly, the security weaknesses of the conventional DW scheme was analysed. Secondly, to overcome the security weakness, a framework of NCSs secure control based on the new DW scheme was designed, which integrated the watermarking as symmetric-key encryption and new DW tests and compensation mechanism. Then, by using the additive distortion power of the closed-loop system, whether attacks can be detected was analysed. Furthermore, the positive correlation between the FDIA detection effectiveness and the watermarking intensity was explored by using the cross covariance of watermarking and residuals and auto covariance of residuals, where zero performance loss from watermarking was yielded thanks to watermarking as symmetric-key encryption. Thirdly, the tolerance capacity of the FDIA against the system was discussed and it was shown that the system performance can be recovered from the FDIA, where the quantitative relationship between the new DW scheme and system performance was studied. Finally, the proposed scheme was applied to a real inverted pendulum system, and experimental results demonstrated its superior performance.

To deal with the weaknesses of the conventional DW scheme, we have studied the secure control of NCSs by a new developed DW scheme. However, there exist some NCSs with many subsystems (e.g., networked multi-agent systems \cite{RFD,NNB}), where information exchange is performed among subsystems via communication networks. If the communication networks between subsystems are attacked, the performance of the subsystems or the whole system goes down. Therefore, it is interesting to investigate the secure control of such NCSs based on the proposed DW scheme.


%


\ifCLASSOPTIONcaptionsoff
  \newpage
\fi



%
%
%
\bibliographystyle{IEEEtran}
\bibliography{IEEEabrv,mybibfile}
\end{document}


%
\title{ Supplementary Materials---Secure Control of Networked Control Systems Using Dynamic Watermarking}
%
%
%

\author{Dajun Du,
        Changda Zhang,
        Xue Li,
        Minrui Fei,
        Taicheng Yang,
        Huiyu Zhou
\thanks{The work of D. Du, C. Zhang, X. Li, M. Fei, T. Yang, and H. Zhou was supported by the National Science Foundation of China under Grant Nos. 92067106, 61773253, 61633016 and 61533010, 111 Project under Grant No.D18003.}
\thanks{D. Du, C. Zhang, X. Li, and M. Fei are with Shanghai Key Laboratory of Power Station Automation Technology, School of Mechatronic Engineering and Automation, Shanghai University, Shanghai 200444, China.}
\thanks{T. Yang is with Department of Engineering and Design, University of Sussex, Brighton BN1 9QT, U.K.}
\thanks{H. Zhou is with School of Informatics, University of Leicester, Leicester LE1 7RH, U.K.}
        }


%
%

\markboth{IEEE Transactions on Cybernetics}%
{Shell \MakeLowercase{\textit{et al.}}: Bare Demo of IEEEtran.cls for IEEE Journals}
%



\maketitle

\begin{abstract}
This is the supplementary document of the paper entitled ``Secure Control of Networked Control Systems Using Dynamic Watermarking'' submitted to IEEE Transactions on Cybernetics. Section \uppercase\expandafter{\romannumeral+1} provides the proofs of limitations and theorems. Supplements of experimental results on the measurement noise and system performance with different watermarking densities under no attacks are given in Section \uppercase\expandafter{\romannumeral+2}.
\end{abstract}

\IEEEpeerreviewmaketitle
\section{Proofs of Limitations and Theorems}

\subsection{Proof of Limitations 1 and 2}
Considering the system (12) in the main text, to analyse the ability of the conventional DW scheme for FDIA detection, we firstly define $\xi_d (k) := {x_{a}}(k) - \hat x_d(k|k - 1)$. The dynamics of $\xi_d (k)$ can be given by
\bea
\left\{ \begin{gathered}
  \xi_d (k) = {\Phi _1}\xi_d (k - 1) + {\Phi _2}{x_{a}}(k - 1) - B w_d(k - 1) \hfill \\
  r_d(k) = C\xi_d (k) \hfill \\
\end{gathered}  \right.
\label{eqAA1}
\eea
where ${\Phi _1}$, ${\Phi _2}$ are the same as (12) in the main text. Then, using (\ref{eqAA1}), the variable $\xi_d (k)$ can be computed recursively from $0$ to $k$ as follows
\bea
\begin{aligned}
  \xi_d (k) &= \Phi _1^k\xi_d (0) \hfill \\
            &+ \sum\limits_{p = 1}^k {\left( {\Phi _1^{p - 1}{\Phi _2}{x_{a}}(k - p) - \Phi _1^{p - 1}B w_d(k - p)} \right)}. \hfill \\
\end{aligned}
\label{eqAA2}
\eea
Considering (\ref{eqAA1}) and $w_d(k - 1) \bot \xi_d (0)$, $w_d(k - 1) \bot {x_{a}}(j)$ for $\forall j$, and $w_d(k - 1) \bot w_d(j)$ for $j \ne k - 1$ in (\ref{eqAA2}), it shows that
\begin{gather}
\mathbb{E}\left[ {w_d(k - 1)L r_d(k)} \right] = -\mathbb{E}\left[ {w_d(k - 1)LCBw_d(k - 1)} \right].
\label{eqAA3}
\end{gather}
The above equation yields (15a) in the main text.

Secondly, we focus on the steady-state behavior of (\ref{eqAA1}). Note that $\mathop {\lim }\limits_{k \to \infty } {x_{a}}(k) = 0$ of the FDIA (11) in the main text; taking the limitation as $k \to \infty$ for (\ref{eqAA1}) yields
\bea
\left\{ \begin{gathered}
\mathop {\lim }\limits_{k \to \infty } \xi_d (k) = \mathop {\lim }\limits_{k \to \infty } {\Phi _1}\xi_d (k - 1) - B w_d(k - 1) \hfill \\
\mathop {\lim }\limits_{k \to \infty } r_d(k) = \mathop {\lim }\limits_{k \to \infty } C\xi_d (k). \hfill
\end{gathered} \right.
\label{eqAA4}
\eea
Denote $\mathbb{E}\left[ {\xi_d (\infty ){\xi_d ^T}(\infty )} \right] := M_d$ and consider $w_d(k-1) \bot \xi_d(k-1)$, then multiplying the both sides of the equal sign by the corresponding transposed vector shown in (\ref{eqAA4}) and taking the expectation yields (15b) in the main text.

Up to now, limitation 1 has been proved. The following is the proof of limitation 2.

To investigate the action of the system state $x(k)$, in the main text, specifically, $\zeta_d(k)=\left[ x(k);\xi_d(k);x_a(k)\right]$, and the matrices $\mathcal{A}_0=\left[ A,{\rm H};0,\Xi\right]$ and $\Xi  = \left[ {{\Phi _1},{\Phi _2};0,{A_a}} \right]$ in (12). Since $\rho (A_a)<1$ and $\rho \left( {{\Phi _1}} \right) < 1$, it follows $\rho (\Xi)<1$, i.e., $\left[ {\xi_d (k);{x_{a}}(k)} \right]$ is exponentially bounded in the mean-square sense. Inequality $\rho (A) >1$ results in $\zeta_d(\infty) \to \infty$. Therefore, it is clear that $x(\infty) \to \infty$, i.e., (16) in the main text. It completes the proof.

\subsection{Proof of Theorem 1}
Since $y_w^{-}(k)$ passes test (20) in the main text, it follows that
\bea
\mathop {\lim }\limits_{T \to \infty } \frac{1}{T}\sum\limits_{k = 0}^{T - 1} {w_{y,i}(k)\left( {\mathcal{D}_{r}(k) + Lr^o(k)} \right)}  = 0.
\label{eqAC1}
\eea
Substituting $\mathop {\lim }\limits_{T \to \infty } \frac{1}{T}\sum\nolimits_{k = 0}^{T - 1} {{w_{y,i}}(k)L{r^o}(k)}  = 0$ into (\ref{eqAC1}) yields
\bea
\mathop {\lim }\limits_{T \to \infty } \frac{1}{T}\sum\limits_{k = 0}^{T - 1} {w_{y,i}(k)\mathcal{D}_{r}(k)}  = 0.
\label{eqAC2}
\eea
Since $y_w^{-}(k)$ also passes test (21) in the main text, it follows that
\bea
\begin{gathered}
  \mathop {\lim }\limits_{T \to \infty } \frac{1}{T}\sum\limits_{k = 0}^{T - 1} {\left( {\mathcal{D}_r (k) + Lr^o (k)} \right){{\left( {\mathcal{D}_r (k) + Lr^o (k)} \right)}^T}}  \hfill \\
   \qquad \qquad \qquad \qquad \qquad \qquad \qquad \qquad \qquad= L{\Sigma _{{o}}}{L^T}. \hfill \\
\end{gathered}
\label{eqAC3}
\eea
Substituting $\mathop {\lim }\limits_{T \to \infty } \frac{1}{T}\sum\nolimits_{k = 0}^{T - 1} {L{r^o}(k){{\left( {L{r^o}(k)} \right)}^T}}  = L{\Sigma _o}{L^T}$ into (\ref{eqAC3}) yields
\bea
\mathop {\lim }\limits_{T \to \infty } \frac{1}{T}\sum\limits_{k = 0}^{T - 1} {\mathcal{D}_r (k){{ {\mathcal{D}^T_r (k)} }} + 2\mathcal{D}_r (k){{\left( L{r^o (k)} \right)}^T}}  = 0
\label{eqAC4}
\eea
which can be re-written as the component-wise form
\bea
\begin{gathered}
  \mathop {\lim }\limits_{T \to \infty } \frac{1}{T}\sum\limits_{k = 0}^{T - 1} {{\mathcal{D}_{r,i}}(k){\mathcal{D}_{r,j}}(k) + {\mathcal{D}_{r,i}}(k)\left( {{L_{j \cdot }}{r^o}(k)} \right)}  \hfill \\
   \qquad \qquad \qquad \qquad \qquad \quad+ {\mathcal{D}_{r,j}}(k)\left( {{L_{i \cdot }}{r^o}(k)} \right) = 0 \hfill \\
\end{gathered}
\label{eqAC4c}
\eea
where $\mathcal{D}_{r,i} (k)$ is the $i$th component of $\mathcal{D}_{r} (k)$.

According to decryption (19) and state estimate calculation (25), (27) in the main text, the decryption and state estimate calculation of the attack-free system can be given as
\begin{align}
&y^{o-}_{w} (k) = y^{o+}_{w} (k)  - w_{y} (k),
\label{eqAC5} \\
&L r^o (k) =  - C\left( {A + BK} \right)\hat x^o (k - 1|k - 1) + y^{o-}_{w} (k)
\label{eqAC6}
\end{align}
where ${y_w^{o+}}(k)$ is the attack-free counterpart of ${y}^{+}_{w}(k)$. Substituting (\ref{eqAC5}) into (\ref{eqAC6}) yields
\bea
\begin{aligned}
  r_{L,i}^{o}(k) &=  - {\left( {C\left( {A + BK} \right)} \right)_{i \cdot }}\hat x^o (k - 1|k - 1) \hfill \\
                 &+ y^{o+}_{w,i} (k) - w_{y,i} (k) \hfill \\
\end{aligned}
\label{eqAC7}
\eea
where the variables $r^{o}_{L,i}(k) := L_{i\cdot}r_o (k)$, and $y^{o+}_{w,i} (k)$ is the $i$th component of $y^{o+}_{w} (k)$. From (\ref{eqAC5})-(\ref{eqAC7}) and let ${\mathcal{F}_k}: = \sigma \left( {{{\left( {\hat x^o } \right)}_{k - 2}},{{\left( {y^{o+}_{w,i} } \right)}_{k - 1}},{{\left( {w_{y,i} } \right)}_{k - 1}}} \right)$, a Markov chain can be formed by
\[\begin{gathered}
  \left( {{{\left( {\hat x^o } \right)}_{k - 2}},{{\left( {y^{o+}_{w,i} } \right)}_{k - 1}},{{\left( {w_{y,i} } \right)}_{k - 1}}} \right) \to  \hfill \\
  \qquad \quad \left( {\hat x^o (k - 1|k - 1),y^{o+}_{w,i} (k)} \right) \to r_{L,i}^{o}(k) \hfill \\
\end{gathered} \]
where the variables
\[{\left( {\hat x^o } \right)_{k - 2}} := \{ \hat x^o (k - 2|,k - 2),\hat x^o (k - 3|,k - 3), \ldots ,\hat x^o (0|0)\}\]
and ${\left( {y^{o+}_{w,i} } \right)}_{k - 1}$, ${\left( {w_{y,i} } \right)}_{k - 1}$ are well defined likewise. According to the above Markov chain, it can be given by
\bea
\begin{aligned}
  \hat r_{L,i}^{o}(k) &= \mathbb{E}\left[ {r_{L,i}^{o}(k)\left| {r_{L,i}^{o}(k) + w_{y,i} (k)} \right.} \right] \hfill \\
                       &= \eta \left( {r_{L,i}^{o}(k) + w_{y,i} (k)} \right) \hfill \\
\end{aligned}
\label{eqAC8}
\eea
where variable $\hat r_{L,i}^{o}(k) := \mathbb{E}\left[ {r_{L,i}^{o}(k)\left| {{\mathcal{F}_k}} \right.} \right]$, and the constant $\eta  := {{{{\left( {{L_{i\cdot}}{\Sigma _{{o}}}L_{i\cdot}^T} \right)}^2}} \mathord{\left/
 {\vphantom {{{{\left( {{L_{i\cdot}}{\Sigma _{{r_o}}}L_{i\cdot}^T} \right)}^2}} {\left( {{{\left( {{L_{i\cdot}}{\Sigma _{{o}}}L_{i\cdot}^T} \right)}^2} + \sigma _{{w_{y,i}}}^2} \right)}}} \right.
 \kern-\nulldelimiterspace} {\left( {{{\left( {{L_{i\cdot}}{\Sigma _{{o}}}L_{i\cdot}^T} \right)}^2} + \sigma _{{w_{y,i}}}^2} \right)}}$.

Let $\tilde r_{L,i}^{o}(k) := r_{L,i}^{o}(k) - \hat r_{L,i}^{o}(k)$ and the following holds
\bea
\tilde r_{L,i}^{o}(k - 1) \in {\mathcal{F}_k}, \mathbb{E}\left[ {\tilde r_{L,i}^{o}(k)\left| {{\mathcal{F}_k}} \right.} \right]=0.
\label{eqAC9}
\eea
From (\ref{eqAC9}), it is clear that $\left\{ {\tilde r_{L,i}^{o}(k)} \right\}$ is a Martingale difference sequence. Applying the Martingale stability theorem \cite[Le. 2]{1} to $\tilde r_{L,i}^{o}(k)$ yields
\bea
\sum\limits_{k = 1}^T {\mathcal{D}_{r,i} (k)\tilde r_{L,i}^{o}(k)}  = o\left( {\sum\limits_{k = 1}^T {{{ {\mathcal{D}^2_{r,i} (k)} }}} } \right) + O(1)
\label{eqAC10}
\eea
where $o(\cdot)$ is the infinitesimal of higher order over an infinitesimal, specially, $o(1)$ is the infinitesimal of higher order over any constant; $O(1)$ is the bounded quantity. Substituting (\ref{eqAC8}) and (\ref{eqAC10}) into (\ref{eqAC4}) yields
\begin{align}
  &\sum\limits_{k = 1}^T {{\mathcal{D}_{r,i}}(k)r_{L,i}^o(k)}  = \sum\limits_{k = 1}^T {{\mathcal{D}_{r,i}}(k)\tilde r_{L,i}^o(k) + {\mathcal{D}_{r,i}}(k)\hat r_{L,i}^o(k)} \nonumber \\
  & \qquad \quad= o\left( {\sum\limits_{k = 1}^T {\mathcal{D}_{r,i}^2(k)} } \right) + O(1) + \eta \sum\limits_{k = 1}^T {{\mathcal{D}_{r,i}}(k)r_{L,i}^o(k)} \nonumber \\
   &\qquad \quad+ \eta \sum\limits_{k = 1}^T {\mathcal{D}_{r,i} (k)w_{y,i} }(k). \label{eqAB11}
\end{align}
Substituting (\ref{eqAC2}) into (\ref{eqAB11}), it follows that
\begin{gather}
\sum\limits_{k = 1}^T {\mathcal{D}_{r,i} (k)r^o_{L,i} (k)}  = o\left( {\sum\limits_{k = 1}^T {{{ {\mathcal{D}^2_{r,i} (k)} }}} } \right) + O(1).
\label{eqAC12}
\end{gather}
Therefore, from (\ref{eqAC12}), the following holds
\bea
\begin{gathered}
  \sum\limits_{k = 1}^T {{{ {\mathcal{D}^2_{r,i} (k)} }} + 2\mathcal{D}_{r,i} (k)r^{o}_{L,i} (k)}  =  \hfill \\
  \qquad \qquad \qquad \qquad \left( {1 + o(1)} \right)\sum\limits_{k = 1}^T {{{ {\mathcal{D}^2_{r,i} (k)} }}}  + O(1). \hfill \\
\end{gathered}
\label{eqAC13}
\eea
Dividing the above equation by $T$, taking the limit as $T \to \infty$, and invoking (\ref{eqAC4c}), we have:
\bea
\mathop {\lim }\limits_{T \to \infty } \frac{1}{T}\sum\limits_{k = 1}^T {{{ {\mathcal{D}^2_{r,i} (k)} }}}  = 0, i=1,2,\ldots,m_{x}.
\label{eqAC14}
\eea
Summing (\ref{eqAC14}) from $i=1$ to $i=m_{x}$ and considering the definition of norm $\left\| {{\mathcal{D}_r}(k)} \right\|$ yields (30) in the main text. It completes the proof.

\subsection{Proof of Theorem 2}
Considering system (28) in the main text, to analyse the ability of the new DW scheme in FDIA detection, we firstly define $\wp (k) :=x_a(k)-\hat x(k|k-1)$. The dynamics of $\wp(k)$ can be given by
\bea
\left\{ \begin{gathered}
  \wp (k+1) = {\Phi _1}\wp (k) + {\Phi _2}{x_a}(k) + (A+BK)Lw_y(k) \hfill \\
  r(k) = C\wp (k) - w_y(k) \hfill \\
\end{gathered}  \right.
\label{eqAB1}
\eea
where ${\Phi _1}$, ${\Phi _2}$ are the same as (12) in the main text. Then, using (\ref{eqAB1}), the variable $\wp (k)$ can be computed recursively from $0$ to $k$ as follows
\bea
\begin{aligned}
  \wp (k) &= \Phi _1^k\wp (0) + \sum\limits_{p = 1}^k {\Phi _1^{p - 1}{\Phi _2}{x_a}(k - p)}  \hfill \\
          &+ \sum\limits_{p = 1}^k {\Phi _1^{p - 1}(A + BK)Lw_y(k - p)}.  \hfill \\
\end{aligned}
\label{eqAB2}
\eea
For (\ref{eqAB2}), considering $w_{y,i}(k) \bot \wp (0)$, $w_{y,i}(k) \bot {x_a}(p)$ for $\forall p$, $w_{y,i}(k) \bot w_{y,i}(p)$ for $p \ne k$, and $w_{y,i}(k) \bot w_{y,j}(p)$ for $\forall p$, the following holds
\begin{align}
  \mathbb{E}\left[ {{w_{y,i}}(k)\wp (k)} \right] &= 0, \hfill \label{eqAB3} \\
  \mathbb{E}\left[ {{w_{y,i}}(k)Lr(k)} \right]   &=  -\mathbb{E}\left[ {{w_{y,i}}(k){L_{\cdot i}}{w_{y,i}}(k)} \right].
\label{eqAB4}
\end{align}
Equation (\ref{eqAB4}) yields (31a) in the main text.

Secondly, we focus on the steady-state behavior of (\ref{eqAB1}). Note that $\mathop {\lim }\limits_{k \to \infty } {x_{a}}(k) = 0$ of the FDIA (11) in the main text; taking the limitation as $k \to \infty$ for (\ref{eqAB1}) produces
\bea
\left\{ \begin{gathered}
\mathop {\lim }\limits_{k \to \infty } \wp (k+1) = \mathop {\lim }\limits_{k \to \infty } {\Phi _1}\wp (k) + (A+BK)Lw_y(k) \hfill \\
\mathop {\lim }\limits_{k \to \infty } r(k) = \mathop {\lim }\limits_{k \to \infty } C\wp (k) - w_y(k). \hfill \\
\end{gathered} \right.
\label{eqAB5}
\eea
Denoting $\mathbb{E}\left[ {\wp (\infty ){\wp ^T}(\infty )} \right]=M$ and considering (\ref{eqAB3}), we multiply both sides of the equal sign by the corresponding transposed vector shown in (\ref{eqAB5}) and take the expectation, yielding (31b) in the main text.

Thirdly, to investigate the action of the system state $x(k)$, let us recall the main text, or more specifically, the variable $\zeta(k)=\left[ x(k);\wp(k);x_a(k)\right]$, and the matrices $\mathcal{A}_0=\left[ A,{\rm H};0,\Xi\right]$ and $\Xi  = \left[ {{\Phi _1},{\Phi _2};0,{A_a}} \right]$ in (28). Since $\rho (A_a)<1$ and $\rho \left( {{\Phi _1}} \right) < 1$, it arrives at $\rho (\Xi)<1$, i.e., $\left[ {\xi_d (k);{x_{a}}(k)} \right]$ is exponentially bounded in the mean-square sense. Inequality $\rho (A) >1$ yields $\zeta_d(\infty) \to \infty$. Therefore, it is clear that $x(\infty) \to \infty$, i.e., (31c) in the main text. It completes the proof.

\subsection{Proof of Theorem 3}
For the switched system (33) in the main text, we first compute recursively $\zeta(k)$ from $t_0$ to $k$ as follows
\bea
\zeta (k) = \mathcal{A}_{{q_{i + 1}}}^{(k - {t_i})}\mathcal{A}_{{q_i}}^{({t_i} - {t_{i - 1}})} \cdots \mathcal{A}_{{q_1}}^{({t_1} - {t_0})}\zeta (t_0) + \varpi (k)
\label{eqAF1}
\eea
where $q_{i}=0,1$; $t_i$ is the switching instant; and
\begin{align}
  &\varpi (k) = \sum\limits_{j = 1}^{i - 1} {\sum\limits_{p = {t_{j - 1}}}^{{t_j} - 1} {\mathcal{A}_{{q_{i + 1}}}^{k - {t_i}} \cdots \mathcal{A}_{{q_{j + 1}}}^{{t_j} - {t_{j - 1}}}\mathcal{A}_{{q_j}}^{{t_j} - p - 1}{\Lambda _{{q_j}}}\psi (p)} }   \hfill \nonumber \\
   &+ \sum\limits_{p  = {t_{i - 1}}}^{{t_i} - 1} {\mathcal{A}_{{q_{i + 1}}}^{(k - {t_i})}\mathcal{A}_{{q_i}}^{{t_i} - p  - 1}{\Lambda _{{q_i}}}\psi (p )} + \sum\limits_{p  = {t_i}}^{k - 1} {\mathcal{A}_{{q_{i + 1}}}^{k - p  - 1}{\Lambda _{{q_{i + 1}}}}\psi (p)}. \label{eqAF2}
\end{align}
Taking the norm of the both sides of the equal sign in (\ref{eqAF1}), using norm triangle equality and taking the expectation yields
\begin{gather}
  \mathbb{E}\left[ {{{\left\| {\zeta (k)} \right\|}}} \right] \leqslant \mathbb{E}\left[ {{{\left\| {\mathcal{A}_{{q_{i + 1}}}^{(k - {t_i})} \cdots \mathcal{A}_{{q_1}}^{({t_1} - {t_0})}\zeta (t_0)} \right\|}}} \right] + \mathbb{E}\left[ {{{\left\| {\varpi (k)} \right\|}}} \right].
\label{eqAF3}
\end{gather}

Secondly, according to the norm consistency principle and (34) in the main text and the definition of $\mathcal{T}_0$, $\mathcal{T}_1$, $N_s(0,k)$,  the first term on the right hand of (\ref{eqAF3}) becomes
\be
\begin{gathered}
  \mathbb{E}\left[ {{{\left\| {\mathcal{A}_{{q_{i + 1}}}^{(k - {t_i})} \cdots \mathcal{A}_{{q_1}}^{({t_1} - {t_0})}\zeta (t_0)} \right\|}}} \right] \leqslant { \lambda _{-}^{g(i + 1)}\lambda _{+}^{{\mathcal{T}_0 }}\lambda _{-}^{{\mathcal{T}_1 }}}\mathbb{E}\left[ {{{\left\| {\zeta (t_0)} \right\|}}} \right]\hfill \\
   \qquad \qquad \qquad \qquad \quad \quad= { \lambda _{-}^g{\lambda _{-}^{g{N_s(t_0,k)}}\lambda _{+}^{{\mathcal{T}_0 }}\lambda _{-}^{{\mathcal{T}_1 }}}}\mathbb{E}\left[ {{{\left\| {\zeta (t_0)} \right\|}}} \right]. \hfill \\
\end{gathered}
\label{eqAF4}
\ee
Substituting (37) in the main text into (\ref{eqAF4}) yields
\bea
\begin{gathered}
  \mathbb{E}\left[ {{{\left\| {\mathcal{A}_{{q_{i + 1}}}^{(k - {t_i})} \cdots \mathcal{A}_{{q_1}}^{({t_1} - {t_0})}\zeta (t_0)} \right\|}}} \right] \leqslant  \hfill \\
  \qquad \qquad \qquad \qquad \qquad \lambda _{-}^{g}\lambda _{-}^{g{N_s }(t_0,k)+{\lambda ^ * }k}\mathbb{E}\left[ {{{\left\| {\zeta (t_0)} \right\|}}} \right]. \hfill \\
\end{gathered}
\label{eqAF5}
\eea
For (\ref{eqAF5}), there are two cases according to the value of $g$:
\begin{enumerate}
  \item
  When $g \geqslant 0$, in (\ref{eqAF5}), it follows that
  \bea
  \mathop {\lim }\limits_{k \to \infty } \lambda _{-}^{g}\lambda _{-}^{g{N_s }(t_0,k) + {\lambda ^ * }k} = 0
  \label{eqAF6}
  \eea
  for any ${N_s }(t_0,k)$, i.e., for any $\tau$ in (36) in the main text. Furthermore, by using the condition $0 < {\lambda ^\dag } < {\lambda ^ * } < 1$, (\ref{eqAF5}) becomes
  \bea
  \begin{gathered}
  \mathbb{E}\left[ {{{\left\| {\mathcal{A}_{{q_{i + 1}}}^{(k - {t_i})} \cdots \mathcal{A}_{{q_1}}^{({t_1} - {t_0})}\zeta (t_0)} \right\|}}} \right] \leqslant  \hfill \\
  \qquad \qquad \qquad \qquad \lambda _{-}^{g}\lambda _{-}^{g{N_s }(t_0,k) + {\lambda ^\dag }k}\mathbb{E}\left[ {{{\left\| {\zeta (t_0)} \right\|}}} \right]. \hfill \\
  \end{gathered}
  \label{eqAF7}
  \eea
  \item
  When $g < 0$, by using (36) in the main text, it has $g{N_s }(t_0,k) + {\lambda ^ * }k \geqslant \iota  + {\lambda ^\dag }k$, i.e.,
  \bea
  {N_s }(t_0,k) \leqslant \frac{\iota }{g} + \frac{{k - {t_0}}}{{{\tau _{ave}}}}
  \label{eqAF8}
  \eea
  where $\iota$ is any constant and $\tau _{ave} = \frac{g}{{{\lambda ^\dag } - {\lambda ^*}}}$. Then, substituting (\ref{eqAF8}) into (\ref{eqAF5}) yields
  \bea
  \begin{gathered}
  \mathbb{E}\left[ {{{\left\| {\mathcal{A}_{{q_{i + 1}}}^{(k - {t_i})} \cdots \mathcal{A}_{{q_1}}^{({t_1} - {t_0})}\zeta (t_0)} \right\|}}} \right] \leqslant  \hfill \\
  \qquad \qquad \qquad \qquad \lambda _{-}^{\iota  + g}\lambda _{-}^{{\lambda ^\dag }(k-t_0)}\mathbb{E}\left[ {{{\left\| {\zeta (t_0)} \right\|}}} \right]. \hfill \\
  \end{gathered}
  \label{eqAF9}
  \eea
\end{enumerate}

Thirdly, by using the norm triangle equality and norm consistency principle, the second term on the right hand of (\ref{eqAF3}) becomes
\begin{align}
  &\mathbb{E}\left[ {\left\| {\varpi (k)} \right\|} \right] \leqslant  \hfill \nonumber \\
  &\mathbb{E}\left[ {\sum\limits_{j = 1}^{i - 1} {\sum\limits_{p = {t_{j - 1}}}^{{t_j} - 1} {\left\| {\mathcal{A}_{{q_{i + 1}}}^{k - {t_i}} \cdots \mathcal{A}_{{q_{j + 1}}}^{{t_j} - {t_{j - 1}}}\mathcal{A}_{{q_j}}^{{t_j} - p - 1}} \right\|\left\| {{\Lambda _{{q_j}}}} \right\|\left\| {\psi (p)} \right\|} } } \right]  \hfill \nonumber \\
  &+ \mathbb{E}\left[ {\sum\limits_{p  = {t_{i - 1}}}^{{t_i} - 1} {\left\| {\mathcal{A}_{{q_{i + 1}}}^{(k - {t_i})}\mathcal{A}_{{q_i}}^{{t_i} - p  - 1}} \right\|\left\| {{\Lambda _{{q_i}}}} \right\|\left\| {\psi (p)} \right\|} } \right] \hfill \nonumber \\
  &+\mathbb{E}\left[ {\sum\limits_{p  = {t_i}}^{k - 1} {\left\| {\mathcal{A}_{{q_{i + 1}}}^{k - p  - 1}} \right\|\left\| {{\Lambda _{{q_{i + 1}}}}} \right\|\left\| {\psi (p )} \right\|} } \right]. \label{eqAF10}
\end{align}
When $g<0$, it is clear that in (\ref{eqAF10}),
\begin{align}
&\left\| {\mathcal{A}_{{q_{i + 1}}}^{(k - {t_i})} \cdots \mathcal{A}_{{q_2}}^{({t_2} - {t_1})}\mathcal{A}_{{q_1}}^{{t_1} - {t_0}}} \right\| \leqslant \lambda _ - ^{\iota  + g}\lambda _ - ^{{\lambda ^\dag }(k-t_0)}, \label{eqAF11} \\
&\left\| {\mathcal{A}_{{q_{i + 1}}}^{(k - {t_i})} \cdots \mathcal{A}_{{q_2}}^{({t_2} - {t_1})}\mathcal{A}_{{q_j}}^{{t_j} - p  - 1}} \right\| \leqslant \lambda _ - ^{\iota  + g}\lambda _ - ^{{\lambda ^\dag }(k - p  - 1)}.
\label{eqAF12}
\end{align}
Substituting (\ref{eqAF11}), (\ref{eqAF12}), and (35) in the main text (i.e., $\mathbb{E}\left[ {\left\| {\psi (p)} \right\|} \right]  \leqslant  \psi_u  < \infty $) into (\ref{eqAF10}) yields
\bea
\begin{aligned}
  \mathbb{E}\left[ {\left\| {\varpi (k)} \right\|} \right] &\leqslant \mathbb{E}\left[ {\sum\limits_{p  = t_0}^{k-1} {\mu \lambda _ - ^{\iota  + g}\lambda _ - ^{{\lambda ^\dag }(k-p-1) }\left\| {\psi (p)} \right\|} } \right] \hfill \\
   &\leqslant \frac{{\mu \lambda _ - ^{\iota  + g}\psi_u }}{{1 - \lambda _ - ^{{\lambda ^\dag }}}}\lambda _ - ^{{\lambda ^\dag }(k - {t_0})}\mathop  \leqslant \limits^{k = {t_0}} \frac{{\mu \lambda _ - ^{\iota  + g}\psi_u }}{{1 - \lambda _ - ^{{\lambda ^\dag }}}} < \infty   \hfill \\
\end{aligned}
\label{eqAF13}
\eea
where $\mu  = \max \left\{ {\left\| {{\Lambda _0}} \right\|,\left\| {{\Lambda _1}} \right\|} \right\}$.

Finally, substituting (\ref{eqAF9}) and (\ref{eqAF13}) into (\ref{eqAF3}) yields
\begin{gather}
\mathbb{E}\left[ {{{\left\| {\zeta (k)} \right\|}}} \right] \leqslant \lambda _{-}^{\iota  + g}\lambda _{-}^{{\lambda ^\dag }(k-t_0)}\mathbb{E}\left[ {{{\left\| {\zeta (t_0)} \right\|}}} \right] + \frac{{\mu \lambda _ - ^{\iota  + g}\psi_u }}{{1 - \lambda _ - ^{{\lambda ^\dag }}}}
\label{eqAF14}
\end{gather}
which is similar to the situation when $g \geqslant 0$. It completes the proof.

\subsection{Proof of Theorem 4}
To analyse the stability and performance of system (29) in the main text, the following Lyapunov-Krasovskii functional candidate is chosen
\begin{align}
  V(k) &= {{\bar \zeta}^T}(k){Z_1}\bar \zeta(k) + \sum\limits_{q  =  - \bar h + 1}^0 {\sum\limits_{p = k - 1 + q }^{k - 1} {{\xi ^T}(p){E^T}{Z_2}E\xi (p)} }  \nonumber \\
       &+ \sum\limits_{p = k - h(k)}^{k - 1} {\bar \zeta(p){E^T}{Z_3}E\bar \zeta(p)} \label{eqAG1}
\end{align}
where $\xi(k) := \bar \zeta(k + 1) - \bar \zeta(k)$. Conducting difference for (\ref{eqAG1}) yields
\begin{align}
  \Delta V(k) &= {{\bar \zeta}^T}(k + 1){Z_1}\bar \zeta(k + 1) - {{\bar \zeta}^T}(k){Z_1}\bar \zeta(k) \hfill \nonumber\\
   &+ \bar h{{\xi}^T}(k){E^T}{Z_2}E\xi(k) + {{\bar \zeta}^T}(k){E^T}{Z_3}E\bar \zeta(k) \hfill \nonumber \\
   &- {{\bar \zeta}^T}(k - h(k)){E^T}{Z_3}E\bar \zeta(k - h(k)) \hfill \nonumber \\
   &- \sum\limits_{p = k - \bar h}^{k - 1} {{{\xi}^T}(p){E^T}{Z_2}E\xi(p)}  \label{eqAG2}
\end{align}
where $\Delta V(k) := V(k+1)-V(k)$.

For convenience, we here define an augmented variable as $\aleph (k) := \left[ {\bar \zeta(k);E\bar \zeta(k - h (k));\bar \psi (k)} \right]$.

According to the defined $\aleph (k)$ and \cite[Le. 1]{2}, (\ref{eqAG2}) becomes
\begin{gather}
\Delta V(k) + {e^T}(k)e(k) - \beta {\tilde \psi ^T}(k)\tilde \psi (k) \leqslant {\aleph ^T}(k)\Pi \aleph (k)
\label{eqAG4}
\end{gather}
where $\Pi  = \left[ {\begin{array}{*{20}{c}}
  \Theta &{\tilde \Omega} \\
   * &{\tilde \Upsilon }
\end{array}} \right]$ and
\[\begin{gathered}
  \tilde \Omega  = \left[ {\begin{array}{*{20}{c}}
  {hW_1^T}&{h{{({A_0} - I)}^T}{E^T}}&{{{(E - {E^c}{A_0})}^T}}&{A_0^T} \\
  {hW_2^T}&{hA_1^T{E^T}}&{{{( - {E^c}{A_1})}^T}}&{A_1^T} \\
  {hW_3^T}&{h\Gamma _0^T{E^T}}&{{{( - {E^c}{\Gamma _0})}^T}}&{\Gamma _0^T}
\end{array}} \right], \hfill \\
  \tilde \Upsilon  = diag\left\{ { - h{Z_2}, - hZ_2^{ - 1}, - I, - Z_1^{ - 1}} \right\}. \hfill \\
\end{gathered}\]
By introducing two matrices $W_4$ and $W_5$ to deal with the nonlinear terms $Z_2^{ - 1}$ and $Z_1^{ - 1}$ \cite{2}, $\Pi <0$ yields (39) in the main text.

As $\Pi <0$, taking the sum of (\ref{eqAG4}) from $0$ to $T-1$ and using the zero-initial condition (i.e., $V(0)=0$) gives
\bea
\sum\limits_{k = 0}^{T - 1} {{e^T}(k)e(k) - \beta {\bar \psi ^T}(k)\bar \psi (k)}  \leqslant  - V(T) \leqslant 0.
\label{eqAG5}
\eea
Dividing (\ref{eqAG5}) by $T$ and seeking the limitation as $T \to \infty$ yield (40) in the main text. It completes the proof.

\section{Supplement on Experimental Results}
\subsection{Statistical Analysis of Measurement Noise}
\begin{figure}[!t]
  \centering
  \subfigure[]{\includegraphics[width=0.48\textwidth]{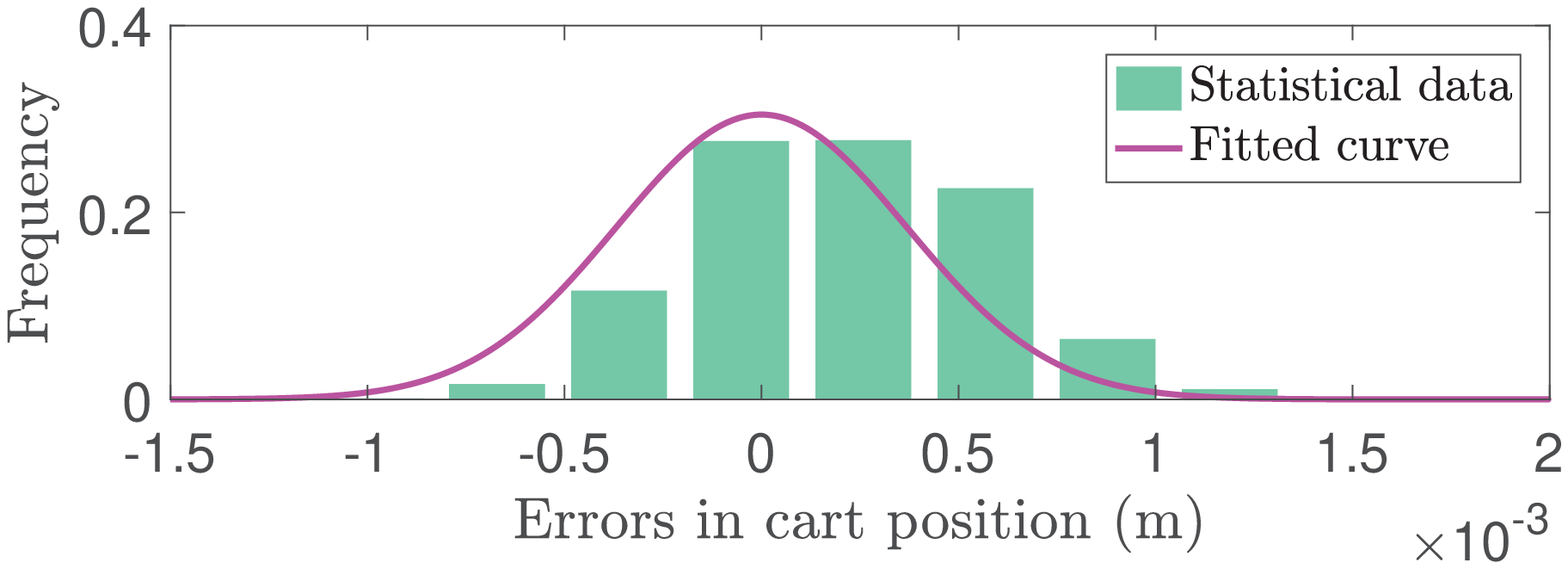}}  \\ \vspace{-0.15in}
  \subfigure[]{\includegraphics[width=0.48\textwidth]{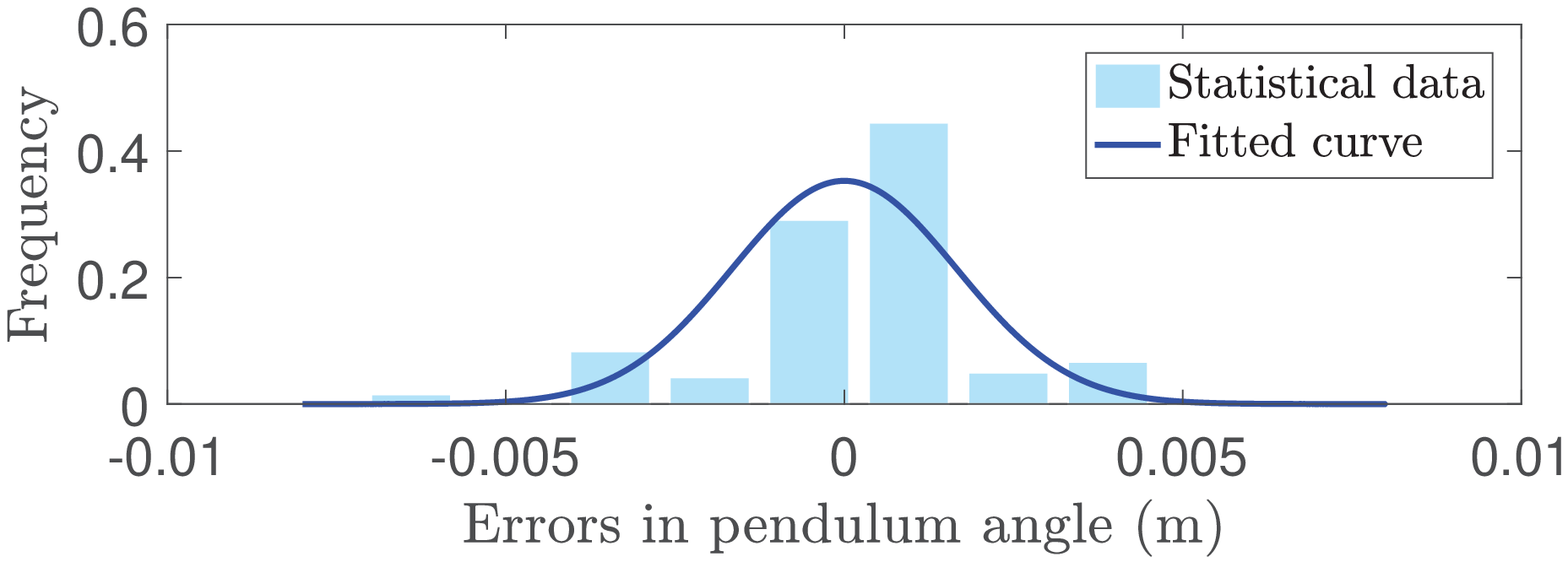}} \\
  \caption{Fitted curves on errors of the cart position and pendulum angle.}
  \label{figA1}
\end{figure}

Using the data from [39] in the main text and the toolbox \emph{cftool} in MATLAB, the fitted curves on the errors of the cart position and the pendulum angle are illustrated in Fig.~\ref{figA1}. For the fitted curves, it is calculated that the variances of $v_1$ and $v_2$ are $2.7\times 10^{-7}$ and $5.5\times 10^{-6}$ with root mean squared errors $0.0572$ and $0.0740$, respectively.

\subsection{Case Comparison of Computational Time Complexity}
\begin{figure}[!t]
  \centering
  \includegraphics[width=0.48\textwidth]{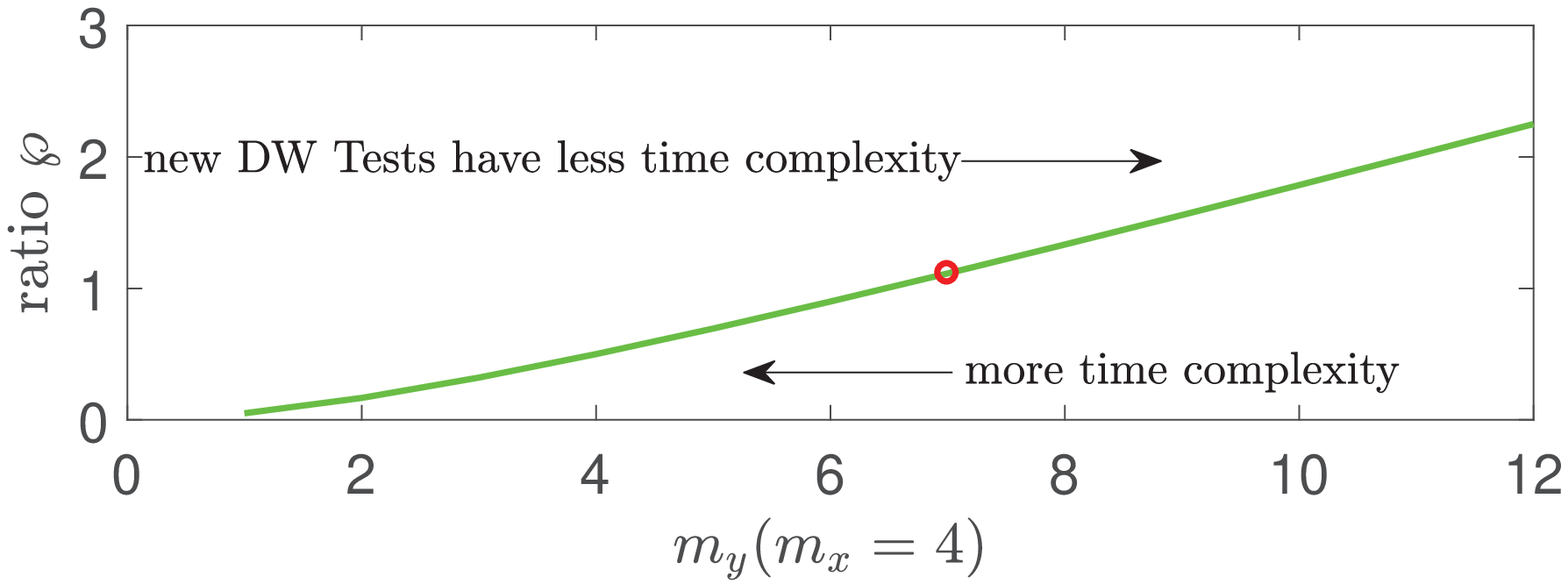}
  \caption{The ratio $\wp$ with $m_x=4$.}
  \label{figratio}
\end{figure}

Fig.~\ref{figratio} presents the ratio $\wp$ with $m_x=4$ between the computational time complexity of residual-based test and new DW Tests. Tab.~\ref{Tabtime} lists computational time complexity of the residual-based test and the new DW tests compared in experiments of Fig.~5 in the main text. Note that the estimator and controller are periodic, i.e., control calculation and attack detection are performed every 10 ms: 1) the experiments of Fig.~5 run 148 steps (where 10 ms every step) and consume 1.48 s; 2) the tests are only calculated every 10 ms and the summations are listed in Tab.~\ref{Tabtime}.
\begin{table}[t]
\centering
\caption{Computational Time Complexity Comparison between the Residual-Based Best and the New DW Tests.}
\label{Tabtime}
\begin{tabular}{lccc}
\toprule
  Test & \tabincell{c}{Computational\\Time\\complexity} & \tabincell{c}{Running time\\for test} & \tabincell{c}{Run time\\of experiments\\in Fig.~5} \\
\midrule
  \tabincell{c}{Residual-based\\test [34]} & $O(m_y^2)$ & 6.237 $\mu$s & 1.48s\\
  New DW Tests &$O(m_x^2+m_xm_y)$ & 34.158 $\mu$s & 1.48s\\
\bottomrule
\end{tabular}
\end{table}

\subsection{Conventional and New DW Schemes for FDIA with Different Initial Instants}
\begin{figure}[!t]
  \centering
  \subfigure[Cart position]{\includegraphics[width=0.48\textwidth]{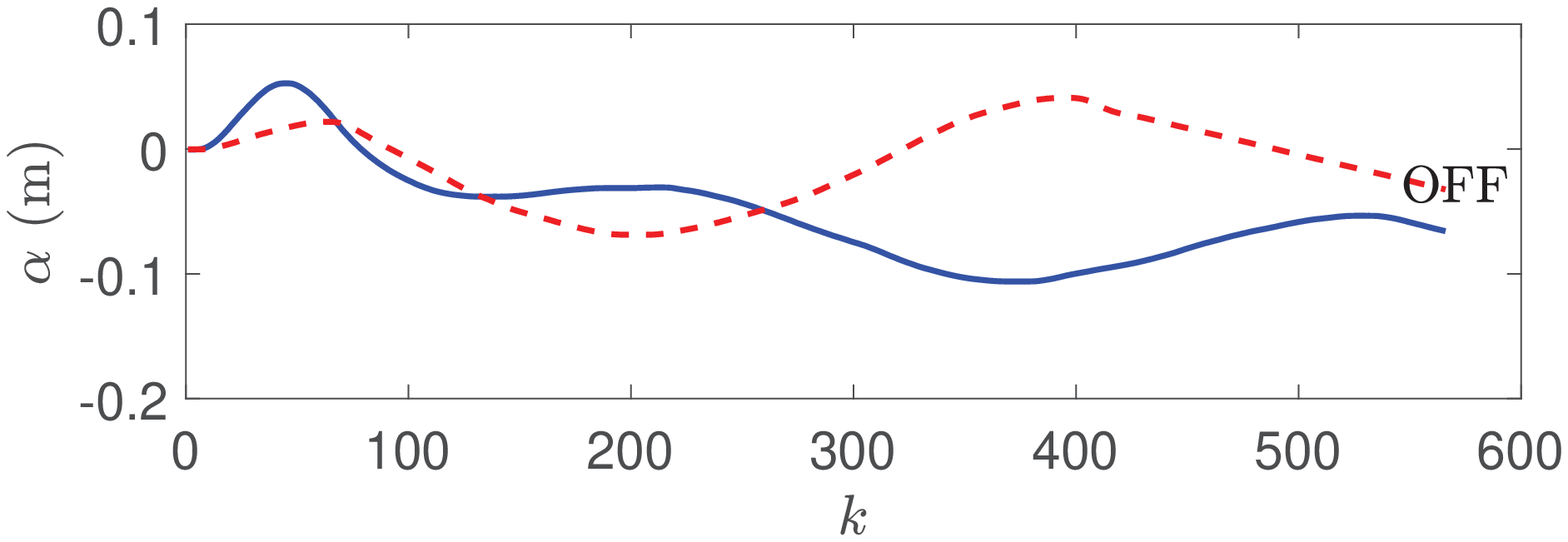}}  \\ \vspace{-0.15in}
  \subfigure[Pendulum angle]{\includegraphics[width=0.48\textwidth]{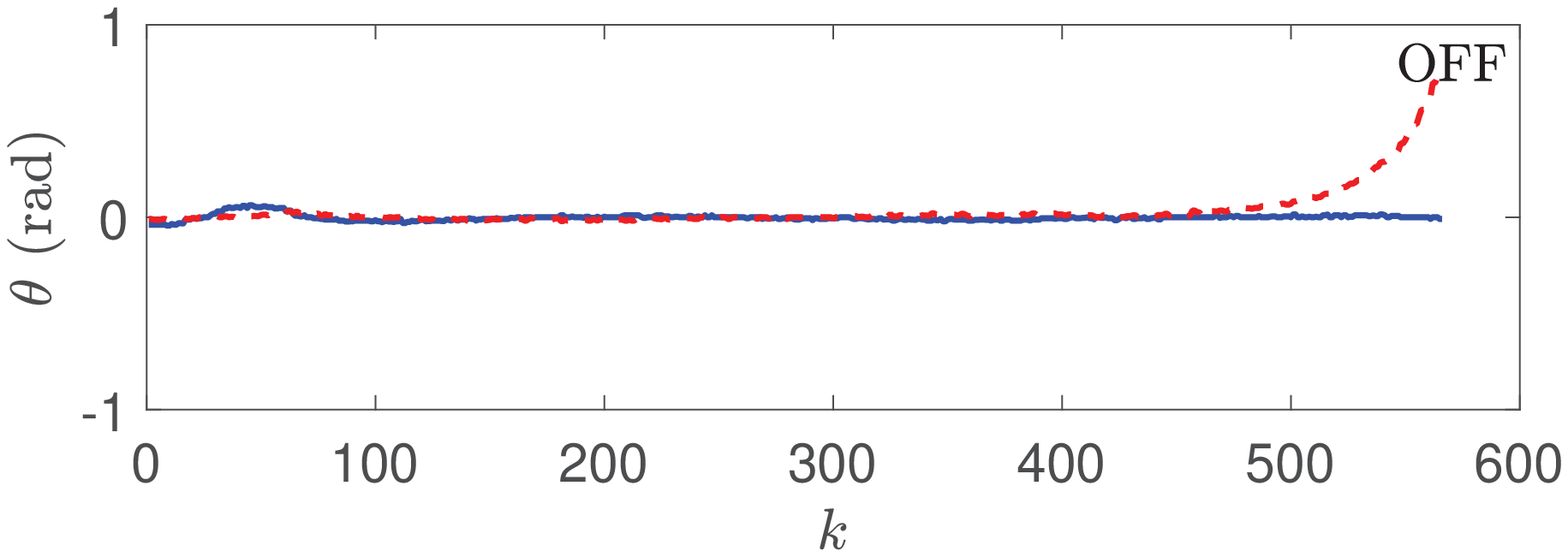}} \\ \vspace{-0.15in}
  \subfigure[DW statistical test 1 $\varphi _{d,1}(k)$]{\includegraphics[width=0.48\textwidth]{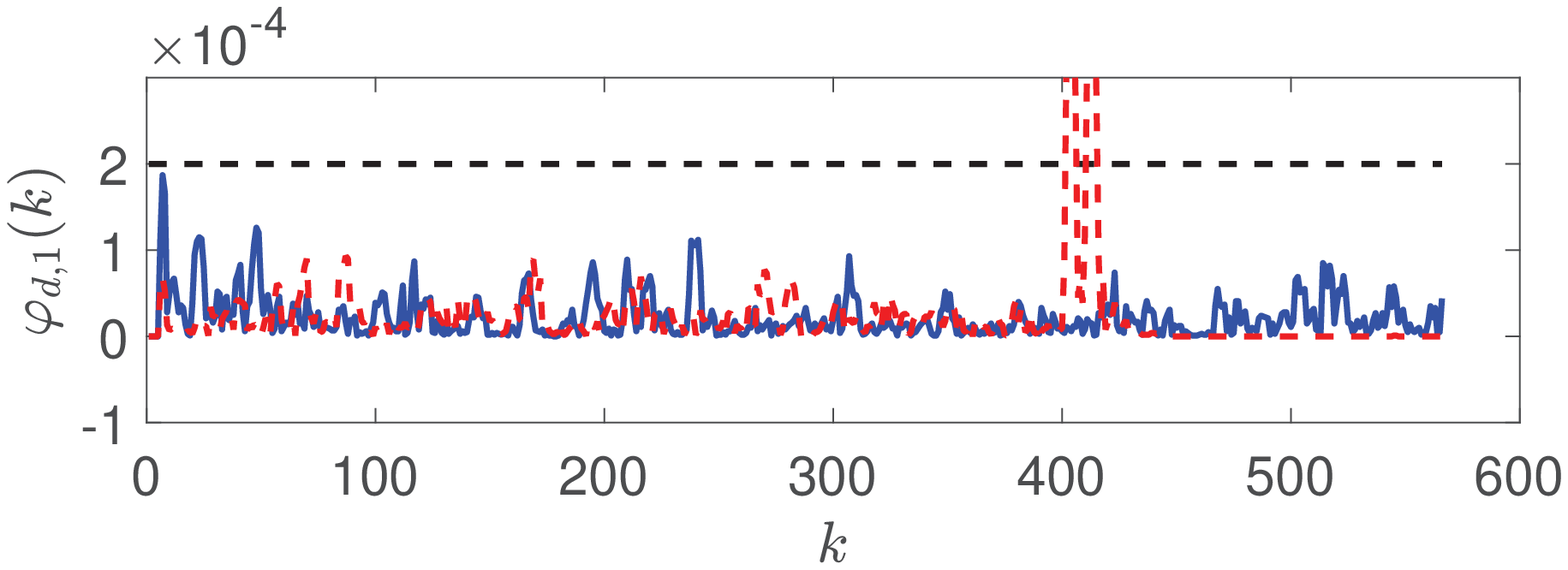}} \\ \vspace{-0.15in}
  \subfigure[DW statistical test 2 $\varphi _{d,2}(k)$]{\includegraphics[width=0.48\textwidth]{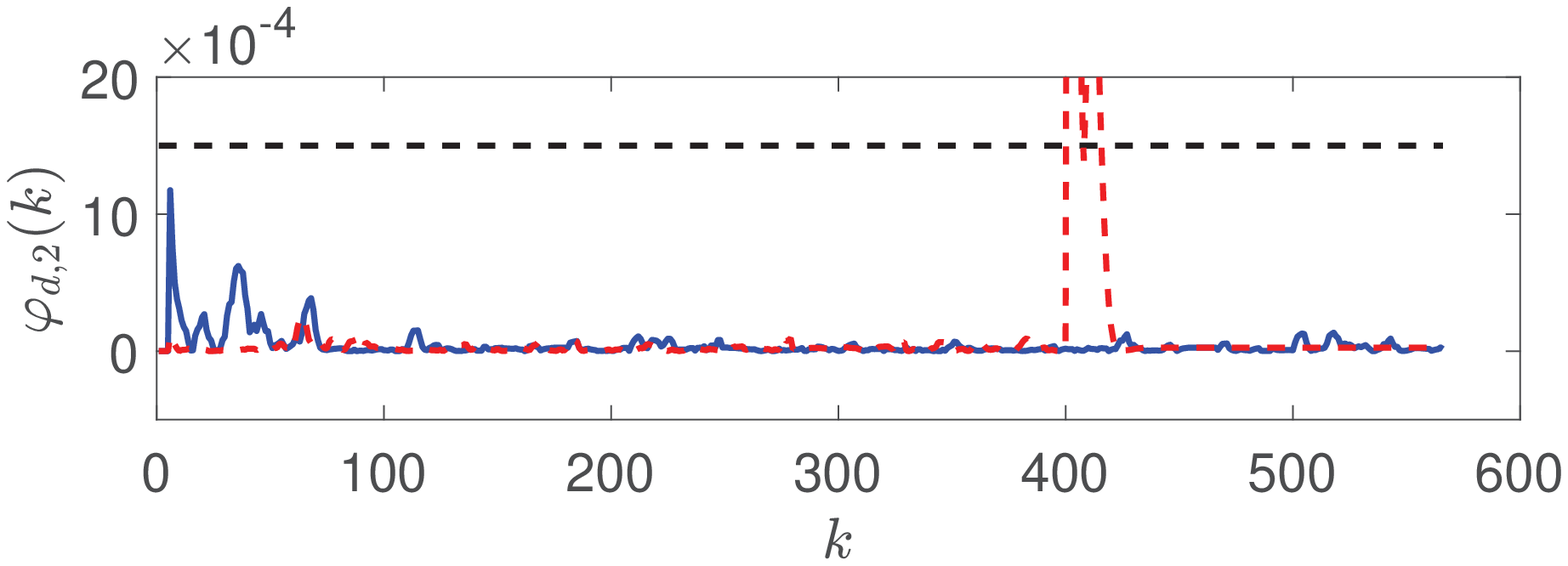}} \\
  \caption{States and the DW statistical tests of NIPVSSs under the FDIA emerging at $k \geqslant 400$ based on the conventional DW scheme with $\sigma^2_{w_d} = 0.0001$. Blue line: Normal system. Red line: System under FDIA. Black line: Detection thresholds. ``OFF'' denotes that the servo is put off.}
  \label{figB1}
\end{figure}

\begin{figure}[!t]
  \centering
  \subfigure[Cart position]{\includegraphics[width=0.48\textwidth]{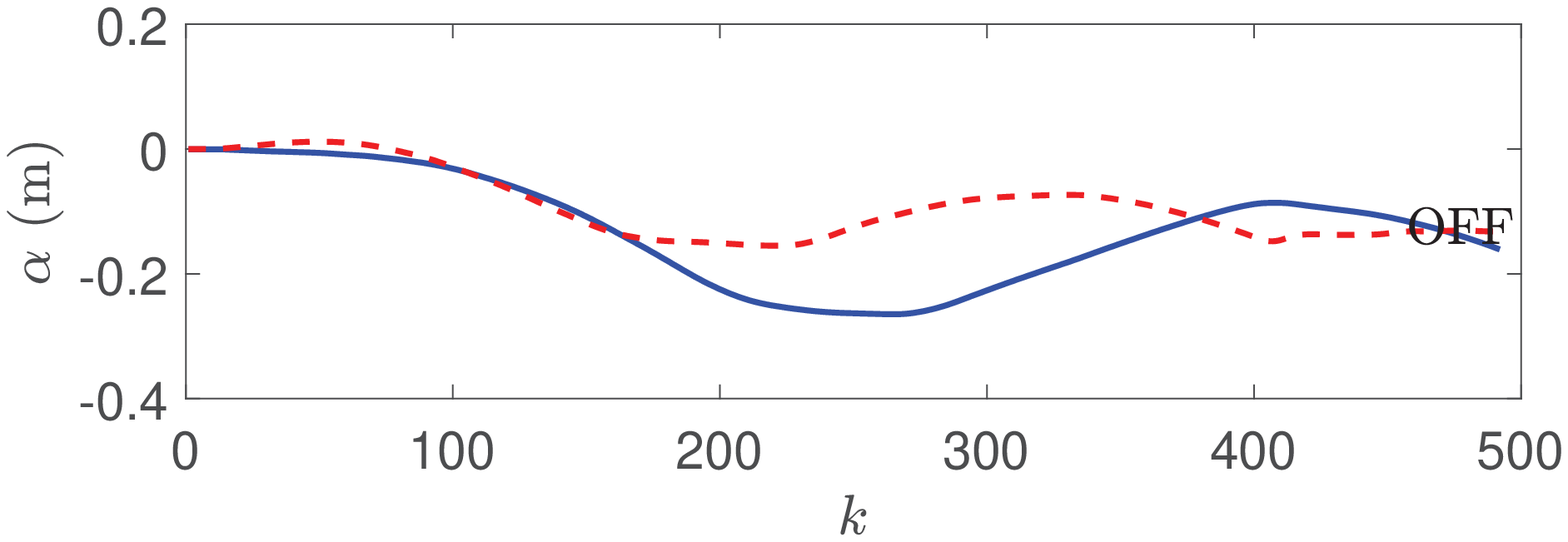}} \\ \vspace{-0.15in}
  \subfigure[Pendulum angle]{\includegraphics[width=0.48\textwidth]{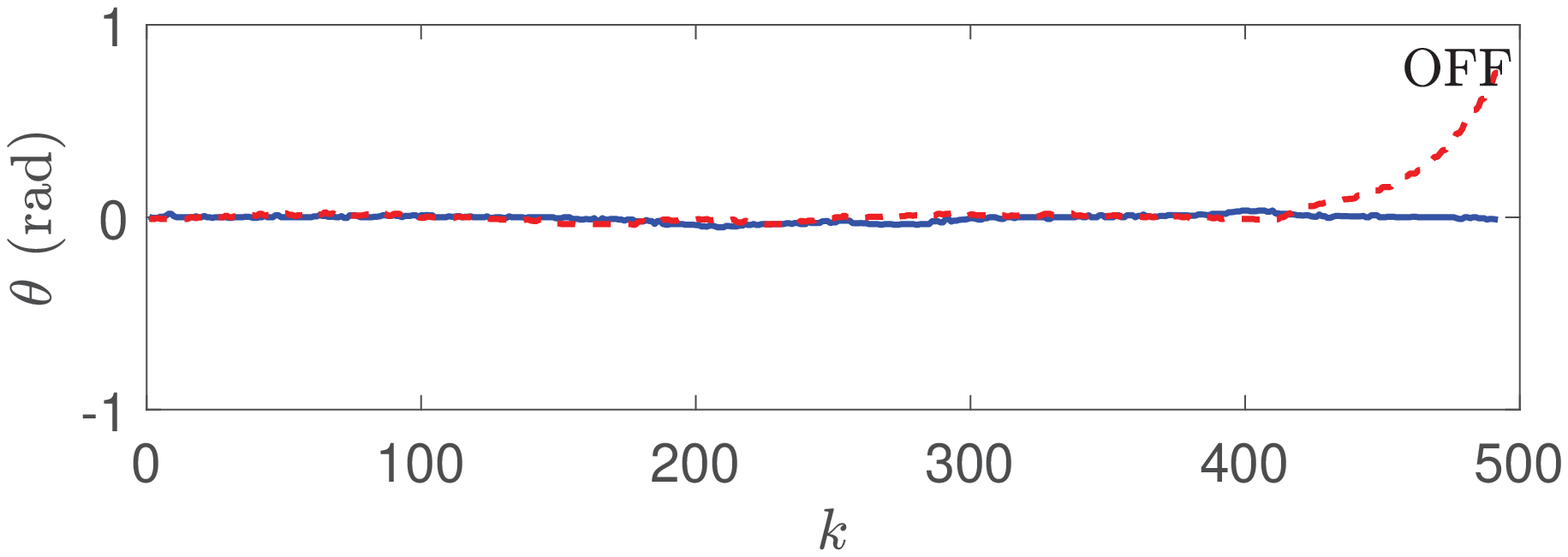}} \\ \vspace{-0.15in}
  \subfigure[New DW statistical test 1 $\varphi _{1,1}(k)$, $\varphi_{1,2}(k)$]{\includegraphics[width=0.48\textwidth]{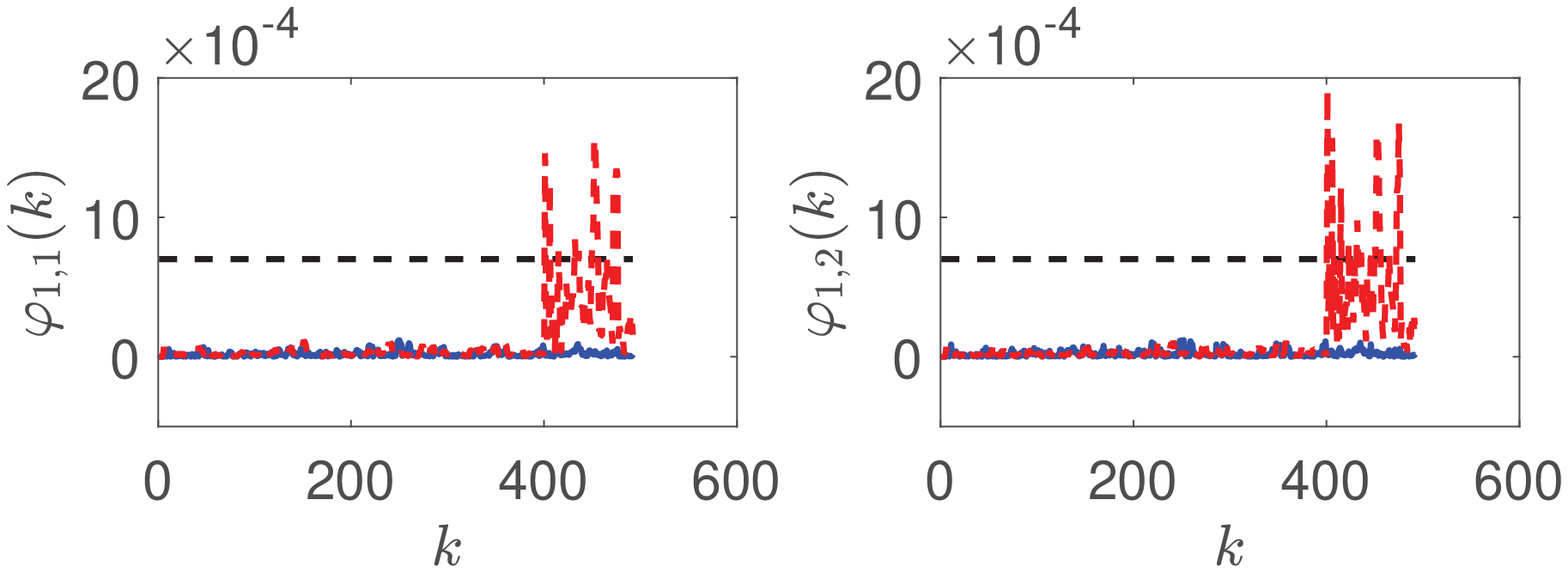}} \\ \vspace{-0.15in}
  \subfigure[New DW statistical test 2 $\varphi _{2}(k)$]{\includegraphics[width=0.48\textwidth]{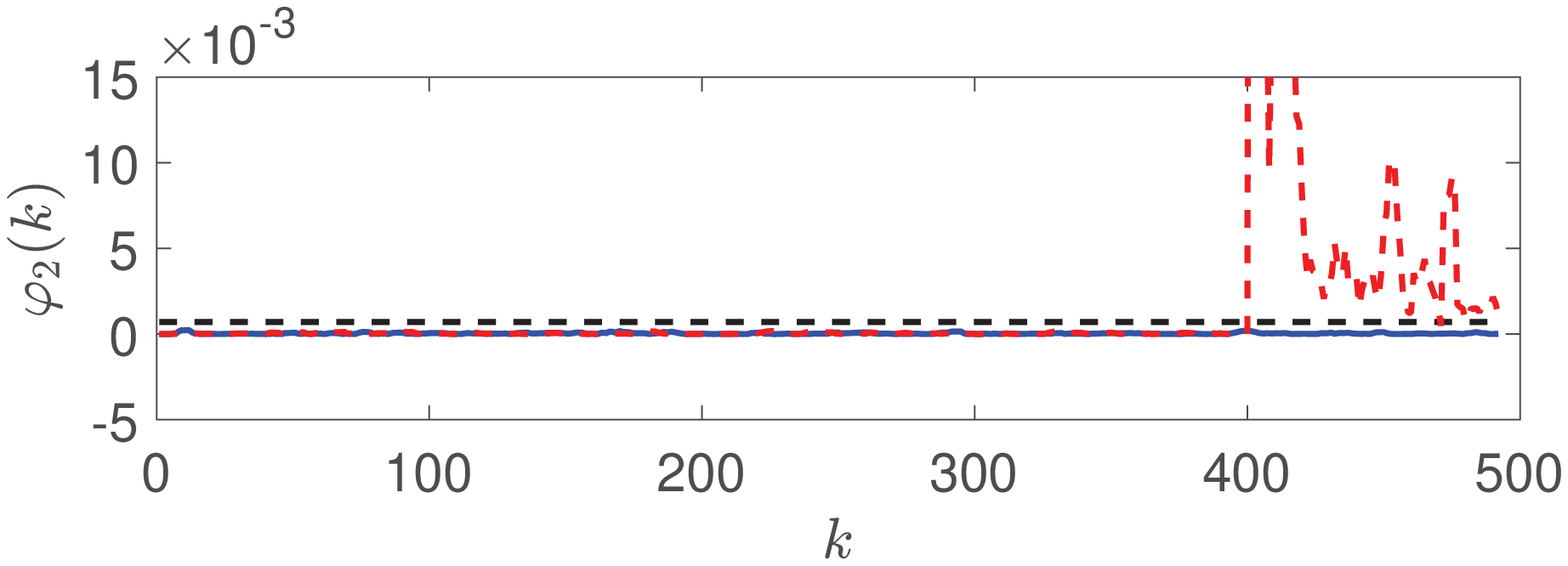}} \\
  \caption{States and the new DW statistical tests of NIPVSSs under the FDIA emerging at $k \geqslant 400$ based on the new DW scheme with $\sigma^2_{w_{y,i}} = 0.0001$. Blue line: Normal system. Red line: System under FDIA. Black line: Detection thresholds. ``OFF'' denotes that the servo is put off.}
  \label{figB2}
\end{figure}

Figs.~\ref{figB1} and \ref{figB2} present experimental results of applying the FDIA at $k \geqslant 400$ to NIPVSSs based on the conventional and the new DW schemes with $\sigma^2_{w_d}=\sigma^2_{w_{y,i}}=0.0001$, respectively.

\subsection{States of NIPVSSs with Different Watermarking Densities under no Attacks}
\begin{figure}[!t]
  \centering
  \subfigure[Cart position with $\sigma^2_{w_d} = \sigma^2_{w_{y,i}} = 0.0001$]{\includegraphics[width=0.48\textwidth]{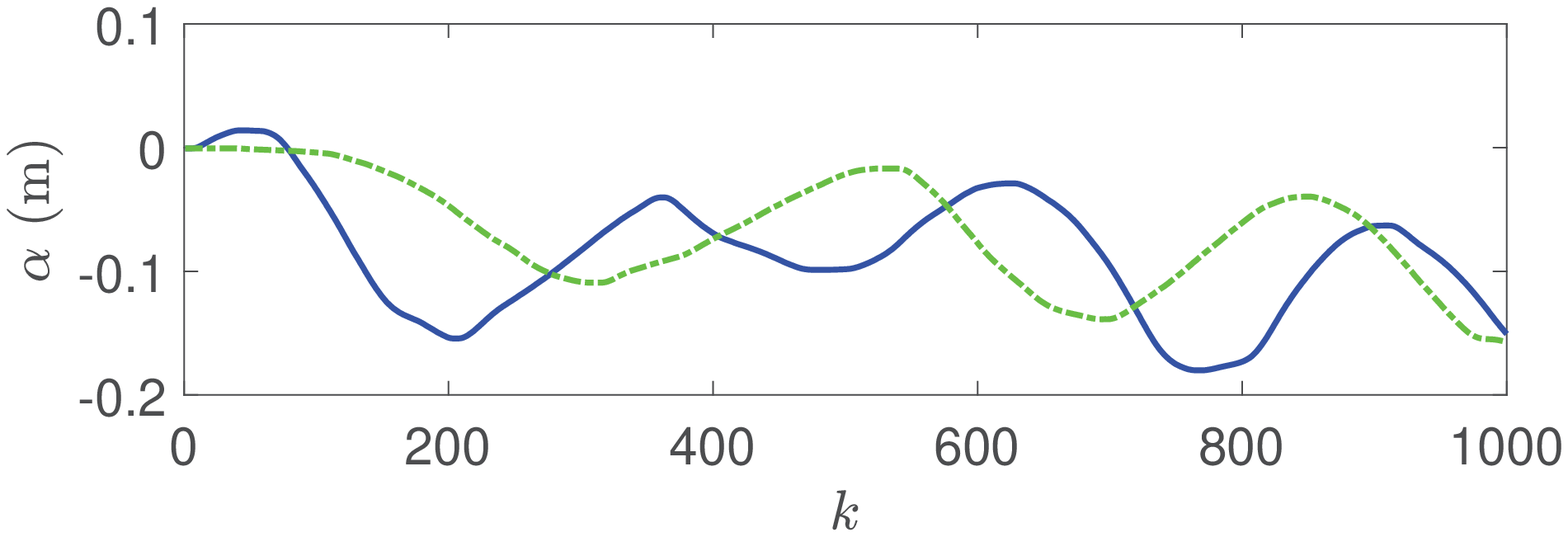}} \\ \vspace{-0.15in}
  \subfigure[Pendulum angle with $\sigma^2_{w_d} = \sigma^2_{w_{y,i}} = 0.0001$]{\includegraphics[width=0.48\textwidth]{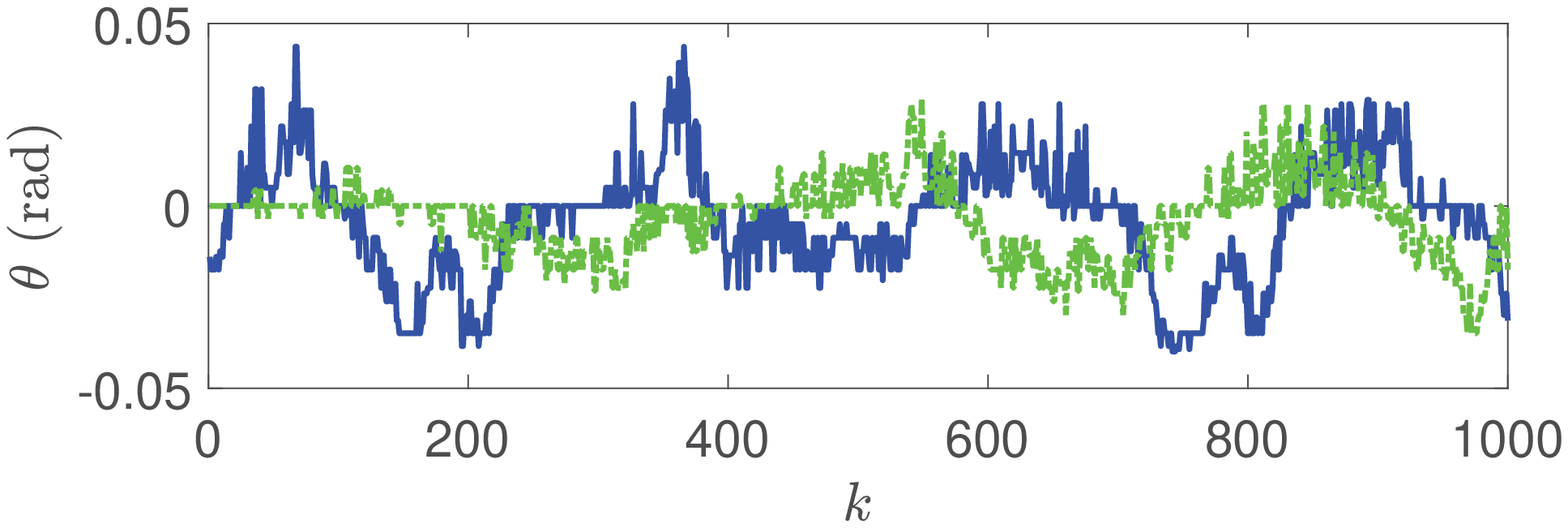}} \\ \vspace{-0.15in}
  \subfigure[Cart position with $\sigma^2_{w_d} = \sigma^2_{w_{y,i}} = 10$]{\includegraphics[width=0.48\textwidth]{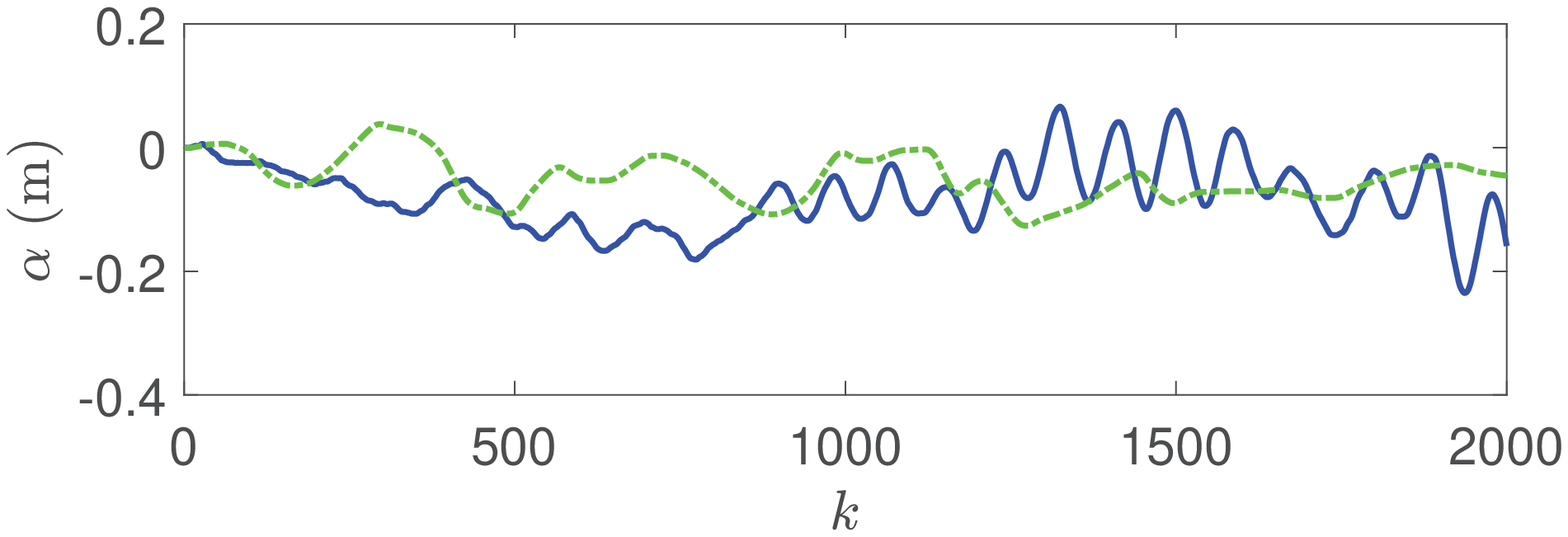}} \\ \vspace{-0.15in}
  \subfigure[Pendulum angle with $\sigma^2_{w_d} = \sigma^2_{w_{y,i}} = 10$]{\includegraphics[width=0.48\textwidth]{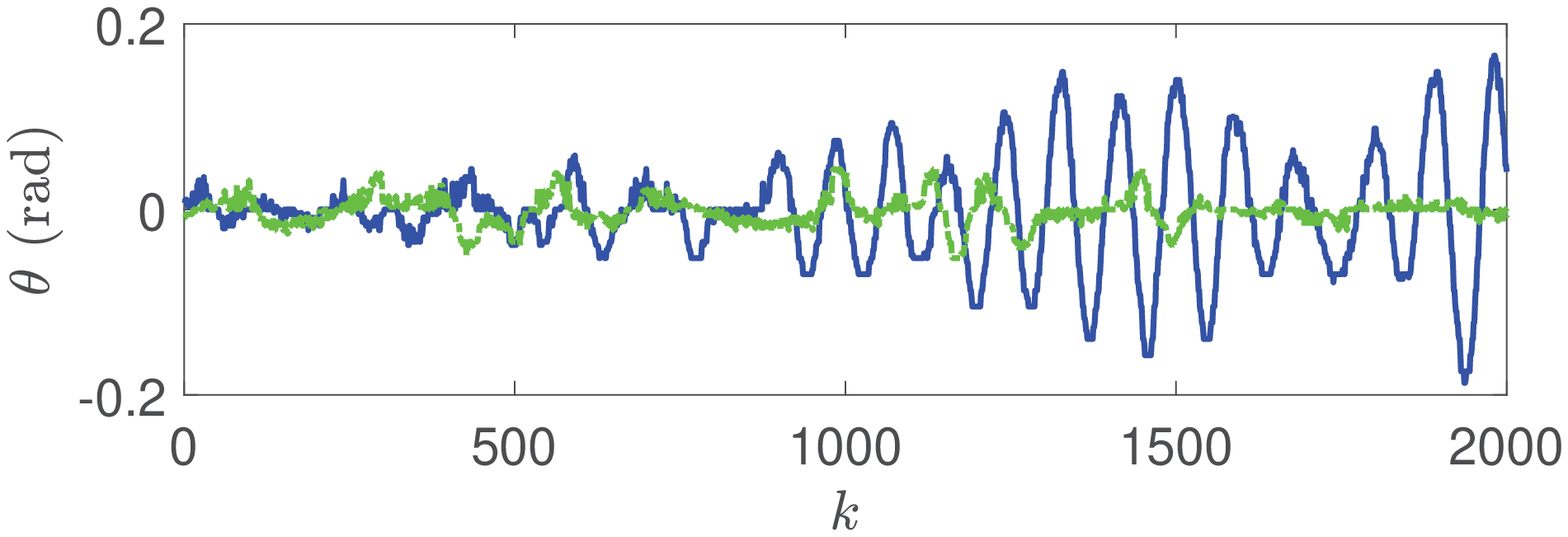}} \\
  \caption{State of the NIPVSS under no attack with different watermark densities. Blue line: System based on conventional DW scheme. Green line: System based on new DW scheme.}
  \label{figC}
\end{figure}

The cart position and pendulum angle are illustrated in Fig.~\ref{figC}, where $\sigma^2_{w_d}=\sigma^2_{w_{y,i}}=0.0001$ for (a) and (b), and $\sigma^2_{w_d}=\sigma^2_{w_{y,i}}=10$ for (c) and (d).

\subsection{Analysis of the Compensated Variable}
\begin{figure}[!t]
  \centering
  \includegraphics[width=0.488\textwidth]{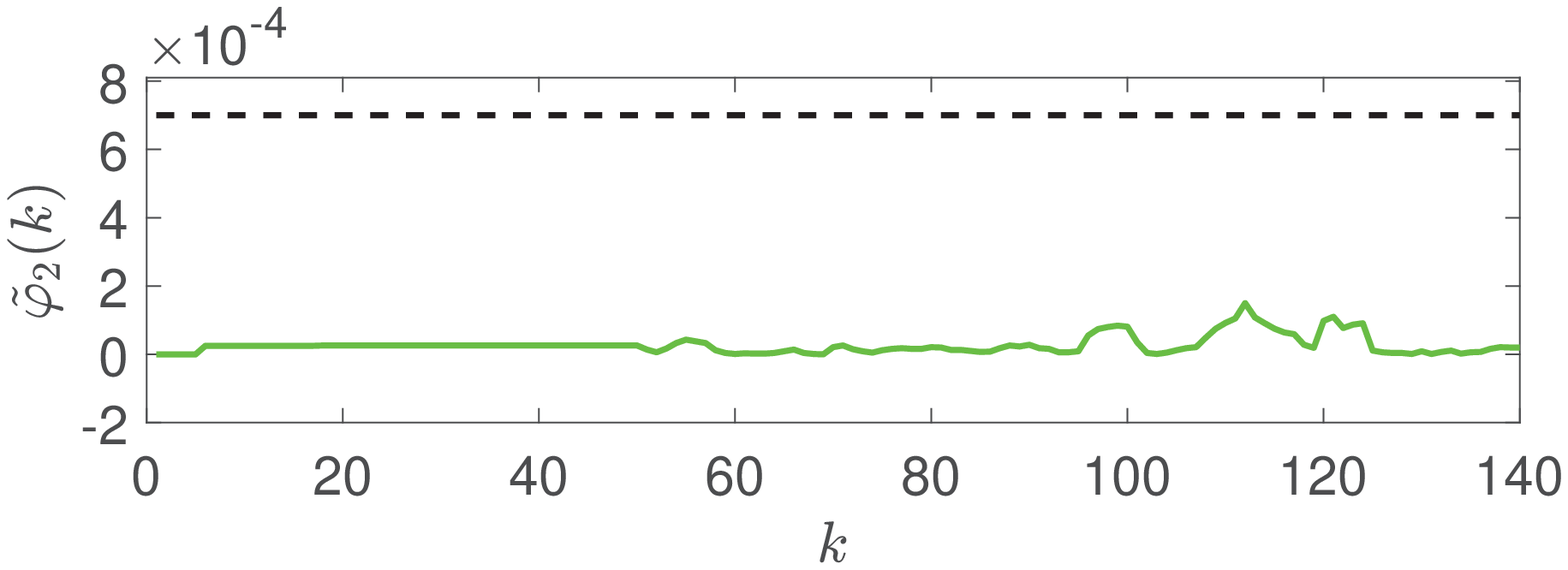}
  \caption{Values of ${{\tilde \varphi }_2}(k)$ on $\tilde r(k)$. Green line: ${{\tilde \varphi }_2}(k)$. Black line: Detection threshold.}
  \label{figE}
\end{figure}

Fig.~\ref{figE} presents the value of compensated detection indicator $\tilde \varphi_2(k)$.
\ifCLASSOPTIONcaptionsoff
  \newpage
\fi